\documentclass[11pt, a4paper, oneside]{Thesis} 

\graphicspath{{Pictures/}} 

\usepackage[square,sort,compress,numbers]{natbib}
\usepackage{cite}
\usepackage{mathtools}

\usepackage{pifont}
\usepackage{cancel}
\usepackage{perpage}
\usepackage{amsmath,latexsym} 
\usepackage{subfigure}
\usepackage{float}
\usepackage{mathpazo}
\usepackage{amsfonts}
\usepackage{bigints} 
\usepackage{mathptmx}
\usepackage{bm}
\usepackage{palatino}
\usepackage{textcomp}
\usepackage{booktabs}
\usepackage{dcolumn}
\usepackage{ragged2e}
\usepackage{hyperref}
\title{\ttitle} 
\DeclareUnicodeCharacter{2212}{-}
\DeclareUnicodeCharacter{2212}{+}

\begin{document}

\setstretch{1.3} 

\fancyhead{} 
\rhead{\thepage} 
\lhead{} 

%

\thesistitle{Study of Cosmic Acceleration of the Universe in the Presence of Bulk Viscous Matter}
\documenttype{\textbf{THESIS}}
\supervisor{\textbf{Prof. Pradyumn Kumar Sahoo}}
\supervisorposition{\textbf{Professor}}
\supervisorinstitute{\textbf{BITS, Pilani, Hyderabad Campus}}
\examiner{}
\degree{Ph.D. Research Scholar}
\coursecode{\textbf{DOCTOR OF PHILOSOPHY}}
\coursename{Thesis}
\authors{\textbf{DHEERAJ SINGH RANA}}
\IDNumber{2021PHXF0451H}
\addresses{}
\subject{}
\university{\texorpdfstring{\href{http://www.bits-pilani.ac.in/} 
                {Birla Institute of Technology and Science, Pilani}} 
                {Birla Institute of Technology and Science, Pilani}}
\UNIVERSITY{\texorpdfstring{\href{http://www.bits-pilani.ac.in/} 
                {BIRLA INSTITUTE OF TECHNOLOGY AND SCIENCE, PILANI}} 
                {BIRLA INSTITUTE OF TECHNOLOGY AND SCIENCE, PILANI}}



\department{\texorpdfstring{\href{http://www.bits-pilani.ac.in/pilani/Mathematics/Mathematics} 
                {Mathematics}} 
                {Mathematics}}
\DEPARTMENT{\texorpdfstring{\href{http://www.bits-pilani.ac.in/pilani/Mathematics/Mathematics} 
                {Mathematics}} 
                {Mathematics}}
\group{\texorpdfstring{\href{Research Group Web Site URL Here (include http://)}
                {Research Group Name}} 
                {Research Group Name}}
\GROUP{\texorpdfstring{\href{Research Group Web Site URL Here (include http://)}
                {RESEARCH GROUP NAME (IN BLOCK CAPITALS)}}
                {RESEARCH GROUP NAME (IN BLOCK CAPITALS)}}
\faculty{\texorpdfstring{\href{Faculty Web Site URL Here (include http://)}
                {Faculty Name}}
                {Faculty Name}}
\FACULTY{\texorpdfstring{\href{Faculty Web Site URL Here (include http://)}
                {FACULTY NAME (IN BLOCK CAPITALS)}}
                {FACULTY NAME (IN BLOCK CAPITALS)}}

\maketitle

\clearpage
\setstretch{1.3} 

\pagestyle{empty} 
\pagenumbering{gobble}

\addtocontents{toc}{\vspace{2em}} 
\frontmatter 
\Certificate
\Declaration
\begin{acknowledgements}
I would like to express my deepest gratitude to my supervisor, \textbf{Prof. Pradyumn Kumar Sahoo}, Professor, Department of Mathematics, BITS, Pilani, Hyderabad Campus, Hyderabad, Telangana. 

I sincerely thank my Doctoral Advisory Committee (DAC) members, \textbf{Prof. Bivudutta Mishra} and \textbf{Prof. Nirman Ganguly}, for their valuable suggestions and constant encouragement to improve my research works.

It is my privilege to thank HoD, DRC convener, Former DRC convener, faculty members, my colleagues and the Department of Mathematics staff for supporting this amazing journey of my Ph.D. career.

  The author acknowledges \textbf{BITS, Pilani, Hyderabad Campus}, for providing me with the necessary facilities and the \textbf{University Grant Commission (UGC), India}, for providing a Research Fellowship (UGC Reference No. 211610106591) to carry out my research work.

I would like to express my heartiest thanks to my parents \textbf{Mrs. Sumitra Rana} and \textbf {Mr. Vijay Singh Rana} for their unconditional love and support.

I would like to express my sincere thanks and gratitude to \textbf{Raja Solanki}, \textbf{Sanjay Mandal}, \textbf{Zinnat Hassan}, \textbf{Lakhan}, \textbf{Moreshwar} and \textbf{Kartik} for their support.

\vspace{1.4 cm}
Dheeraj Singh Rana,\\
ID: 2021PHXF0451H
\end{acknowledgements}

\begin{abstract} 
In this thesis, we investigated the role of bulk viscous matter in cosmic evolution of the Universe by considering different forms of the bulk viscosity coefficient in the framework of modified theories of gravity.

In chapter \ref{Chapter1}, we provide a concise overview of the basics of General Relativity (GR), which is based on Riemannian geometry and its alternative formulations based on non-Riemannian geometry, namely the "Teleparallel Equivalent of GR (TEGR)" and the "Symmetric Teleparallel Equivalent of GR (STEGR)". Furthermore, we discuss the benefits and drawbacks of the standard model of cosmology. Further, we briefly discuss the modified theories of gravity and the introduction of the bulk viscosity in cosmology. Lastly, we briefly discuss dynamical system analysis.

In chapter \ref{Chapter2}, we investigate the role of bulk viscosity in understanding the accelerated expansion of the Universe within the framework of $f(Q)$ gravity. We considered a power law $f(Q)$ model with a bulk viscosity coefficient $\zeta= \zeta_1 H + \zeta_0 \rho H^{-1}$ and obtained an analytic solution for the corresponding field equations. Further, we obtained the constraint values of the free parameters using the combined CC+Pantheon+SH0ES dataset. In addition, we explore the evolutionary behavior of significant cosmological parameters. The effective equation of state
(EoS) parameter predicts the accelerating behavior of the cosmic expansion phase. Furthermore, by the statefinder and $Om(z)$ diagnostic test, we found that our viscous model favors quintessence-type behavior.

In chapter \ref{Chapter3}, we analyze the viscous fluid cosmological model within the framework of $f(Q)$ gravity in the presence of viscosity. We considered a power law $f(Q)$ model with a bulk viscosity coefficient $\zeta = \zeta_0 \sqrt{\Omega} + \zeta_1 \Omega H $ and obtained an analytic solution for field equations. Additionally, we used the combined CC+Pantheon dataset to obtain the estimated values of the free parameters in the analytic solution. Further, we use dynamical system analysis to analyze the asymptotic behavior of our model.

In chapter \ref{Chapter4}, we analyze the viscous fluid cosmological model within the framework of $f(Q)$ gravity in the presence of viscosity. We considered a linear $f(Q)$ model along with a bulk viscosity coefficient $\zeta =\zeta_{0}+\zeta_{1}\left( \frac{\dot{a}}{a}\right) +\zeta_{2}\left( \frac{{
\ddot{a}}}{\dot{a}}\right) =\zeta_{0}+\zeta_{1}H+\zeta_{2}\left( \frac{\dot{H}}{
H}+H\right) $   and obtained an analytic solution for the field equations. In addition, we obtained the estimated values of the free parameters in the analytic solution using the combined $H(z)$+Pantheon+BAO observational dataset. Lastly, we use dynamical system analysis to analyze the asymptotic behavior of  a linear $f(Q)$ model and a non-linear model, specifically $f(Q)=-Q+\beta Q^2$.

In chapter \ref{Chapter5}, we used the dynamical system method to analyze the asymptotic behavior of viscous fluid cosmological models in the framework of $f(Q)$ gravity with a bulk viscosity coefficient $\zeta = \Bar{\zeta}_0 {\Omega^s_m}H $,
where $\bar{\zeta}_0 = \frac{\zeta_0}{{\Omega^s_{m_0}}} $. In particular, we considered two different models, $f(Q)=-Q+\alpha Q^{-1}$ and $f(Q)=-Q+\alpha Q^{2}$ and we found that $f(Q)=-Q+\alpha Q^{-1}$ describes the late-time accelerated expansion of the Universe without the transition phase and, on the other hand, $f(Q)=-Q+\alpha Q^{2}$ gives a better description of matter and the radiation era along with the transition phase.

Chapter \ref{Chapter6} offers a brief overview of the key results of this thesis and suggests possible avenues for future investigation.
\end{abstract}

\Dedicatory{\bf \begin{LARGE}
Dedicated to
\end{LARGE} 
\\
\vspace{0.2cm}
\it My Family Members}



\lhead{\emph{Contents}} 
\tableofcontents 
\addtocontents{toc}{\vspace{1em}}
\addtocontents{toc}{\vspace{1em}}
\lhead{\emph{List of Tables}}
\listoftables 
\addtocontents{toc}{\vspace{1em}}
\lhead{\emph{List of Figures}}
\listoffigures 
\addtocontents{toc}{\vspace{1em}}



\lhead{\emph{List of symbols and Abbreviations}}
 
\listofsymbols{ll}{
\begin{tabular}{ll}
$g_{\mu\nu}$: & Metric tensor \\
$g$: & Determinant of $g_{\mu\nu}$ \\
$\tilde{\Gamma}^\lambda_{\mu\nu}$: & General affine Connection \\
$\Gamma^\lambda_{\mu\nu}$: & Levi-Civita connection \\
$\overset{\circ}{\Gamma}{}^\lambda_{\mu\nu}$: & Teleparallel connection \\
$\mathring{\Gamma}^\lambda_{\mu\nu}$: & Symmetric teleparallel connection \\
$\nabla_\mu$: & Covariant derivative w.r.t. Levi-Civita connection \\
$\tilde{\nabla}_\mu$: & Covariant derivative w.r.t. general affine connection \\
$\overset{\circ}{\nabla}_\mu$: & Covariant derivative w.r.t. teleparallel connection \\
$\mathring{\nabla}_\mu$: & Covariant derivative w.r.t. symmetric teleparallel connection \\
$(\mu\nu)$: & Symmetrization over the indices $\mu$ and $\nu$ \\
$[\mu\nu]$: & Anti-symmetrization over the indices $\mu$ and $\nu$ \\
$\zeta$: & Bulk viscosity coefficient \\
GR: & General Relativity \\
$\Lambda$CDM: & $\Lambda$ Cold Dark Matter \\
EoS: & Equation of State \\
SN Ia: & Type Ia Supernovae \\
CC: & Cosmic Chronometer\\
CMB: & Cosmic Microwave Background \\
BAO: & Baryon Acoustic Oscillations \\
DE: & Dark Energy \\
MCMC: & Markov Chain Monte Carlo \\
\end{tabular}}

\addtocontents{toc}{\vspace{2em}}

%
%


\clearpage 





\mainmatter 

\pagestyle{fancy} 


\chapter{Preliminaries} 
\label{Chapter1}

\lhead{Chapter 1. \emph{Preliminaries}} 

\clearpage
\pagebreak

 \section{The General Relativity}
The theory of GR formulated by Albert Einstein in 1915, represents a profound extension of the special theory of relativity by incorporating the effects of gravitation into the fabric of spacetime. Unlike Newtonian gravity which treats gravity as a force acting instantaneously at a distance, GR describes it as a manifestation of the curvature of spacetime caused by the distribution of mass and energy. In this framework, 
massive bodies distort the geometry of spacetime and other objects move along the resulting curved paths known as geodesics. Essentially, GR is built on the Equivalence Principle asserting that, in a local frame gravitational effects cannot be distinguished from those arising due to acceleration. This principle leads to the realization that spacetime is dynamic and responsive to the presence of matter and energy, as mathematically expressed by the Einstein field equations, 
\begin{equation}\label{9}
G_{\mu\nu}\equiv R_{\mu\nu}-\frac{1}{2}g_{\mu\nu}R=\frac{8\pi G}{c^4} {T_{\mu\nu}}.
\end{equation}
where $G_{\mu \nu}$, $R_{\mu\nu}$, $R$ and $T_{\mu \nu}$ denote the Einstein tensor,  the Riemannian tensor, the Ricci scalar and the energy-momentum tensor of matter respectively. These relations constitute the foundation of the theory, establishing a direct connection between spacetime geometry and physical matter–energy.
 GR has been remarkably successful in explaining and predicting a wide range of astrophysical and cosmological phenomena, including the precession of Mercury’s orbit, gravitational redshift, deflection of light by massive bodies and the existence of gravitational waves, which were detected directly a century after Einstein’s prediction. Moreover, GR provides the theoretical foundation for modern cosmology, leading to the development of models that describe the expansion of the Universe, the formation of black holes, and the large-scale structure of spacetime.
Despite its success, GR faces limitations in extreme regimes, such as at singularities or the Planck scale, where quantum effects become significant. This motivates ongoing research into quantum gravity and modified gravitational theories, aimed at unifying GR with quantum mechanics and extending our understanding of the Universe at both cosmological and microscopic scales.

  \subsection{Basic mathematical elements}
GR is a geometric framework that can be expressed in terms of invariant structures  defined on a smooth differentiable manifold $\mathcal{M}$, which subsequently represents the spacetime manifold. The structure of the manifold enables the introduction of local coordinate systems and the use of differential calculus in the neighborhood of each point on the manifold. The class of manifolds relevant in this context is Lorentzian manifolds, which are locally equivalent to a flat Minkowski spacetime.
In GR the manifold is taken to be $4$-dimensional. however, for the sake of generality, we consider a manifold with an arbitrary number of dimensions. The key geometric entities are tensors, defined as multi-linear maps belonging to the tensor product spaces constructed from the tangent and cotangent spaces at each point $p$ of the manifold. The space for a rank $(m,n)$ tensor is, 
\begin{equation}
    T^{(m,n)}_p(\mathcal{M})= \underbrace{T_p(\mathcal{M})\otimes T_p(\mathcal{M})\otimes...\otimes T_p(\mathcal{M})}_{m\,\,times}\otimes\underbrace{T_p^*(\mathcal{M})\otimes...\otimes T_p^*(\mathcal{M})}_{n\,\,times} \text{.}
\end{equation}
where $T_p(\mathcal{M})$ and $T_p^*(\mathcal{M})$ are the tangent and cotangent spaces at any point $p$ on the manifold, respectively. The set of all such tensor product spaces at every point  $p\in \mathcal{M}$ forms the tensor bundle 
$T^{(m,n)}_p(\mathcal{M})$, and the tensor fields arise as sections of this bundle.
Therefore, tensor fields are genuine geometric entities whose definitions are independent of any particular choice of coordinates. Nevertheless, for practical calculations, it is often convenient to represent these objects using coordinate-based expressions. In a local coordinate system $x^{\mu}$, a tensor of rank $(m,n)$ may be represented as
\begin{equation}
    T^{(m,n)}_p(\mathcal{M})=T^{\mu_1\mu_2...\mu_m}_{\,\,\,\,\,\,\,\,\,\,\,\,\,\,\,\,\,\,\,\,\,\,\,\,\,\,\,\,\nu_1\nu_2...\nu_n}\frac{\partial}{\partial x^{\mu_1}}\otimes \frac{\partial}{\partial x^{\mu_2}}\otimes...\otimes\frac{\partial}{\partial x^{\mu_m}}\otimes dx^{\nu_1}\otimes dx^{\nu_2}\otimes ...\otimes dx^{\nu_n} .
\end{equation}
where the symbols $\frac{\partial}{\partial x^{\mu}}$ and $dx^{\nu}$ denote the vector and one-form coordinate bases, respectively.\\
A key feature of geometric objects is the manner in which their components transform under a change from one coordinate system 
${x^{\mu}}$ to another ${y^{\mu}}$. Under a coordinate transformation $x^{\mu}\to\hat{x}^{\mu}(x)$, the components of a tensor obey the following transformation law,
\begin{equation}
T^{\mu_1\mu_2...\mu_m}_{\,\,\,\,\,\,\,\,\,\,\,\,\,\,\,\,\,\,\,\,\,\,\,\,\,\,\,\,\nu_1\nu_2...\nu_n}(x)\longmapsto \hat{T}^{\mu_1\mu_2...\mu_m}_{\,\,\,\,\,\,\,\,\,\,\,\,\,\,\,\,\,\,\,\,\,\,\,\,\,\,\,\,\nu_1\nu_2...\nu_n}(\hat{x})=\frac{\partial \hat{x}^{\mu_1}}{\partial x^{\alpha_1}}...\frac{\partial \hat{x}^{\mu_m}}{\partial x^{\alpha_m}} \frac{\partial x^{\beta_1}}{\partial \hat{x}^{\nu_1}}...\frac{\partial x^{\beta_n}}{\partial \hat{x}^{\nu_n}}T^{\alpha_1\alpha_2...\alpha_m}_{\,\,\,\,\,\,\,\,\,\,\,\,\,\,\,\,\,\,\,\,\,\,\,\,\,\,\,\,\beta_1\beta_2...\beta_n}(x).
\end{equation}
Combined with the relevant transformation rules for their bases, the object $T^{(m,n)}$ remains invariant under coordinate changes.

\subsection{Metric tensor}
A fundamental component in the geometry of curved spacetime is the metric tensor, which provides a way to define the squared interval between infinitely close points. Consequently, the metric tensor $g_{\mu\nu}$, which is a symmetric rank-
$(0,2)$ tensor, together with its inverse $g^{\mu\nu}$ of rank $(2,0)$, must be non-degenerate. In four dimensions, this symmetry implies that the metric possesses ten independent components. The metric tensor establishes an inner product in the tangent space $T_p(\mathcal{M})$ at every point $P$ in the manifold $\mathcal{M}$.
For any two vectors $U,V\in T_p(\mathcal{M})$, the metric acts as a map $g(V,U)\in \mathbb{R}$, defining their inner product, which is symmetric so that $g(U,V)=U.V=V.U=g(V,U)$. This map can be viewed as a generalization of the Euclidean dot product, providing a natural way to define distances on the manifold. The metric, often referred to as the spacetime interval, can be expressed in local coordinates as follows,
\begin{equation}
    ds^2=g_{\mu\nu}\,dx^{\mu}\,dx^{\nu}.
\end{equation}
Moreover, the metric tensor $g_{\mu\nu}$ can be represented as a matrix, whose determinant is given by $g=det(g_{\mu\nu})$.\\
The magnitude or norm of a vector $V^{\mu}$ is obtained by taking its inner product with it and is defined as
$l^2=g_{\mu\nu}\,V^{\mu}\,V^{\nu}$. A spacetime vector $V^{\mu}$ is classified as 
\begin{equation}
    g_{\mu\nu}\,V^{\mu}\,V^{\nu} \begin{cases}
        <0,\,\,\,\,  V^{\mu}\,\, \text{is timelike}, \\
        =0,\,\,\,\,  V^{\mu}\,\, \text{is null or lightlike},\\
        >0,\,\,\,\,  V^{\mu}\,\, \text{is spacelike.}
    \end{cases}
\end{equation}

Another important tensor is the Kronecker delta, written as \(\delta^{\mu}_{\sigma}\), which is a rank-
\((1,1)\) tensor. It can be expressed in terms of the metric tensor through the relation \(g^{\mu\nu}g_{\nu\sigma}=g_{\sigma\nu}g^{\nu\mu} = \delta^{\mu}_{\sigma}\).

 \subsection{Metric: The spacetime line element}
In GR, points in curved spacetime are located using a $4$-coordinate system $x^\mu$. The line element represents the spacetime interval for two points with coordinate separations $dx^{\mu}$ and the line element is defined as follows,
\begin{equation}\label{e25}
 ds^2 = g_{\mu\nu}(x) dx^\mu dx^\nu \text{,}
\end{equation}
Here, $g_{\mu\nu}(x)$ is a metric or a metric tensor, which is a $4 \times 4$ symmetric matrix  that depends on the coordinates. A metric tensor that is positive definite is described as a Euclidean or  Riemannian metric and the associated geometry is termed Euclidean or  Riemannian. However, a metric with a negative time component and positive spatial components is known as a Lorentzian or pseudo-Riemannian metric.
The mathematical description of spacetime in modern physics particularly in relativity relies on a metric with a Lorentzian signature. A fundamental property of a curved spacetime is that there is no coordinate system in which the metric $g_{\mu\nu}(x)$ is equal to the flat Minkowski metric 
$\eta_{\mu\nu}(x)$ globally. However, at any given point $P$ in spacetime, it is possible to define a coordinate system $x^{\mu'}$ as follows,
\begin{equation}\label{e26}
g_{\mu\nu}(x^{\mu'}_P)= \eta_{\mu\nu} = diag(-1,1,1,1) \:\: \text{and} \:\: \left( \frac{\partial g_{\mu\nu}}{\partial x^{\mu'}} \right)_{x=x_P}= 0 .
\end{equation}
This coordinate system is known as the local inertial frame at the point 
$P$. Einstein's key realization was that, according to the equivalence principle gravity's effects can be removed within a sufficiently small neighborhood of spacetime. Within this local region where gravitational effects vanish, GR coincides with special relativity and preserves the same local light-cone structure at each point, for instance $ds^2 > 0$ denotes spacelike intervals, $ds^2 < 0$ timelike intervals and $ds^2 = 0$ corresponds to lightlike trajectories.

 \subsection{The Levi-Civita connection}
A manifold $\mathcal{M}$ models a curved spacetime geometry that possess a complex global topology. However, it is defined by the property that in the vicinity of any point, its structure is locally equivalent to that of the Euclidean space $\mathbb{R}^n$. Formally, for a given open subset $S\subset \mathcal{M}$, there exists a homeomorphism $\phi :S \xrightarrow{}R^n$ that maps $S$ to an open set within $\mathbb{R}^n$. The pair $(S,\phi)$ is called a chart or coordinate system. A smooth atlas on a manifold $M$ is a family of charts $\{ (S_i,\phi_i)\}_{i\in I}$ such that the domains $\{S_i\}_{i\in I}$ form an open cover of $M$, and the charts are compatible with $C^\infty$. This compatibility requires that for every non-empty intersection $S_i \cap S_j$, the transition map $\phi_i \circ \phi_j^{-1}$, which connects the two coordinate systems on the overlap, is an infinitely differentiable $(C^\infty)$ function between the open sets $\phi_j(S_i \cap S_j) $ and $ \phi_i(S_i \cap S_j)$. A set M equipped with a maximal smooth atlas is defined as a smooth $n$-dimensional manifold. This mathematical structure formally captures the notion of a space that is locally modeled on $\mathbb{R}^n$ \cite{B2}.

To define a meaningful derivative on a manifold, we introduce the concept of parallel transport. This is the process of moving a vector $A^\mu$ from a point 
 $P(x^\mu)$ to a neighboring point $Q(x^\mu+\delta x^\mu)$ in such a way that the vector remains locally parallel to itself. If $dA^\mu$  represents the net change in the components of vector and $\delta A^\mu$ is the change attributable solely to parallel transport, then we obtain the following relation,
\begin{equation}\label{e27}
dA^\mu = A^\mu(x^\mu+\delta x^\mu) - A^\mu(x^\mu) \approx  A^\mu(x^\mu) + \delta x^\eta \frac{\partial A^\mu}{\partial x^\eta}  - A^\mu(x^\mu) = \delta x^\eta \frac{\partial A^\mu}{\partial x^\eta}\text{,}
\end{equation}
and 
\begin{equation}\label{e28}
\delta A^\mu = - \Gamma^\mu_{\sigma \eta} A^\sigma \delta x^\eta \text{.} 
\end{equation}

When a vector $A^\mu$ is parallel transported along a path, the physical change it undergoes is not simply its coordinate differential $dA^\mu$, but rather that value corrected by the transport adjustment $\delta A^\mu $, resulting in a net change of $dA^\mu - \delta A^\mu $. The covariant derivative of $A^\mu$  is defined as the limit of its change under parallel transport as $\delta x^\eta $ approaches zero, expressed by the following relation,
\begin{equation}\label{e29}
 \nabla_\eta A^\mu = \frac{\partial A^\mu}{\partial x^\eta} + \Gamma^\mu_{\sigma \eta} A^\sigma = \partial_\eta A^\mu + \Gamma^\mu_{\sigma \eta} A^\sigma  \text{.}
\end{equation}
In the geometry of curved spacetime, the components $\Gamma^\mu_{\sigma \eta}$ are known as Christoffel symbols, which collectively form the Levi-Civita connection essential to describe how vectors change along a manifold. An affine connection that is both symmetric $(\Gamma^\mu_{\sigma \eta} = \Gamma^\mu_{\eta \sigma })$ and metric-compatible $(\nabla_\eta g_{\mu\nu} = 0)$
is termed a Riemannian connection, and the corresponding geometric framework is known as Riemannian geometry. With the help of these two conditions, we can derive the Levi-Civita connection in terms of the metric tensor as follows, 
\begin{equation}\label{e30}
 \Gamma_{\mu \sigma \eta} = \frac{1}{2} \left[\partial_\mu g_{\sigma \eta} + \partial_\sigma g_{\eta \mu} - \partial_\eta g_{\mu \sigma} \right]  \text{.}
\end{equation}

 \subsection{Einstein field equations}
The affine connection provides the necessary structure to determine the spacetime curvature, which is described by the Riemann curvature tensor,
\begin{equation}\label{e31}
R^{\rho}_{\sigma\mu\nu}=\partial_{\mu}\Gamma^{\rho}_{\nu\sigma}-\partial_{\nu}\Gamma^{\rho}_{\mu\sigma}+\Gamma^{\rho}_{\mu\lambda} \Gamma^{\lambda}_{\nu\sigma}-\Gamma^{\rho}_{\nu\lambda} \Gamma^{\lambda}_{\mu\sigma} \text{.}
\end{equation}
Furthermore, the Riemann curvature tensor satisfies the Bianchi identity given as follows,
\begin{equation}\label{e32}
\nabla_\alpha R^{\rho}_{\sigma\mu\nu} + \nabla_\mu R^{\rho}_{\sigma\nu\alpha} + \nabla_\nu R^{\rho}_{\sigma\alpha\mu} = 0 \text{.}
\end{equation}
The condition for a Riemannian space to be flat or curvature-free is the vanishing of the Riemann curvature tensor i.e. $R^{\rho}_{\sigma\mu\nu}=0$.
The Ricci tensor is obtained by contracting the Riemann curvature tensor and is expressed as follows,
\begin{equation}\label{e33}
R_{\sigma\mu} = R^{\nu}_{\sigma\mu\nu} \text{.}
\end{equation}
A subsequent contraction of the Ricci tensor gives rise to the Ricci curvature scalar, which is defined as follows,
\begin{equation}\label{e34}
R = g^{\sigma\mu} R_{\sigma\mu} \text{.}
\end{equation}
In the framework of GR, A central object in describing spacetime geometry is the Einstein tensor, defined  as follows.
\begin{equation}\label{e35}
G_{\mu\nu} \equiv R_{\mu\nu}-\frac{1}{2}g_{\mu\nu}R \text{.} 
\end{equation}
Furthermore, the Bianchi identity implies that the covariant divergence of the Einstein tensor vanishes,
\begin{equation}\label{e36}
\nabla^\mu G_{\mu\nu}  = 0 \text{.}
\end{equation}
According to relativity, mass is fundamentally a type of energy (rest energy). Since energy and momentum are interrelated they can transform into each other simply by changing the reference frame of the observer. Within the framework of relativity, the mass density $\rho$ is represented in a more general form through the energy–momentum tensor, which is denoted by ${T_{\mu\nu}}$. This tensor is symmetric, resulting in only ten independent components. Among these, the time-time component $T_{00}$ corresponds to the density of matter-energy $\rho c^2$, while the diagonal components $T_{ii}$ represent pressure. The other off-diagonal components correspond to energy, momentum flux and momentum density as well as shear stresses within the medium.
It is important to note that the Newtonian formulation of the gravitational field is described by,
\begin{equation}\label{e37}
\nabla^2 \Phi = 4\pi G_N \rho \text{.}  
\end{equation}
In other terms, the Laplacian $\nabla^2$ of the gravitational potential $\Phi$ is proportional to the mass density $\rho$ i.e. $\nabla^2 \Phi \propto \rho$. As discussed previously, the Levi-Civita connection $\Gamma$ involves the first derivative of the metric tensor and consequently the Riemann curvature tensor expressed as $d\Gamma + \Gamma \Gamma$ contains second-order derivatives of the metric. This leads to a relativistic generalization of the Newtonian field equation maintaining a similar structure, the left-hand side is the Einstein tensor $G_{\mu\nu}$, which incorporates second derivatives of the metric tensor $g_{\mu\nu}$ and represents the relativistic gravitational potential while the right-hand side is the energy–momentum tensor ${T_{\mu\nu}}$, which describes the distribution of matter and energy. Hence, the Einstein field equation of GR is expressed as follows \cite{R26},
\begin{equation}\label{e38}
G_{\mu\nu} \propto {T}_{\mu\nu} \:\: \text{i.e.} \:\: G_{\mu\nu} = \kappa {T}_{\mu\nu} \text{.}
\end{equation}
Here, $\kappa = \frac{8\pi G}{c^4}$ serves as the proportionality or coupling constant. The Einstein field equations can also be derived by applying the variational principle to the Einstein-Hilbert action,
\begin{equation}\label{e39}
S=\frac{1}{2\kappa}\int R \sqrt{-g} d^4x +\int L_m \sqrt{-g} d^4x \text{.}
\end{equation}
In this expression, $\kappa = \frac{8\pi G}{c^4}$ denotes the coupling constant, $L_m$ represents the matter Lagrangian density and $d^4x\sqrt{-g}$ corresponds to the coordinate-invariant volume element in the $4$-dimensional spacetime manifold, where $g = \det(g_{\mu\nu})$.

 \section{$\Lambda$CDM cosmology}
The previous section explored the theoretical foundations of GR. We now proceed to explore how these formulations can be applied to explain different cosmological observations.

\subsection{Comoving distances}
According to the cosmological principle, the large-scale Universe can be modeled as a fluid with individual galaxies serving as its basic constituent particles. A fluid element is a region large enough to contain many galaxies, yet small enough to be considered a point relative to the total Universe.
Hence, the motion of a cosmic fluid element represents the averaged or collective motion of the galaxies contained within it. This concept provides the foundation for the comoving coordinate system, a reference frame that is carried along by the expansion of the Universe.
The comoving distance $x$ and the physical distance 
$r$ are related as follows,
\begin{equation}\label{e40}
 r = a(t) x \text{.}  
\end{equation}
where $a(t)$ denotes the cosmic scale factor which characterizes the expansion of the Universe. Let us consider a galaxy moving away with a recessional velocity 
$v$, then
\begin{equation}\label{e41}
v= \frac{dr}{dt}  = \dot{r} = \dot{a} x = \frac{\dot{a}}{a} a x = \frac{\dot{a}}{a} r \text{.}
\end{equation}
Comparing this expression with Hubble's Law $(v=Hr)$, we get the following equation,
\begin{equation}\label{e42}
 H = \frac{\dot{a}}{a}    
\end{equation}
Here, $H$ is known as the Hubble parameter. 
\subsection{FLRW Universe}
In the large-scale structure, the Universe can only take one of three forms which are both isotropic and homogeneous, known as open, closed or flat. For a flat Universe, the spatial curvature is zero at all points. A closed Universe possesses a constant positive curvature, while an open Universe is characterized by a constant negative curvature. The line element describing all three possible geometries for an isotropic and homogeneous Universe is given by,


\begin{equation}\label{e45}
ds^2=-c^2dt^2+a^2(t)\left[\frac{dr^2}{1-k r^2}+r^2(d\theta^2+sin^2\theta d\phi^2)\right]\text{.}
\end{equation}
The evolution of a homogeneous and isotropic Universe is governed by this line element through the change of scale factor $a(t)$. This line element is widely referred to as the Friedmann–Lemaître–Robertson–Walker (FLRW) metric \cite{R27,R28}. We now switch to a unit system where $c=1$. The parameter $k$ denotes the spatial curvature. An open Universe geometry is characterized by $k=−1$, a flat Universe by $k=0$, and a closed Universe by $k=+1$.

\subsection{Evolution of the cosmological constant ($\Lambda$)}
The components of Einstein's equation \eqref{e38} are derived from the line element \eqref{e45} as follows,
\begin{equation}\label{e46}
\left(\frac{\dot{a}}{a}\right)^2 + \frac{k}{a^2}=\frac{8\pi G}{3}\rho ,
\end{equation}
and
\begin{equation}\label{e47}
2\frac{\ddot{a}}{a} + \left(\frac{\dot{a}}{a}\right)^2 + \frac{k}{a^2} = - 8\pi G p \text{.}
\end{equation}
The two equations above can be combined to give the acceleration equation,
\begin{equation}\label{e48}
\frac{\ddot{a}}{a} = - \frac{4\pi G}{3}(\rho + 3p) \text{.} 
\end{equation}
When GR was first developed the dominant scientific belief was that the Universe was static, which means that the scale factor $a(t)$ remained constant over time. For a constant scale factor $a(t)$ the field equations imply the following,
\begin{equation}\label{e49}
\rho=-3p= \frac{3k}{8\pi G a^2} \text{.} 
\end{equation}
The requirement for a positive energy density $\rho$ leads to a negative pressure term. Alternatively, if the pressure is assumed to be zero $(p=0)$, the equations force the energy density to also be zero ($\rho = 0$). In either case, the resulting conclusions are physically inconsistent. To solve this problem, Einstein later introduced an additional term $\Lambda$ (cosmological constant) into his field equations. Hence, the modified set of equations that describe a static Universe are follows,
\begin{equation}\label{e50}
\left(\frac{\dot{a}}{a}\right)^2 + \frac{k}{a^2}=\frac{8\pi G}{3}\rho + \frac{\Lambda}{3},
\end{equation}
and
\begin{equation}\label{e51}
2\frac{\ddot{a}}{a} + \left(\frac{\dot{a}}{a}\right)^2 + \frac{k}{a^2} = - 8\pi G p + \Lambda \text{.}
\end{equation}
The cosmological constant 
$\Lambda$ was first introduced by Einstein in 1917 as a modification of his original field equations to obtain a static model of the Universe \cite{R29}. 
Einstein later rejected the cosmological constant, famously calling it his greatest blunder after astronomical observations in 1931 confirmed that the Universe was not static but expanding \cite{R30}. In 1967, Zel'dovich reintroduced the concept of the cosmological constant $\Lambda$, interpreting it in terms of vacuum fluctuations \cite{R31}. In 1987, Weinberg proposed the existence of a small but non-zero cosmological constant $\Lambda$ \cite{R32}. Ultimately, in 1998, the SN Ia observation of the accelerating expansion of the Universe revived interest in the cosmological constant $\Lambda$, identifying it as a possible form of dark energy (DE) responsible for this acceleration.

 \subsection{The standard cosmological model}
In an isotropic and homogeneous Universe, the energy-momentum tensor for a perfect fluid is given by the following,
\begin{equation}\label{e52}
{T}_{\mu\nu}=(\rho+p)u_{\mu}u_{\nu}+ p g_{\mu\nu} .
\end{equation}
with the four-velocity vector of the fluid given by $u_\mu = (1,0,0,0)$ in a moving coordinate system. The energy-momentum tensor satisfies a vanishing covariant divergence i.e. $\nabla^\mu {T}_{\mu\nu} = 0 $ implies
\begin{equation}\label{e53}
\dot{\rho}+3 \frac{\dot{a}}{a}(\rho+p)=0  . 
\end{equation}
Substituting the barotropic Equation of State (EoS) $p=\omega \rho$ into the above equation, we get 
\begin{equation}\label{e54}
\rho \propto a^{-3(1+\omega)} .
\end{equation}

It is important to note that the value of the EoS parameter $\omega$ determines the evolution of energy density, characterizing different eras of the Universe as shown below,
\begin{itemize}
\item $\omega=1/3 \implies \rho_r = \rho_{r_0}a^{-4}$  radiation era.
\item $\omega=0 \implies \rho_m = \rho_{m_0}a^{-3}$ matter era.
\item $\omega=-1 \implies \rho_\Lambda = constant$  cosmological constant case as DE.
\end{itemize}
Combining equation \eqref{e50} with the definitions for the total density $\rho = \rho_m + \rho_r$ and the DE density $\rho_\Lambda = \frac{\Lambda}{8 \pi G}$, we get
\begin{equation}\label{e55}
 H^2 = \frac{8 \pi G}{3} \left[ \rho_{m_0}a^{-3} + \rho_{r_0}a^{-4}  + \rho_{\Lambda} \right] - \frac{k}{a^2} \text{.} 
\end{equation}
Here, $\rho_m = \rho_b + \rho_{cdm}$.
 In this context $\rho_b$ represents the density of ordinary baryonic matter and $\rho_{cdm}$ is the density associated with cold dark matter. The dimensionless density parameters for the various components are defined below,
\begin{equation}\label{e56}
\Omega_{m_0} = \frac{\rho_{m_0}}{\rho_{crit_0}}, \:\: \Omega_{r_0} = \frac{\rho_{r_0}}{\rho_{crit_0}},  \:\: \Omega_{\Lambda} =\Omega_{\Lambda_0} = \frac{\rho_{\Lambda}}{\rho_{crit_0}}=\frac{\Lambda}{3H_0^2}, \:\: \text{and} \:\: \Omega_{k_0} = -\frac{k}{H_0^2} \text{.} 
\end{equation}
Here, $a_0=1$ (conventional assumption) and $\rho_{crit_0}=\frac{3H_0^2}{8 \pi G}$.
By substituting the scale factor-redshift relation $a^{-1} = 1+z$ into equation \eqref{e55}, we get
\begin{equation}\label{e57}
H^2 = H_0^2  \left[ \Omega_{m_0}(1+z)^{3} + \Omega_{r_0}(1+z)^{4}  + \Omega_{\Lambda_0} + \Omega_{k_0}(1+z)^2 \right] \text{.} 
\end{equation}
The model described above is commonly referred to as the standard cosmological model also known as the $\Lambda$CDM model. According to the Planck 2018 results \cite{R2}, the constraints on the free parameters of the standard model were determined as $H_0 = 67.4 \pm 0.5  \text{km/s/Mpc}$, $\Omega_{m_0} = 0.315 \pm 0.007$, and $\Omega_{k_0} = 0.001 \pm 0.002$, indicating that the Universe is spatially flat.

 \subsection{Drawbacks of the $\Lambda$CDM model}
The concepts of vacuum energy and the cosmological constant are frequently considered interchangeable in cosmological literature. But in quantum field theory the concept of vacuum energy holds deeper and more significance. The cosmological constant problem is highlighted by the immense difference between the vacuum energy density calculated from quantum field theory $\rho_\Lambda(\text{QFT}) = 3.7873 \times 10^{73} \:\: \text{GeV}^{4}$ and the value obtained from observation $\rho_\Lambda = \Omega_{\Lambda_0} \rho_{\text{crit}_0} = 3.7151 \times 10^{-47} \:\: \text{GeV}^{4}$. The discrepancy in these two values constitutes a major theoretical challenge. Furthermore, the cosmological parameters derived from the Planck base $\Lambda$CDM model show strong consistency with independent observations including Baryon Acoustic Oscillations (BAO), Type Ia supernovae (SN Ia) and several  galaxy lensing studies. Nevertheless, a minor inconsistency exists with the findings from the DE Survey, which focuses on galaxy clustering. Furthermore, a significant $3.6 : \sigma$ tension exists with local determinations of the Hubble constant, which generally indicate a higher value \cite{R2}.
The $\Lambda$CDM model serves as the standard cosmological model because it aligns precisely with a wide range of observations and is computationally straightforward. The standard $\Lambda$CDM model is built on three basic postulates \cite{R33},
\begin{itemize}
    \item The inflationary scenario is typically described by a slow-rolling scalar field with minimal coupling.  
     \item Cosmological models describe dark matter as a cold, pressureless fluid of non-interacting particles.
    \item The cosmological constant $\Lambda$ serves as the primary candidate for DE.
\end{itemize}

To be fully consistent with the fundamental laws of physics the $\Lambda$CDM model requires a better theoretical foundation than it currently possesses. So, it is essential not to rely too rigidly on the model as future high-precision and extensive observational data may reveal deviations from the $\Lambda$CDM framework. Future observations, which are becoming increasingly precise and numerous, are likely to reveal inconsistencies with the $\Lambda$CDM model. Several unresolved theoretical questions encourage exploration beyond the standard cosmological model, including
\begin{itemize}
    \item Why do dark matter and DE exhibit density of the same order in the present Universe$?$
    \item Is there a possible connection between inflation and DE$?$
    \item What possible mechanism or theory can resolve the Big Bang singularity$?$
    \item Does a complete formulation of the standard model require a theory of quantum gravity that unifies GR with quantum field theory$?$
   \item What is the underlying cause of the tensions observed in the measurements of $H_0$, $S_8$ and $f_{\sigma_8}$?
\end{itemize}
Considering the aforementioned persistent challenges associated with the standard $\Lambda$CDM model, it is unrealistic to regard it as the ultimate cosmological framework capable of addressing all longstanding problems of the Universe. Therefore, the next critical step in cosmology is to establish a coherent theoretical foundation that aligns with increasingly precise data and resolves the model's current shortcomings.

 \section{Modified gravity theories}
The primary reason behind exploring modified gravity theories is to seek viable alternatives for the three major unresolved aspects of the standard cosmological model, namely, inflation, dark matter and DE. GR can be modified through several approaches, which are generally classified into two main categories. One method introduces additional fields into the theory whereas the other alters the underlying geometric structure of spacetime. In this study, our attention is directed toward the geometrical modifications of GR. Therefore, throughout this work, the term modified gravity will specifically refer to such geometrical alterations. Certain modified theories of gravity have demonstrated the potential to resolve $H_0$ tension, particularly by providing late-time cosmological solutions. Modified gravity theories provide a useful framework for analyzing the late-time Universe, as they can generate the observed accelerated expansion without requiring DE \cite{R34}.

 \subsection{Geometric trinity}
A spacetime manifold is fundamentally characterized by an affine connection $\hat{\Gamma}$ and a metric structure $g$ and is denoted $(M, g, \hat{\Gamma})$.
The metric tensor defines the notions of distance, inner products and the correspondence between covariant and contravariant tensor components while the affine connection governs the parallel transport and covariant differentiation, allowing for the comparison of vectors from distinct points on the manifold. A general affine connection $\hat{\Gamma}$ is neither metric-compatible nor symmetric due to its non-zero torsion and non-metricity components. The framework of a spacetime manifold featuring both a metric and an affine connection that lacks symmetry and metric compatibility is known as non-Riemannian geometry. The curvature tensor (Riemann tensor) is defined in terms of a general affine connection as follows,
\begin{equation}\label{e63}
\hat{R}^\lambda_{\gamma\alpha\beta} =  2\partial_{[\alpha}\hat{\Gamma}^\lambda_{|\gamma|\beta]} + 2\hat{\Gamma}^\lambda_{\rho[\alpha}\hat{\Gamma}^\rho_{|\gamma|\beta]} \text{.} 
\end{equation}
Here, the vertical bar represents the index that is excluded from the anti-symmetrization process. The antisymmetric part of the generic affine connection $\hat{\Gamma}$ defines the torsion tensor according to the following relation,
\begin{equation}\label{e64}
\hat{T}^\lambda_{\alpha\beta} = -2\hat{\Gamma}^\lambda_{[\alpha\beta]} \text{.} 
\end{equation}
Since the affine connection 
is assumed to be non-metric compatible, the non-metricity tensor is introduced to quantify this deviation from the metricity condition, defined as follows,
\begin{equation}\label{e65}
\hat{Q}_{\lambda\alpha\beta} = \hat{\nabla}_\lambda g_{\alpha\beta} \text{.} 
\end{equation}
The affine connection $\hat{\Gamma}$ can be written as the sum of components in the following decomposition \cite{R36,R37},
\begin{equation}\label{e66}
\hat{\Gamma}^{\lambda}_{\mu\nu}= \frac{1}{2}g^{\lambda \beta}\left(\partial_{\mu}g_{\beta \nu}+\partial_{\nu}g_{\beta \mu}-\partial_{\beta}g_{\mu \nu}\right)  + \frac{1}{2}g^{\lambda \beta}\left(  \hat{T}_{\nu\beta\mu} + \hat{T}_{\mu\beta\nu} - \hat{T}_{\beta\mu\nu} \right) + \frac{1}{2}g^{\lambda \beta}\left(  -\hat{Q}_{\mu\nu\beta} - \hat{Q}_{\nu\beta\mu} + \hat{Q}_{\beta\mu\nu} \right) \text{.} 
\end{equation}
In a more concise form, the above decomposition can be rewritten as
\begin{equation}\label{e67}
\hat{\Gamma}^{\lambda}_{\mu\nu}= \Gamma^{\lambda}_{\mu\nu} + K^{\lambda}_{\mu\nu} + L^{\lambda}_{\mu\nu} \text{.} 
\end{equation}
Here, the first term is the usual Christoffel symbols,
\begin{equation}\label{e68}
 \Gamma^{\lambda}_{\mu\nu} =   \frac{1}{2}g^{\lambda \beta}\left(\partial_{\mu}g_{\beta \nu}+\partial_{\nu}g_{\beta \mu}-\partial_{\beta}g_{\mu \nu}\right) \text{,} 
\end{equation}
The second term is the contortion tensor,
\begin{equation}\label{e69}
 K^{\lambda}_{\mu\nu} = \frac{1}{2}g^{\lambda \beta}\left(  \hat{T}_{\nu\beta\mu} + \hat{T}_{\mu\beta\nu} - \hat{T}_{\beta\mu\nu} \right) \text{,} 
\end{equation}
and the last term is disformation tensor, given as follows,
\begin{equation}\label{e70}
 L^{\lambda}_{\mu\nu}  =  \frac{1}{2}g^{\lambda \beta}\left(  -\hat{Q}_{\mu\nu\beta} - \hat{Q}_{\nu\beta\mu} + \hat{Q}_{\beta\mu\nu} \right) \text{.}  
\end{equation}
This decomposition often called the geometric trinity of gravity separates the generic affine connection $\hat{\Gamma}$ into a Riemannian component consisting of the Levi-Civita connection and a non-Riemannian component that incorporates the effects of torsion and non-metricity.

 \subsection{Geometric interpretation of torsion}
Let us take two curves $C$ and $\bar{C}$, represented by the parameterizations $x^\mu(\tau)$ and $\bar{x}^\mu(\tau)$ respectively. The corresponding tangent vectors are given by $u^\mu = \frac{dx^\mu}{d\tau}$ and $\bar{u}^\mu = \frac{d\bar{x}^\mu}{d\tau}$.
The first-order change in the components of a vector 
$u^\sigma$, when transported parallel along the curve $\bar{C}$ by a displacement, $d\bar{x}^\mu$ produces the new vector $u'^\sigma$ as follows,
\begin{equation}\label{e71}
 u'^\sigma = u^\sigma + (\partial_\mu u^\sigma)d\bar{x}^\mu \text{.} 
\end{equation}
As the vector $u^\sigma$ undergoes parallel transport along the curve $\bar{C}$, it satisfies the condition,
\begin{equation}\label{e72}
 \frac{d\bar{x}^\mu}{d\tau} (\hat{\nabla}_\mu u^\sigma) = 0 \implies \frac{d\bar{x}^\mu}{d\tau} (\partial_\mu u^\sigma + \hat{\Gamma}^\sigma_{\nu\mu} u^\nu) = 0 \implies (\partial_\mu u^\sigma)d\bar{x}^\mu = - \hat{\Gamma}^\sigma_{\nu\mu} u^\nu \bar{u}^\mu d\lambda \text{.} 
\end{equation}
Using equations \eqref{e71} and \eqref{e72}, we obtain
\begin{equation}\label{e73}
u'^\sigma = u^\sigma - \hat{\Gamma}^\sigma_{\nu\mu} u^\nu \bar{u}^\mu d\lambda \text{.} 
\end{equation}
Similarly, parallel transport of the vector $u'^\sigma$ along the curve $C$ by a displacement $dx^\mu$ gives a new vector as follows,
\begin{equation}\label{e74}
\bar{u}'^\sigma = \bar{u}^\sigma - \hat{\Gamma}^\sigma_{\mu\nu} \bar{u}^\mu u^\nu d\lambda \text{.} 
\end{equation}
Subtracting equation \eqref{e74} from \eqref{e73}, we get
\begin{equation}\label{e75}
 (u'^\sigma + \bar{u}^\sigma) - (u^\sigma + \bar{u}'^\sigma)  =  - ( \hat{\Gamma}^\sigma_{\nu\mu} - \hat{\Gamma}^\sigma_{\mu\nu} ) u^\nu \bar{u}^\mu d\lambda = - \hat{T}^\sigma_{\mu\nu} u^\nu \bar{u}^\mu d\lambda  \text{.} 
\end{equation}
A symmetric connection, such as the Levi-Civita connection $\hat{\Gamma} = \Gamma$ in GR, causes the right-hand side of the above equation to be zero. As a result, the requirement for the existence of an infinitesimal parallelogram is fulfilled.
 However, for a general affine connection $\hat{\Gamma}$ as given in \eqref{e67}, this situation changes the infinitesimal parallelogram no longer closes and instead deforms into a pentagon which arises from the presence of torsion $\hat{T}^\sigma_{\mu\nu}$  in the underlying spacetime geometry. Further we can assume a vector 
$V^\sigma =  - \hat{T}^\sigma_{\mu\nu} u^\nu \bar{u}^\mu$ , which quantifies the extent of deviation caused by the breaking of the parallelogram.

\subsection{Teleparallel gravity}
In recent decades, teleparallel gravity theories have attracted the growing attention of researchers. Teleparallel gravity offers an alternative geometric formulation in which the connection $\tilde{\Gamma}$  is flat, free from curvature and metric-compatible with torsion serving as the only geometric quantity. Since the teleparallel connection $\tilde{\Gamma}$ is flat, the parallel transport of a vector is path-independent, maintaining a consistent definition of parallelism even between distant points. This feature gives rise to the teleparallel theory \cite{R39}. Since the connection for teleparallel theory is metric-compatibility and curvature-free, we get 
\begin{equation}\label{e80}
\tilde{R}^{\rho}_{\sigma\mu\nu}=\partial_{\mu}\tilde{\Gamma}^{\rho}_{\nu\sigma}-\partial_{\nu}\tilde{\Gamma}^{\rho}_{\mu\sigma}+\tilde{\Gamma}^{\rho}_{\mu\lambda} \tilde{\Gamma}^{\lambda}_{\nu\sigma}-\tilde{\Gamma}^{\rho}_{\nu\lambda} \tilde{\Gamma}^{\lambda}_{\mu\sigma} = 0 \text{,} 
\end{equation} 
and 
\begin{equation}\label{e81}
Q_{\sigma\mu\nu} =  \tilde{\nabla}_\sigma g_{\mu\nu}  = 0  \text{.} 
\end{equation}
Furthermore, the torsion tensor, which is the anti-symmetric part of the connection $\tilde{\Gamma}$ is given by the following expression,
\begin{equation}\label{e82}
T^\lambda_{\alpha\beta} = -2\tilde{\Gamma}^\lambda_{[\alpha\beta]} = \tilde{\Gamma}^\lambda_{\beta\alpha}  - \tilde{\Gamma}^\lambda_{\alpha\beta} \neq 0 \text{.}
\end{equation}  
In Einstein's original formulation, the tetrad defined by the relation $e^i = e^i_\mu dx^\mu$ serves as the main fundamental variable. This tetrad field serves as the fundamental object from which both the metric and teleparallel connection components are derived,
\begin{equation}\label{e83}
g_{\mu\nu}=\eta_{ij}e^i_{\mu}e^j_{\nu} \text{,}
\end{equation}
and 
\begin{equation}\label{e84}
\tilde{\Gamma}^{\sigma}_{\mu\nu}\equiv e^{\sigma}_i\partial_{\nu}e^i_{\mu} \text{.}
\end{equation}
Here, $e_i = e_i^\mu \partial_\mu$ stands for the inverse tetrad satisfying the conditions $e^i_\mu e_i^\nu = \delta^\nu_\mu$ and $e^i_\mu e_j^\mu = \delta^i_j$.
This teleparallel connection defined by the coefficients of the tetrad field is called the Weitzenb$\ddot{\text{o}}$ck connection \cite{R40}. We now proceed to establish the link between the Ricci scalar constructed from the Levi-Civita connection and the torsion scalar associated with the teleparallel connection. In process to verify this, the contortion tensor is introduced as the difference between the Weitzenb$\ddot{\text{o}}$ck teleparallel connection and the Levi-Civita connection given by,
\begin{equation}\label{e85}
K^\sigma_{\mu\nu} =  \tilde{\Gamma}^\sigma_{\mu\nu} - \Gamma^\sigma_{\mu\nu} = \frac{1}{2}
\left( T_\mu^\sigma{}_\nu + T_\nu^\sigma{}_\mu - T^\sigma_{\mu\nu} \right) \text{.}
\end{equation}
Using the above relation, the Riemann tensor associated with the Levi-Civita connection can be expressed as follows,
\begin{equation}\label{e86}
R^{\rho}_{\sigma\mu\nu}  =  \tilde{R}^{\rho}_{\sigma\mu\nu} - \nabla_\mu K^\rho_{\sigma\nu} + \nabla_\nu K^\rho_{\sigma\mu} -  K^\rho_{\alpha\mu} K^\alpha_{\sigma\nu} + K^\rho_{\alpha\nu} K^\alpha_{\sigma\mu} \text{.}
\end{equation}
Employing the condition $\tilde{R}^{\rho}_{\sigma\mu\nu}=0$ given in equation \eqref{e80}, along with the antisymmetric property $K^{\rho\sigma\mu} = 0$, leads to the following formulation for the Ricci scalar corresponding to the Levi-Civita connection,
\begin{equation}\label{e87}
R = -2\nabla_\rho K^{\rho\sigma}{}_\sigma + K^{\mu\rho}{}_\rho  K_{\mu\sigma}{}^\sigma - K^{\mu\rho\sigma} K_{\mu\sigma\rho}   \text{.}
\end{equation}
We introduce another key quantity, known as the torsion scalar, which is constructed by contracting the torsion tensor in the following manner,
\begin{equation}\label{e88}
T =  T^\mu{}_{\rho\sigma} S_\mu{}^{\rho\sigma} \text{.}
\end{equation}
Here, 
\begin{equation}\label{e89}
S_\mu{}^{\rho\sigma} = \frac{1}{2} \left( K^{\rho\sigma}{}_\mu -\delta^\rho_\mu T_\alpha{}^{\alpha\sigma} + \delta^\sigma_\mu T_\alpha{}^{\alpha\rho}  \right)    
\end{equation}
is the superpotential tensor. Now, using the above definitions in equation \eqref{e87}, we get
\begin{equation}\label{e90}
 R = -T + 2\nabla_\rho T_\sigma{}^{\sigma\rho} = -T + B  \text{.}
\end{equation}
Therefore, the Ricci scalar 
$R$ of the Levi-Civita connection and the torsion scalar $T$ of the teleparallel Weitzenb$\ddot{o}$ck connection are related by a total divergence known as the boundary term 
$B$. Since this boundary term does not contribute to the field equations, it can be neglected.
Consequently, GR can be equivalently described by an action \cite{R41}, which uses the torsion scalar instead of the Ricci scalar. The action is as follows,
\begin{equation}\label{e91}
S=-\frac{1}{2\kappa}\int T e d^4x + \int e L_m d^4x  \text{.}
\end{equation}
Here, $e=\sqrt{-g}$ and $L_m$ is a matter Lagrangian. This torsion-based reformulation of GR is commonly referred to as the TEGR.

 \subsection{Geometric interpretation of non-metricity}
Let $a^\mu$ and $b^\mu$ be two vectors parallel transported along a curve $C$ described by $x^\mu(\tau)$. The total covariant derivative $\hat{D}$ of their inner product $a.b = a^\mu b^\nu g_{\mu\nu}$ along $C$ is expressed as
\begin{equation}\label{e76}
\frac{\hat{D}}{d\tau}(a.b) = \frac{\hat{D}}{d\tau}(a^\mu b^\nu g_{\mu\nu}) = \frac{dx^\sigma}{d\tau} (\hat{\nabla}_\sigma a^\mu) b_\mu + \frac{dx^\sigma}{d\tau} (\hat{\nabla}_\sigma b^\nu) a_\nu  + \frac{dx^\sigma}{d\tau} (\hat{\nabla}_\sigma g_{\mu\nu} ) a^\mu b^\nu \text{.} 
\end{equation}
Since the vectors $a^\mu$ and $b^\mu$ are parallel transported along this curve $C$, we have
\begin{equation}\label{e77}
 \frac{dx^\sigma}{d\tau} (\hat{\nabla}_\sigma a^\mu) = 0 \:\: \text{and}  \:\:  \frac{dx^\sigma}{d\tau} (\hat{\nabla}_\sigma b^\nu) = 0 \text{.} 
\end{equation}
Now using the above fact and the expression 
 $\hat{Q}_{\sigma\mu\nu}=\hat{\nabla}_\sigma g_{\mu\nu} $, the equation \eqref{e76} becomes,
\begin{equation}\label{e78}
\frac{\hat{D}}{d\tau}(a.b) = \hat{Q}_{\sigma\mu\nu}  \frac{dx^\sigma}{d\tau} a^\mu b^\nu \text{.} 
\end{equation}
Assuming $b^\mu=a^\mu$, we get
\begin{equation}\label{e79}
\frac{\hat{D}}{d\tau}(|a|^2) =  \hat{Q}_{\sigma\mu\nu}  \frac{dx^\sigma}{d\tau} a^\mu a^\nu \text{.} 
\end{equation}
With the help of equations \eqref{e78} and \eqref{e79}, It can be seen that, during parallel transport along a given curve, both the inner product and the magnitude of a vector undergo variation. 
Using the Levi-Civita connection 
 $(\hat{\Gamma} = \Gamma)$  from GR, the right-hand sides of \eqref{e78} and \eqref{e79} become zero, ensuring the invariance of a vector's length during parallel transport. However, for a general affine connection $\hat{\Gamma}$ as defined in \eqref{e67}, this property no longer holds; the length of a vector varies due to the non-metricity condition (metric incompatibility), which arises from the presence of the non-metricity tensor $\hat{Q}_{\sigma\mu\nu}$ \cite{R38}.

  \subsection{Modified teleparallel gravity: the $f(T)$ gravity}
Among curvature-based modified theories, $f(R)$ gravity is the most basic modification in the setting of curvature \cite{R42,R43}. This method extends the Einstein–Hilbert action by replacing it with a general function of the Ricci scalar, allowing a wider array  of gravitational behaviors to be investigated. Approximately ten years ago, the $f(T)$ framework was proposed \cite{R44}. It is an extension of TEGR by allowing the gravitational action to depend on nonlinear functions of the torsion scalar. It is important to note that while the TEGR reproduces Einstein’s field equations, their modifications, such as $f(R)$ and $f(T)$ gravity do not remain equivalent to each other. The action of $f(T)$ gravity is given as follows,
\begin{equation}\label{e92}
S= \frac{1}{2\kappa}\int f(T) \sqrt{-g} d^4x + \int  L_m\sqrt{-g} d^4x  \text{.}
\end{equation}
Varying the gravitational action with respect to the tetrad field, one arrives at the corresponding field equations for the $f(T)$ gravity framework,
\begin{equation}\label{e93}
 e^{-1}\partial_{\mu}(ee^{\gamma}_i S_{\gamma}{}^{\mu
\nu})f_T-f_Te^{\gamma}_i T^{\gamma}_{\mu\lambda}S_{\gamma}{}^{\lambda
\mu} +e^{\gamma}_i S_{\gamma}{}^{\mu\nu}\partial_{\mu}(T)f_{TT}+
\frac{1}{4}e^{\nu}_i f(T)= 4\pi G e^{\gamma}_i{T}^\nu_\gamma  \text{.}
\end{equation}
Here, $f_T=df(T)/dT$ and $f_{TT}=d^2f(T)/dT^2$.
It should be noted that while the combination $R=-T+B$ constitutes a Lorentz scalar, neither the torsion scalar $T$ nor the boundary term $B$ is individually a Lorentz scalar. Consequently, while GR and its curvature-based generalization, $f(R)$ gravity maintain local Lorentz invariance, the standard formulation of $f(T)$ gravity does not share this property. So, choosing an appropriate tetrad is crucial in $f(T)$ cosmological models, since different tetrad choices yield different field equations and consequently, different physical solutions.
A tetrad is considered good when it does not impose any constraints on the possible functional forms of $f(T)$ \cite{R45}. This problem was solved in an alternative, invariant formulation of the $f(T)$ theory, commonly referred to as the covariant formulation of $f(T)$ gravity \cite{R46}. For a comprehensive discussion of $f(T)$ gravity and its cosmological implications, the reader may consult the relevant references \cite{R47,R47aa,R47ab,R47ac}.

 \subsection{Symmetric teleparallel gravity}
The symmetric teleparallel gravity framework has been attracting significant interest among the cosmology community. Like teleparallel gravity, it is formulated using a flat, curvature-free connection $\mathring{\Gamma}$. However, the connection $\mathring{\Gamma}$ is metric-incompatible and torsionless. Its flat nature guaranties that parallel transport does not depend on the chosen path, maintaining the concept of absolute parallelism across large distances. Therefore, this framework is also described as a form of teleparallel gravity.
Furthermore, the absence of torsion ensures that the connection $\mathring{\Gamma}$ is symmetric in its final two indices. As a result, $\mathring{\Gamma}$ is termed symmetric teleparallel connection, and the associated gravitational theory is designated symmetric teleparallel gravity \cite{R48}. Since the symmetric teleparallel connection is free of curvature and torsion while obeying the non-metricity condition, it follows that
\begin{equation}\label{e94}
\mathring{R}^{\rho}_{\sigma\mu\nu}=\partial_{\mu}\mathring{\Gamma}^{\rho}_{\nu\sigma}-\partial_{\nu}\mathring{\Gamma}^{\rho}_{\mu\sigma}+\mathring{\Gamma}^{\rho}_{\mu\lambda} \mathring{\Gamma}^{\lambda}_{\nu\sigma}-\mathring{\Gamma}^{\rho}_{\nu\lambda} \mathring{\Gamma}^{\lambda}_{\mu\sigma} = 0 \text{,}
\end{equation} 
and 
\begin{equation}\label{e95}
T^\lambda_{\alpha\beta} = -2\mathring{\Gamma}^\lambda_{[\alpha\beta]} = \mathring{\Gamma}^\lambda_{\beta\alpha}  - \mathring{\Gamma}^\lambda_{\alpha\beta} = 0 \implies \mathring{\Gamma}^\lambda_{\alpha\beta} = \mathring{\Gamma}^\lambda_{\beta\alpha} \text{.}
\end{equation}
Additionally, we introduce the non-metricity tensor, which emerges from the metric incompatibility associated with the symmetric teleparallel connection
$\mathring{\Gamma}$,
\begin{equation}\label{e96}
Q_{\sigma\mu\nu} =  \mathring{\nabla}_\sigma g_{\mu\nu} = \partial_\sigma g_{\mu\nu} - \mathring{\Gamma}^\rho_{\sigma\mu} g_{\rho\nu} - \mathring{\Gamma}^\rho_{\sigma\nu} g_{\rho\mu} \neq 0  \text{.}
\end{equation}
It can be shown that the symmetric teleparallel connection may be represented using a general element of the group $GL(4, \mathbb{R})$ in the following way \cite{R49}, 
\begin{equation}\label{e97}
\mathring{\Gamma}^\sigma_{\mu\nu}  = A_\lambda{}^\sigma \partial_\mu A^\lambda{}_\nu  \text{.}
\end{equation}
Here, $A^\mu{}_\nu$ denotes a $4 \times 4$ matrix with inverse $A_\nu{}^\mu$. Since the connection $\mathring{\Gamma}$
is required to be torsionless, the matrix $A^\sigma{}_\mu$ must take the specific form of a Jacobian, namely $A^\sigma{}_\mu = \partial_\mu \xi^\sigma$ for some arbitrary vector field $\xi^\sigma$. So we get
\begin{equation}\label{e98}
\mathring{\Gamma}^\sigma_{\mu\nu}  =  \frac{\partial x^\sigma}{\partial \xi^\rho} \frac{\partial^2 \xi^\rho}{\partial x^\mu \partial x^\nu} \text{.}
\end{equation}
Hence, from the above relation it becomes clear that an appropriate coordinate transformation can always be chosen so that the general symmetric teleparallel connection $\mathring{\Gamma}$ vanishes identically.
This particular choice of coordinate transformation is called the coincident gauge \cite{R49}. In this specific case where $\mathring{\Gamma}$ vanishes, the non-metricity tensor simplifies to $Q_{\sigma\mu\nu} = \partial_\sigma g_{\mu\nu}$. This reduction also causes the covariant derivative to coincide with the partial derivative, leading to a significant simplification of the resulting field equations. In case of general symmetric teleparallel connection, a direct relationship can be established between the standard Ricci scalar from the Levi-Civita connection and the non-metricity tensor of the symmetric teleparallel connection. To analyze this relation, we introduce the disformation tensor, defined as the difference between the symmetric teleparallel connection and the Levi-Civita connection, given by
\begin{equation}\label{e99}
L^\sigma_{\mu\nu} =  \mathring{\Gamma}^\sigma_{\mu\nu} - \Gamma^\sigma_{\mu\nu} = \frac{1}{2}
\left( Q^\sigma{}_{\mu\nu} - Q_\mu{}^\sigma{}_\nu - Q_\nu{}^\sigma{}_\mu \right) \text{.}
\end{equation}
Using the above relation, the Riemann tensor corresponding to the Levi-Civita connection can be written as
\begin{equation}\label{e100}
R^{\rho}_{\sigma\mu\nu}  =  \mathring{R}^{\rho}_{\sigma\mu\nu} - \nabla_\mu L^\rho_{\sigma\nu} + \nabla_\nu L^\rho_{\sigma\mu} -  L^\rho_{\alpha\mu} L^\alpha_{\sigma\nu} + L^\rho_{\alpha\nu} L^\alpha_{\sigma\mu}  \text{.}
\end{equation}
We also introduce a key quantity known as the non-metricity scalar $Q$, which is derived by contracting the non-metricity tensor as shown below,
\begin{equation}\label{e101}
Q =  - Q_{\sigma\mu\nu} P^{\sigma\mu\nu}  \text{.}
\end{equation}
where 
\begin{equation}\label{e102}
P^\sigma_{\mu\nu} = \frac{1}{4} \left( -Q^\sigma\:_{\mu\nu} + 2Q_{(\mu}\:^\sigma\:_{\nu)} + (Q^\sigma - \bar{Q}^\sigma) g_{\mu\nu} - \delta^\sigma_{(\mu}Q_{\nu)} \right) \text{.}
\end{equation}
$P^\sigma_{\mu\nu}$ is called the superpotential tensor. In this expression, $Q_\alpha = Q_\alpha\:^\mu\:_\mu $ and $ \bar{Q}_\alpha = Q^\mu\:_{\alpha\mu} $ 
represent two distinct non-metricity vectors. Furthermore, applying these definitions along with the condition $\mathring{R}^{\rho}_{\sigma\mu\nu}=0$ 
from equation \eqref{e94} to equation \eqref{e100}, we arrive at the following significant relation,
\begin{equation}\label{e103}
 R = Q - 2\nabla_\rho \left( Q^\rho - \bar{Q}^\rho \right) = Q - B  \text{.}
\end{equation}
Hence, the Ricci scalar $R$ of the Levi-Civita connection and the non-metricity scalar $Q$ of the symmetric teleparallel connection are related through a total derivative, known as the boundary term $B$. Since this boundary term does not influence the field equations, it is typically discarded. Consequently, a formulation equivalent to GR can be constructed by using the non-metricity scalar in place of the Ricci scalar, as defined by the following action,
\begin{equation}\label{e104}
S=-\frac{1}{2\kappa}\int Q  \sqrt{-g}d^4x + \int  L_m \sqrt{-g}d^4x  \text{.}
\end{equation}
where $L_m$ denotes the standard matter Lagrangian. This alternative non-metricity based representation of GR is commonly referred to as the STEGR \cite{R50}.

 \subsection{Modified symmetric teleparallel gravity: the $f(Q)$ gravity}
The extended theory for the STEGR framework was proposed in 2018 \cite{R51}, which is commonly known as the $f(Q)$ theory of gravity. $f(Q)$ theory includes arbitrary non-linear functions of the non-metricity scalar. Although STEGR reproduces the equivalent field equations as GR, its modified extension known as $f(Q)$ gravity, is not equivalent to either $f(R)$ gravity or $f(T)$ gravity. The $f(Q)$ gravity action is given as follows,
\begin{equation}\label{e105}
S= \frac{1}{2\kappa}\int f(Q) \sqrt{-g} d^4x + \int  L_m\sqrt{-g} d^4x  \text{.}
\end{equation}
By performing a variation of the generic action \eqref{e105} with respect to the metric, the corresponding metric field equations for $f(Q)$ gravity are obtained as
\begin{equation}\label{e106}
\frac{2}{\sqrt{-g}}\nabla_{\sigma}\left( \sqrt{-g}f_Q {P^{\sigma}}_{\mu\nu}\right)+\frac{1}{2}g_{\mu\nu}f
+f_Q\left(P_{\mu\sigma\rho}{Q_{\nu}}^{\sigma\rho}-2Q_{\sigma\rho\mu}{P^{\sigma\rho}}_{\nu} \right)=-{T}_{\mu\nu} \text{.} 
\end{equation}
Here, $f_Q=\frac{df}{dQ}$. Similarly, varying the general action \eqref{e105} with respect to the symmetric teleparallel connection leads to the following connection field equation for $f(Q)$ gravity as below,
\begin{equation}\label{e107}
\nabla_{\mu}\nabla_\nu \left( \sqrt{-g}f_Q {P^{\mu\nu}}_{\sigma} \right) = 0 \text{.} 
\end{equation}
\section{Introduction of bulk viscosity in cosmology}
Modern cosmology aims to explain the large-scale dynamics of the Universe including its early-time inflationary phase, the radiation- and matter-dominated eras and the current accelerated expansion. Although the standard cosmological model successfully describes a wide range of observations, it often relies on idealized assumptions, such as treating cosmic fluids as perfect and non-dissipative. However, realistic cosmic media may exhibit dissipative processes, among which bulk viscosity plays a significant role.
Bulk viscosity naturally arises in fluids that deviate from local thermodynamic equilibrium during cosmic expansion. In an expanding Universe, particle interactions, decay processes and phase transitions can generate internal friction, leading to an effective viscous pressure. Unlike shear viscosity and heat conduction, which vanish in the FLRW spacetime, bulk viscosity is fully compatible with cosmological symmetry and can therefore influence the background evolution. Consequently, bulk viscosity becomes the only significant dissipative coefficient in the cosmic fluid. Bulk viscosity functions as the internal restorative pressure required to maintain thermal stability as the fluid expands alongside the fabric of space-time. In this context, the modification of GR accounts for the geometric expansion, while the viscosity coefficients contribute to the pressure term, effectively propelling cosmic acceleration. This mechanism has spurred significant recent research into how bulk viscosity shapes the long-term evolution of our Universe \cite{IB-1,IB-2,IB-3,IB-4,IB-5,JM,AVS,MAT}.

Initially, the study of viscosity in the cosmic fluid began as a way to investigate the early inflation era of cosmic evolution without the assumption of a DE component. The theoretical framework for this is divided into two main approaches:
\begin{itemize}
    
\item The Eckart Theory: Introduced by Carl Eckart, this is a non-causal theory that considers first-order deviations from thermodynamic equilibrium \cite{C.E.}. Although influential, it has limitations with respect to the speed of signal propagation.

\item The Israel-Stewart Theory: To address the causality issues of the Eckart model, Werner Israel and John Stewart developed a causal theory based on second-order deviations from equilibrium \cite{W.I.,W.I.-2,W.I.-3}. This second-order analysis reveals two distinct types of viscosity, shear viscosity and bulk viscosity.
\end{itemize}

Mathematically, the bulk viscosity modifies the equilibrium pressure $P$ into an effective pressure $P_{v}$ as follows,
\begin{equation}
P_{v}=P+\Pi \text{.}
\end{equation}
Here, $\Pi$ represents the bulk viscous stress. It is related to the Hubble parameter $H$ and the bulk viscosity coefficient $\zeta$ as follows,
\begin{equation}
\Pi=-3\zeta H \text{.}
\end{equation}

Overall, the inclusion of bulk viscosity enriches the cosmological modeling of the Universe by accounting for non-equilibrium processes and dissipative effects. Its implications span from early-Universe physics to late-time cosmic acceleration, making bulk viscous cosmology an active and important area of research in theoretical and observational cosmology.

 \section{Modeling and statistics}
In the preceding section, we examined various cosmological observations along with their enduring challenges. In the present section, we explore how the underlying geometrical frameworks can account for the observable Universe and yield statistically meaningful predictions.
\subsection{Modeling the Universe}
To develop a cosmological framework capable of representing the physical Universe, we proceed through the following essential steps.
\begin{itemize}
    \item \textbf{Specifying the underlying geometry:}  We start by defining a spacetime manifold equipped with an appropriate connection and a metric which together establish the concepts of distance and covariant differentiation.
    \item \textbf{Model assumptions:} 
    Once the geometric framework is established, we impose physically motivated assumptions on the model. This includes specifying the functional form of the modified gravity theory and adopting a suitable EoS. These choices render the model capable of describing particular cosmological epochs in a realistic manner.
    \item \textbf{Solving the model:} These suppositions of the model also play a crucial role in facilitating the derivation of analytical solutions. so, the analytical solution of the chosen cosmological model is obtained by applying appropriate differential equation methods.
  \item \textbf{Estimation of parameters:}
  After deriving the exact solution of the model, the resulting expression typically contains a set of free parameters within the resulting equations. Therefore, the subsequent step involves determining the values of these free parameters using various observational datasets and established statistical techniques.

    \item \textbf{Model prediction and comparison:} Using the parameter values fixed by observational constraints, cosmological quantities such as the deceleration parameter, the statefinder parameters and  the EoS parameter can be analyzed. The profile of these parameters reveals how the proposed model behaves over time, allowing its predictions to be compared with those of well-established cosmological models.
\end{itemize}

\begin{figure}[H]
    \centering
    \includegraphics[width=14 cm, height= 11 cm]{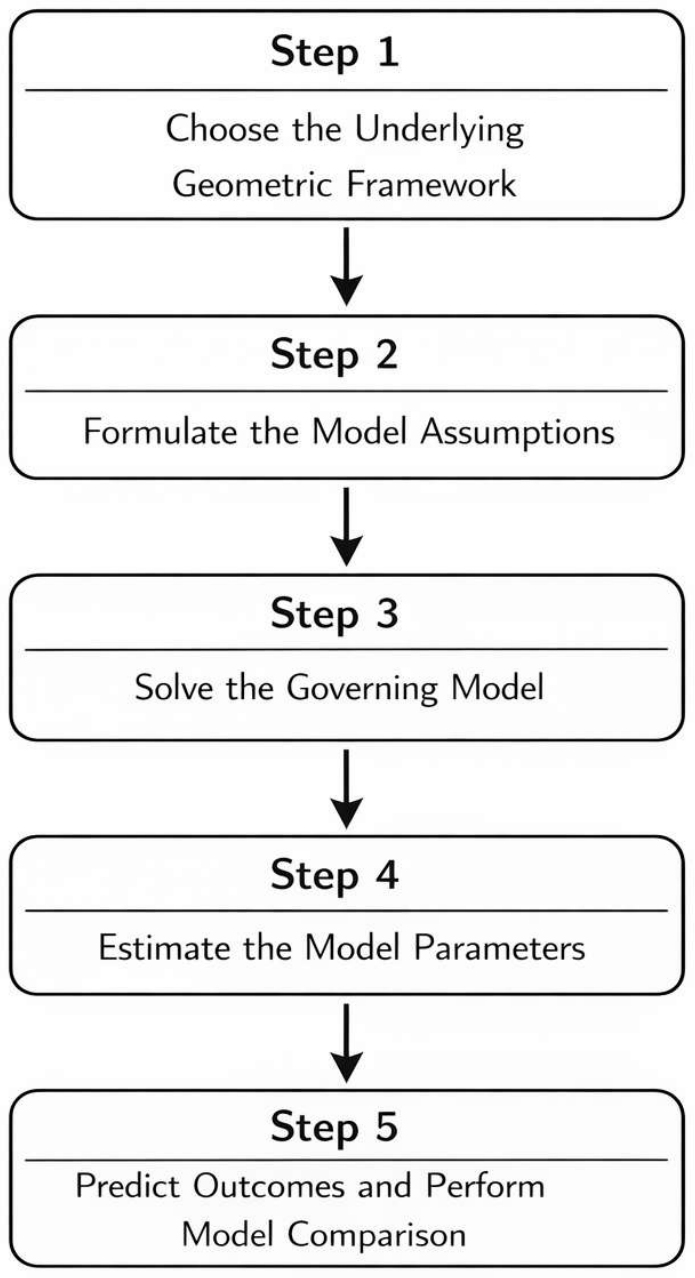}
    \caption{Flowchart representing the steps to analyze a cosmological model.}
    \label{fig:int-1}
\end{figure}

The general methodology for analyzing any cosmological model can be clearly understood by the flow chart shown in Figure
\eqref{fig:int-1}.

\subsection{$\chi^2$ minimization}
Let us assume a model function $f_{model}(x,\theta)$, which depends on the independent variable $x$ and a set of free parameters $\theta$. If the collection $\{f_{k,obs}(x_{k,obs})\}_{k=1}^n$ denotes $n$ independent observational data points with corresponding standard deviations $\sigma_{k,obs}$, then the $\chi^2$ statistic is defined as \cite{R52},
\begin{equation}\label{e108}
\chi^2(\theta) = \sum_{k=1}^n  \frac{\left( f_{model}(x_{k,obs},\theta) - f_{k,obs} \right)^2}{\sigma_{k,obs}^2} \text{.} 
\end{equation}
Furthermore, if $n$ observations  are correlated, a covariance matrix $C$ replaces the standard deviation. In this scenario, the $\chi^2$ function is expressed as
\begin{equation}\label{e109}
\chi^2(\theta) = \sum_{i,j=1}^n X_i C^{-1}_{ij} X_j \text{.} 
\end{equation}
Here, $X_i = f_{i,obs} - f_{model}(x_{i,obs},\theta) $.
A key statistical quantity, known as the likelihood function, can be defined in the following way,
\begin{equation}\label{e110}
\mathcal{L}(\theta) = P(\theta| D = Data) \text{.} 
\end{equation}
The relation between these two aforementioned statistical quantities can be expressed through the following expression,
\begin{equation}\label{e111}
\mathcal{L}(\theta) \propto exp\left( -\frac{1}{2} \chi^2(\theta) \right) \text{.}  
\end{equation}
Our main objective is to determine the optimal set of parameters $\theta$ such that the model function $f_{model}(x,\theta)$ provides the closest match to the observational data. To determine this, we need to minimize the function $\chi^2(\theta)$, which is equivalent to maximizing the likelihood, and the set of parameters that yields the minimum value $\chi^2_{min}$ denoted by $\theta_{min}$ is identified as the best-fit estimate. Various methods have been developed to address this optimization problem, including Gradient Descent, Newton’s method, and the Random Walk algorithm \cite{R53,R54}. The main drawback of these methods is their tendency to converge on a local minimum rather than the global minimum when the parameter space contains numerous local optima. In contrast, a novel and effective algorithm can be developed by combining the strengths of these existing methods. The Markov Chain Monte Carlo (MCMC) method is one of that kind, which is extensively used across computational science for this purpose.

 \subsection{The MCMC approach }
In recent years, probabilistic data analysis, especially Bayesian inference, has significantly transformed scientific investigation. This method operates by sampling from either the posterior probability density function (PDF) or the likelihood function. Although the optimal values of these functions can often be obtained using standard algorithms, a comprehensive analysis of the posterior PDF is frequently necessary for a deeper understanding. MCMC algorithms were explicitly designed to effectively draw samples from the posterior PDF, enabling effective exploration even within high dimensional parameter spaces \cite{R55}.
A Markov chain is a mathematical model that represents a sequence of potential events, randomly generated from a set of possible states. The defining property is that the probability of transitioning to a given future state depends only on the system's current state, with no memory of the steps that preceded it.
Moreover, the Monte Carlo method enables a broad analysis of the entire parameter space. The fundamental principle of MCMC is to construct a Markov chain that samples the model’s parameter space according to a chosen probability distribution. The chain itself is a sequence of values of the sampled parameters. A new value is proposed at each step based on the current value, following specific transition rules of a chosen proposal distribution. A new parameter value is proposed via the proposal distribution and its approval is determined by evaluating the posterior probability, which incorporates the observational data and the chosen prior probability. Once the Markov chain has reached convergence, the posterior parameter distribution can be approximated by computing the frequency with which different parameter values appear in the chain. This resulting posterior distribution allows one to determine the most likely parameter values along with their uncertainties, which in turn provides a foundation for predictive analysis of related observables. Among MCMC techniques, the Metropolis–Hastings algorithm, which uses a random walk proposal mechanism, is widely used. For more complex sampling problems, the emcee ensemble sampler (colloquially called "The MCMC Hammer" and introduced by Mackey) represents a significant advancement \cite{R56}. The emcee ensemble sampler outperforms traditional MCMC methods. A key feature is its ability to efficiently employ hundreds of parallel walkers during the sampling process. In principle, increasing the number of walkers does not pose any disadvantage, except when computational constraints begin to limit performance. The present work does not undertake a comprehensive analysis comparing various MCMC algorithms to the emcee ensemble sampling method.

\section{Dynamical system method}

The dynamical systems theory offers a robust mathematical way to understand how complex systems behave over long periods, whether in fluid mechanics or molecular biology. This mathematical framework is particularly useful in cosmology, where it models the evolution of the universe using autonomous differential equations that track changes in scale factor, density parameter, and curvature. This approach is particularly effective for studying asymptotic behavior and global dynamics. This framework is especially valuable for categorizing the different phases of the Universe, including early inflation, the radiation-dominated epoch, and the current phase of accelerated expansion.  Furthermore, 
by calculating fixed points and assessing their stability, researchers can determine if the Universe is heading towards a stable equilibrium or transitioning between different phases of expansion. For detailed overviews of dynamical system methods in cosmology, one can check the references \cite{da49}.

\subsection{ An overview to dynamical systems}
There are two different categories of dynamical systems: differential equations and iterated maps. Differential equations model systems with a continuous time variable, which is especially relevant in cosmological contexts. On the other hand, iterated maps treat time as a discrete parameter and will not be considered here. Our primary focus is on differential equations that depend on a single independent variable, most notably time. Such equations are known as ordinary differential equations.
Suppose, $x_1,x_2,...,x_n \in X \subseteq R^n$ are coordinates in a $n-$dimensional phase space. we also defined $t \in R$ as an independent variable, which does not necessarily have to represent time. Dynamical systems can be expressed within this context using ordinary differential equations as follows \cite{da50}, 
\begin{center}
\begin{equation}\label{sys}
\begin{aligned}
\dot{x}_1 &= f_1(x_1, x_2, \ldots, x_n), \\
\dot{x}_2 &= f_2(x_1, x_2, \ldots, x_n), \\
&\ \vdots \\
\dot{x}_n &= f_n(x_1, x_2, \ldots, x_n).
\end{aligned}
\end{equation}
\end{center}
Here, the over-dot denotes differentiation with respect to the variable $t$ i.e. $\dot{x}_i=\frac{dx_i}{dt}$ and $f_i$ are functions such that $f_i:X \to X$. The system \eqref{sys} is said to be autonomous when it does not have an explicit dependence on the independent variable $t$. The system of equations \eqref{sys} can be written as follows,
\begin{equation}\label{dsa11}
\dot{\mathbf{x}} = \mathbf{f}(\mathbf{x}) \text{.}
\end{equation}
This formulation describes a system of differential equations in which each time derivative $\dot{x_i}$ depends on all variables $x_1,x_2,...,x_n$ and the function $ \mathbf{f(x)}=(f_1(\mathbf{x}),f_2(\mathbf{x}),...,f_n(\mathbf{x})) $ can be regarded as a vector field in 
$n-$dimensional real space $R^n$. Our analysis will consider only those regions of phase space in which $f(x)$ is smooth and take real values. The methods discussed here are not valid in areas of the phase space where these conditions do not hold. 
For a given initial condition, a solution of equation \eqref{dsa11} traces a curve $\phi(t)$ in phase space, which is referred to as a trajectory or orbit.
As a result, the phase space contains a collection of trajectories corresponding to a range of initial conditions. The behavior of the system can be understood by analyzing these geometric representations and the corresponding flow of trajectories in phase space.

\subsubsection{Analysis of equilibrium points and stability}
\textbf{Definition:} A point $\mathbf{x}=\mathbf{x_*}$ is an equilibrium point or fixed point or critical point of an autonomous system if and only if it satisfies the condition $\mathbf{f(\mathbf{x_*})=0}$ for the autonomous system given by equation \eqref{dsa11} \cite{da51}.

In an autonomous dynamical system, trajectories that begin at a fixed point stay unchanged and motionless throughout time. In order to study how small deviations evolve near this point and how neighboring trajectories behave, one must define the stability of the fixed point.

\textbf{Definition:} Lyapunov Stable Fixed Point or Stable Fixed Point: A fixed point $\mathbf{x_*}$ is said to be stable (or Lyapunov stable) if, for every small $\epsilon >0 $, there exists a $\delta>0$ such that whenever $||\phi(t_0)-\mathbf{x_*}||< \delta$, the corresponding solution satisfies $||\phi(t)-\mathbf{x_*}||< \epsilon$ for all $ t \ge t_0$ \cite{da49}.

A fixed point is stable if small perturbations from equilibrium remain bounded as time evolves. In the context of cosmology, stable fixed points represent self-regulating phases of cosmic evolution. For example, the late-time de-sitter phase, where the accelerated expansion of the Universe persists without end ($t \to \infty$). Although, points inside this neighborhood do not need to converge to a stable fixed point, which motivates the introduction of a stronger stability concept.

\textbf{Definition:} Asymptotically Stable Fixed Point: A fixed point $x_*$ is said to be asymptotically stable if it is stable and if there exists a $\delta>0$ such that, whenever $||\phi(t_0)-\mathbf{x_*}||< \delta$, the corresponding solution $\phi(t)$ satisfies $\lim_{t \to \infty}\phi(t)=\mathbf{x_*}$.

When a fixed point is asymptotically stable, any trajectory entering a sufficiently small neighborhood of that point will eventually converge to it. In cosmological contexts, this form of equilibrium is particularly important because stable fixed points are typically also asymptotically stable. However, this definition does not prescribe how long it takes for a trajectory to approach an asymptotically stable fixed point. In contrast, an unstable fixed point is simply an equilibrium point that fails to be stable.

\subsubsection{Linear stability theory}
The behavior of trajectories near a critical point is examined using linear stability theory. The non-linear dynamics of complex systems, described by $\dot{\mathbf{x}} = \mathbf{f}(\mathbf{x})$, can be approximated by linearizing the system about a critical point $\mathbf{x_*}$. This approximation holds under the assumption that 
$f(x)$ is sufficiently smooth. Expanding 
$f(x)$ in a Taylor series on the critical point $\mathbf{x_*}$, we get
\begin{equation}
\mathbf{f(x)=f(x_*)+(x-x_*) \left.\frac{\partial f}{\partial x}\right|_{x = x_{*}}+... } 
\end{equation}
In this context, only the first order partial derivatives are required \cite{da50}. Since $\mathbf{f({x_*})=0}$, the dynamics of the perturbations $\mathbf {({x-x_*})}$ is determined by the Jacobian matrix computed at the critical point.

\begin{equation}\label{jacobian}
\left.\mathbf{ J } \right|_{ \mathbf{x = x_{*}}} 
=
\left. \mathbf{ \frac{\partial f}{\partial x} }\right|_{\mathbf{x = x_{*}}}
=
\begin{pmatrix}
\left. \dfrac{\partial f_1}{\partial \mathbf{x_1}} \right|_{\mathbf{x = x_{*}}} & \cdots & \left. \dfrac{\partial f_1}{\partial \mathbf{x_n}} \right|_{\mathbf{x = x_{*}}} \\
\vdots & \ddots & \vdots \\
\left. \dfrac{\partial f_n}{\partial \mathbf{x_1}} \right|_{\mathbf{x = x_{*}}} & \cdots & \left. \dfrac{\partial f_n}{\partial \mathbf{x_n}} \right|_{\mathbf{x = x_{*}}} 
\end{pmatrix} \text{.}
\end{equation}
The stability of a critical point $x_*$
 is determined from the eigenvalues of the Jacobian (or stability) matrix. The eigenvalues can be computed manually or utilize computational tools for more complicated systems. The same methodology can be employed to determine the system's fixed point. A point is considered unstable or a repeller when all its eigenvalues have positive real parts. In this case, trajectories tend to move away from that point. In contrast, a point is considered stable or a attractor when all its eigenvalues have negative real parts, with trajectories converging toward it. A point is identified as a saddle point when there are at least two eigenvalues that have real parts with opposite signs, in this case, trajectories approach it from some directions while diverging from it in others. Most of the fixed points in cosmological systems are described by these three classes. while in two dimensions, nonzero imaginary parts of eigenvalues may produce spiral behavior. The stability of these spirals is determined by the real parts of the eigenvalues, they may be either stable or unstable. Although it is possible to classify a wider range of critical points \cite{da50}. Linear stability analysis is valid provided that none of the eigenvalues have zero real parts.

\section{Conclusions}
In this chapter, several important cosmological observations were presented, along with a discussion of their implications. After establishing the basics of relativity, the discussion shifted to modified gravity frameworks, which serve as an alternative to DE models.
This thesis seeks to investigate the role of bulk viscosity in the evolution of the Universe by taking different forms of the bulk viscous coefficient in the framework of modified gravity theories. Furthermore, statistical analysis and estimation of model parameters are carried out using the MCMC framework, supported by the emcee sampler and Bayesian inference techniques. Lastly, the dynamical system approach is discussed. The detailed investigation and its findings are presented in the chapters that follow.


\chapter{Observational constraints on viscous fluid cosmological model in $f(Q)$ gravity} 

\label{Chapter2} 

\lhead{Chapter 2. \emph{Observational constraints on viscous fluid cosmological model in $f(Q)$ gravity}} 

\vspace{10 cm}
* The work, in this chapter, is covered by the following publications: \\
 
\textit{Cosmological constraints in symmetric teleparallel gravity with bulk viscosity}, General Relativity and Gravitation \textbf{56}, 82 (2024).

\clearpage
\pagebreak

In this chapter, we explore the accelerated expansion of the Universe within the framework of modified $f(Q)$ gravity. The investigation focuses on the role of bulk viscosity in understanding the accelerated expansion of the Universe. Specifically, a cosmological model dominated by bulk viscous matter is considered, with the bulk viscosity coefficient expressed as $\zeta= \zeta_0 + \zeta_1 H $. We consider the power law function $f(Q)=\alpha Q^n $, where $\alpha$ and $n$ are arbitrary constants, and derive analytical solutions for the field equations corresponding to a flat FLRW metric. Subsequently, we used the combined CC+Pantheon+SH0ES dataset to estimate the free parameters of the analytic solution obtained. We perform Bayesian statistical analysis to estimate the posterior probability employing the likelihood function and the MCMC random sampling technique, along with the AIC and BIC statistical assessment criteria. In addition, we explore the evolutionary behavior of significant cosmological parameters. The effective EoS parameter predicts the accelerating behavior of the cosmic expansion phase. Furthermore, by the statefinder and $Om(z)$ diagnostic test, we found that our viscous model favors quintessence-type behavior and can successfully describe the late-time scenario.\\


\section{Introduction}\label{sec1m}

In this study, we will present our analysis and findings within the context of the newly introduced $f(Q)$ gravity framework \cite{R51}. The standard GR describes spacetime with a non-zero curvature tensor, while non-metricity and torsion both disappear. In contrast, the TEGR describes spacetime with a non-zero torsion tensor, while non-metricity and curvature both disappear. Additionally, the STEGR describes spacetime with a non-zero non-metricity tensor while torsion and curvature both disappear. The simple extensions of GR, TEGR and STEGR are represented by $f(R)$, $f(T)$, and $f(Q)$ gravity respectively \cite{R36}.
An important benefit of $f(Q)$ gravity lies in its field equations, which remain second order with respect to the scale factor.  This characteristic allows for the inclusion of higher-order Lagrangian without encountering complications. An additional benefit of $f(Q)$ gravity is its automatic fulfillment of Bianchi's identity. In contrast, in $f(T)$ gravity, the existence of the anti-symmetric component poses challenges. Moreover, various extensions of $f(Q)$ theory have been introduced in the literature, including the commonly referenced $f(Q,T)$ theory \cite{m1}, the Weyl-type $f(Q,T)$ theory \cite{m2} and $f(Q,L_m)$ gravity \cite{FOPK}. For more on symmetric teleparallel gravity and its extension, one can refer to \cite{ADD1,ADD2,ADD3,ADD4,ADD5,ADD6}.
In this work, we investigate the significance of a well motivated viscosity parameterization under the coincident $f(Q)$ gravity formalism in order to probe the late time accelerating behavior of the expansion phase of the Universe. We calculate the exact solution of the assumed model and then confront it from the cosmological observations. In addition, we predict the results of the investigation with the observed phenomenon and then provide a thorough comparison with the existing literature.
The structure of this chapter is as follows.  In sec \ref{cha2sec2}, the flat FLRW Universe in symmetric teleparallel cosmology with bulk viscous matter is presented. In sec \ref{cha2sec3}, we estimate the median value of model parameters and bulk viscous parameters, utilizing the CC+Pantheon+SH0ES dataset. Moreover, to assess the robustness of the MCMC process, we employ AIC and BIC statistical assessment. Further in sec \ref{cha2sec4},  we investigate the evolutionary behavior of cosmological parameters. Lastly, in sec \ref{cha2sec5}, we conclude our findings.

\section{The $f(Q)$ cosmological model in FLRW background with viscous matter}\label{cha2sec2}
 \justify
We assume spatial flatness, homogeneity, and isotropy in the Universe, and consequently use the FLRW metric to define the line element,
\begin{equation}\label{FLRW}
ds^2= -dt^2 + a^2(t)[dx^2+dy^2+dz^2] \text{.}
\end{equation}
Here, $a(t)$ is the scale factor of the Universe. Now coincident gauge coordinates are utilized in the gauge taken into account in the line element \eqref{FLRW}, suggesting that the metric is the only fundamental variable.
However, using a different gauge creates a generic connection, which leads to a non-trivial contribution to the field equations, via the non-metricity scalar in particular \cite{Hohmann2,fQfT1}.  As we have fixed the coincident gauge, the connection is trivial, and hence the metric is only the fundamental variable. Since we have used diffeomorphisms to fix the coincident gauge, one could think that we are not allowed to select any particular lapse function. However, the special case of $f(Q)$ theories allows for this because $Q$ retains a residual time-reparameterization invariance, as already explained in \cite{R51}, so we will use this symmetry to set the lapse function $N=1$. The line element \eqref{FLRW} has the non-metricity scalar $Q$ as follows
\begin{equation}\label{3b}
 Q= 6H^2  \text{.}
\end{equation}

In the literature, the bulk viscosity coefficient $\zeta$ is often assumed to depend on factors such as the expansion rate, its time derivative, and the energy density. However, in general $\zeta$ can be a function of all these variables, i.e., $\zeta = \zeta (H,\dot{H},\rho)$. Recently, a new form of the bulk viscosity coefficient $\zeta \sim \zeta_i H^{1-2i} \rho^i$ \cite{LGG} has been introduced in the literature. This particular value $i=0$ corresponds to a bulk viscosity coefficient of the form $\zeta \propto H$ that is commonly attributed to describe unified viscous models along with dissipative dark matter, whereas case $i=1$ corresponds to a bulk viscosity coefficient of the form $\zeta \propto H^{-1}\rho$ which could drive accelerated expansion \cite{LGG}. In this section, we consider the following form of the bulk viscosity coefficient for our analysis \cite{ADD10},
\begin{equation}
 \zeta= \zeta_1 H + \zeta_0 \rho H^{-1}  \text{.}  
\end{equation}
Note that $\zeta_0$ is the dimensionless quantity, whereas $\zeta_1$ has the unit $ML^{-1}$.\\
The corresponding energy momentum tensor is,
\begin{equation}\label{3d}
{T}_{\mu\nu}=(\rho+ p_v)u_\mu u_\nu + p_v g_{\mu\nu}  \text{.}
\end{equation}
Here, $\rho$ stands for the density of matter-energy, with $p_v = p - 3\zeta H$ as the pressure due to viscosity and $u^\mu = (1,0,0,0)$ representing the four-velocity vector. In the case of bulk viscosity, the field equations are expressed as \cite{LZ},
\begin{equation}\label{fqfe1}
3H^{2}= \rho_{eff} = \frac{1}{2f_{Q}}\left(-\rho +\frac{f}{2}\right)  \text{,}
\end{equation}
and
\begin{equation}
\dot{H}+3H^{2}+\frac{\dot{f_{Q}}}{f_{Q}}H=\frac{1}{2f_{Q}}\left({p_v}+
\frac{f}{2}\right) \text{.} \label{fqfe2}
\end{equation}
Furthermore, the matter conservation equation is as follows,
\begin{equation}\label{3f}
\dot{\rho} + 3H\left(\rho+{p_v}\right)=0 \text{.}
\end{equation}

Since, we focus only on the dust scenario for further analysis, the effective pressure will become ${p_v}=-3\zeta H$.
\justify We assume the following $f(Q)$ power law model \cite{RSU},  
\begin{equation}\label{fqp}
f(Q)=\alpha Q^n,\ \ \ \alpha \neq 0  \text{.}
\end{equation}
In particular, for $\alpha=-1$ and $n=1$, the standard Friedmann equations for GR can be obtained.
The field equations \eqref{fqfe1} and \eqref{fqfe2} for the considered $f(Q)$ model \eqref{fqp} become,
\begin{equation}\label{3i}
2\rho=\left(1-2n \right) \alpha 6^n H^{2n}  \text{,}
\end{equation}
and
\begin{equation}\label{3j}
\dot{H} + \frac{3}{4n}{(2-3 \zeta_0)} H^2 = \frac{3 \zeta_1 H^{4-2n}}{2n(2n-1)\alpha 6^{n-1}H^{2n-2}}     \text{.}
\end{equation}
The solution for this differential equation is the following,

\begin{equation}\label{3oa}
H(z)= \Biggl\{H^{2n-2}_0(1+z)^{\frac{3 (n-1) (2-3 \zeta_0)}{2n}}  +\frac{2\zeta_1} {\alpha (2n-1) 6^{n-1}(2-3 \zeta_0)} \left[(1+z)^{\frac{3 (n-1) (2-3 \zeta_0)}{2n}}-1\right]\Biggl\}^{\frac{1}{2n-2}}\text{.}
\end{equation}

\section{Estimation of parameter using observational data}\label{cha2sec3}
\justify

In this segment, we focus on utilizing Cosmic Chronometer (CC) and Pantheon Supernovae datasets to estimate the values for parameters that appear in the expression of $H(z)$ for our assumed model. We utilize the MCMC random sampling technique in conjunction with Bayesian analysis and the Scipy optimization method in the Python package emcee \cite{Mackey/2013} to estimate the median values of the model parameters and viscosity parameters.

\subsubsection{Cosmic chronometer datasets}
\justify
In this section, we have used CC dataset which includes 31 Hubble points, obtained by using the method of differential age (DA) in the redshift range given as $ 0.07 \leq z \leq 2.41 $. The complete list of 31 data points for $H(z)$ measurements using the DA method is listed in \cite{RS}. The $\chi^2$ function for the $H(z)$ dataset considered is given as follows,
\begin{equation}\label{4acc}
\chi_{CC}^{2}=\sum\limits_{k=1}^{31}
\frac{[H_{th}(z_{k},\theta)-H_{obs}(z_{k})]^{2}}{
\sigma _{H(z_{k})}^{2}}.  
\end{equation}
In this context, $H_{obs}$ refers to the Hubble parameter value obtained from observational data. On the other hand, $H_{th}$ represents its theoretical value calculated at $z_{k}$ within the parameter space $\theta$, and $\sigma_{H(z_{k})}$ denotes the associated error.
\subsubsection{Pantheon+SH0ES datasets}
\justify

In this current study, we analyze a recently published  revised Pantheon+SH0ES  dataset of Pantheon supernovae. The dataset comprises 1701 supernova samples, each associated with its observed distance modulus $\mu^{obs}$ in the redshift range $z \in [0.001, 2.3]$.
The Pantheon+SH0ES datasets surpass previous compilations of SNIa, integrating the latest available observations. In recent times, several collections of Type Ia supernova data have surfaced, such as Union \cite{R30}, Union2 \cite{R31}, Union2.1 \cite{R32}, JLA \cite{R33}, Pantheon \cite{R34}, and the most recent one, Pantheon+SH0ES \cite{R35}.

The $\chi^2$ function associated with the Pantheon+SH0ES data points is as follows,
\begin{equation}\label{4ca}
\chi^2_{SN}=  D^T C^{-1}_{SN} D.
\end{equation}
Here, $C_{SN}$ denotes the covariance matrix \cite{R35} and the vector $D$ is described as $D=m_{Bi}-M-\mu^{th}(z_i)$, where $m_{Bi}$ and $M$ are the apparent magnitude and the absolute magnitude, respectively.
Furthermore, the theoretical value of the distance modulus is given by
\begin{equation}\label{4d}
\mu(z)= 5log_{10} \left[ \frac{D_{L}(z)}{1 Mpc}  \right]+25 \text{,} 
\end{equation}
where 
\begin{equation}\label{4e}
D_{L}(z)= c(1+z) \int_{0}^{z} \frac{ dx}{H(x,\theta)} \text{.}
\end{equation}
Here, $\theta$ stands for the parameter space.
Unlike the Pantheon dataset, the Pantheon+SH0ES dataset effectively addresses the degeneracy between the parameters $H_0$ and $M$ by redefining the vector $D$ as follows,
\begin{equation}\label{4f}
\bar{D} = \begin{cases}
     m_{Bi}-M-\mu_i^{Ceph} & i \in \text{Cepheid hosts} \\
     m_{Bi}-M-\mu^{th}(z_i) & \text{otherwise}
    \end{cases}   \text{.}
\end{equation}
Here, $\mu_i^{Ceph}$ independently estimated using Cepheid calibrators. Hence, the relation \eqref{4ca} becomes $\chi^2_{SN}=  \bar{D}^T C^{-1}_{SN} \bar{D} $.\\

\subsubsection{CC+Pantheon+SH0ES dataset}
\justify
We used the combined CC+Pantheon+SH0ES dataset to get the estimated values of the model and the bulk viscous parameters. The chi-squared function for the CC+Pantheon+SH0ES dataset is $\chi_{total}$=$\chi_{CC}^{2}+\chi_{SN}^{2}$.

\begin{figure}[H]
\centering
\includegraphics[width=16cm,height=17cm]{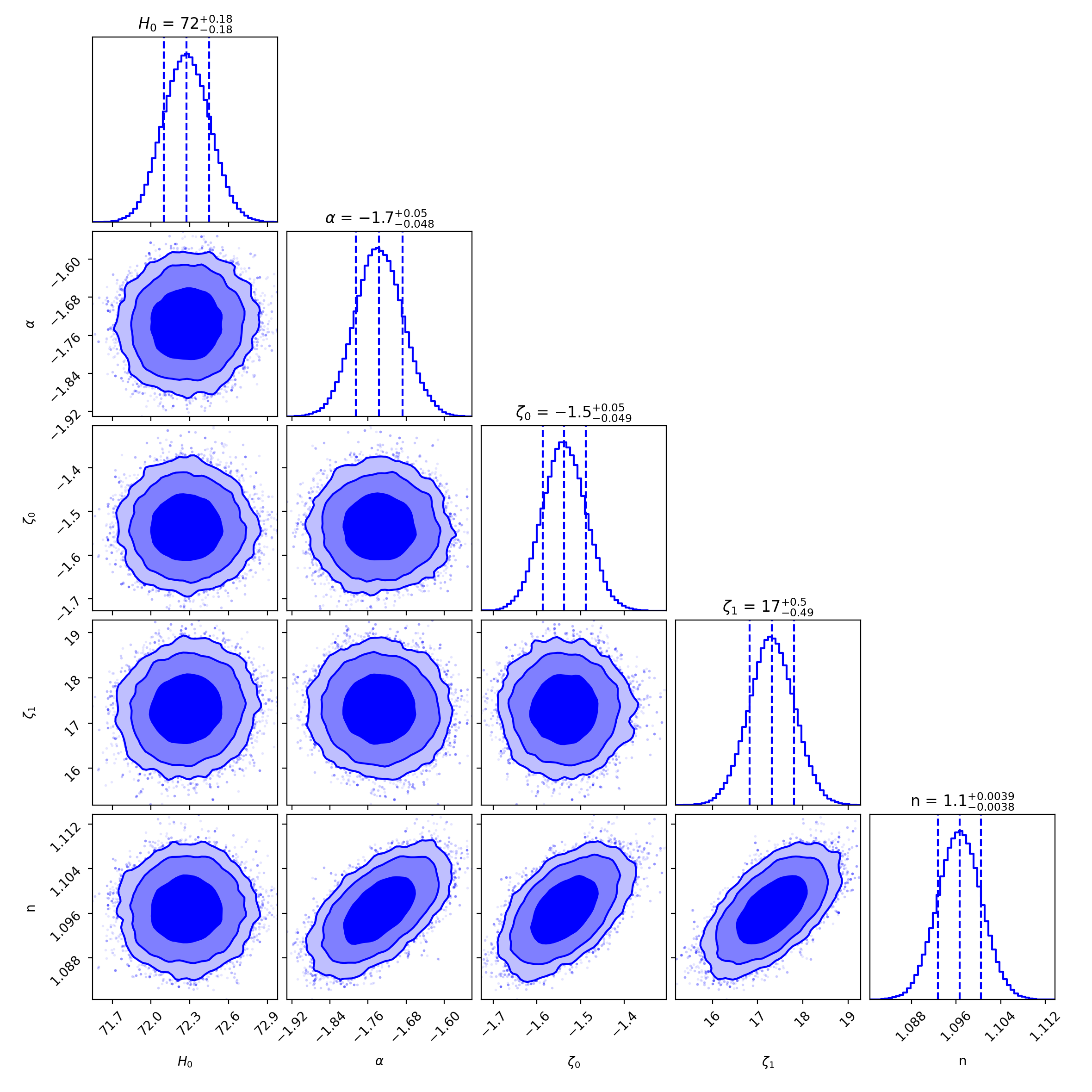}
\caption{Constraints on parameters at $1-\sigma$, $2-\sigma$ and $3-\sigma$ confidence interval using combined CC+Pantheon+SH0ES dataset.}
\label{cha2f1}
\end{figure}

The estimated values obtained for the model parameter and the viscosity parameters using the CC+Pantheon+SH0ES dataset are $H_0=72^{+0.18} _{-0.18}$, $\alpha=-1.7^{+0.05} _{-0.048}$, $\zeta_0=-1.5^{+0.05} _{-0.049}$, $\zeta_1=17^{+0.05} _{-0.049}$, $n=1.1^{+0.0039} _{-0.0038}$ and we got the minimum value of ${\chi^2}_{total} $ as ${\chi^2}_{min} = 1633.683$.
\subsubsection{Model comparison}
\justify

It is essential to conduct a statistical analysis using the Akaike Information Criterion (AIC) and Bayesian Information Criterion (BIC) to assess the robustness of our MCMC analysis. The expression for AIC is as follows,

\begin{equation}
 AIC={\chi^2}_{min}+2d \text{.}
\end{equation}
Here, $d$ indicates the number of parameters in the specified model. To compare our model with the standard model $\Lambda CDM$, we define $\Delta AIC = AIC_{Model} - AIC_{\Lambda CDM}$. The value of $\Delta AIC $ less than 2 indicates substantial support for the proposed theoretical model, while the range $4 < \Delta AIC < 7 $ indicates moderate support. Furthermore, if the value of $ \Delta AIC $ is greater than 10, it indicates a lack of support for the proposed model. The second criterion, BIC, can be described as follows,
\begin{equation}
 BIC={\chi^2}_{min}+dln(N) \text{.}
\end{equation}
In this context, $N$ denotes the number of data samples utilized in the MCMC analysis. Similarly, if $\Delta BIC$ is less than 2, it indicates a high support for the proposed theoretical model, and for $2 < \Delta BIC < 6 $, it indicates a moderate support. The values obtained from $AIC$ and $BIC$ for the theoretical model are $AIC_{Model}=1643.683$ and $BIC_{Model}=1670.968$. Also, we got $\Delta AIC=0.515 $ and $\Delta BIC=15.856 $. Here, The $\Lambda CDM$ values of $AIC$ and $BIC$ are taken as $AIC_{\Lambda}=1644.198$ and $BIC_{\Lambda}=1655.112$.
Thus, the value $\Delta AIC$ clearly indicates strong support for the proposed theoretical $f(Q)$ model. However, as is often known, a high $\Delta BIC$ value can be offset by a large number of parameters.

\section{Evolutionary behavior of cosmological parameters}\label{cha2sec4}
\justifying

We analyze the evolutionary behavior of several key cosmological parameters, including the effective EoS, statefinder parameters, and the Om diagnostic parameter at constraints values of parameters obtained from the combined CC+Pantheon+SH0ES dataset.

\begin{figure}[H]
\centering
\includegraphics[width=11cm,height=6cm]{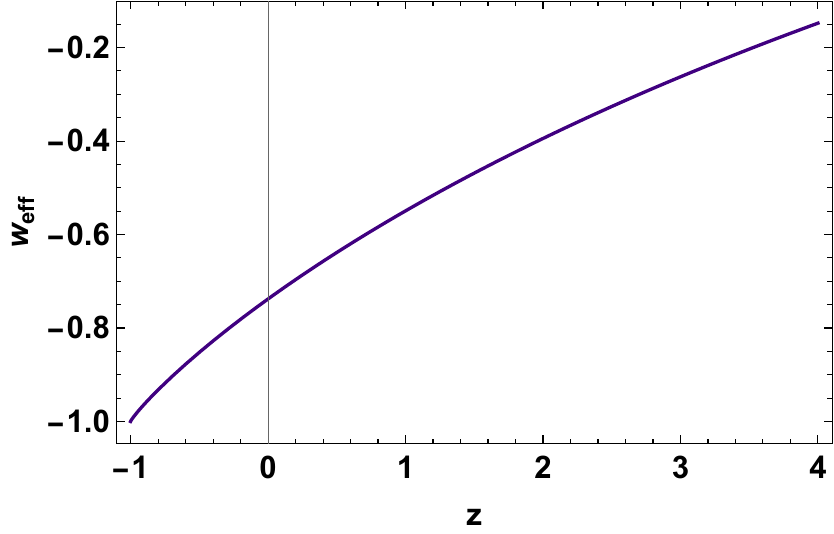}
\caption{The figure represent the behavior of the EoS parameter.}
\label{f7a}
\end{figure}

The effective equation for the behavior of the state parameter, illustrated in Figure \eqref{f7a}, indicates the accelerating expansion phase of the Universe. The present value of the EoS parameter for our model is obtained as $\omega_0 \approx -0.7378 $ for the combined CC + Pantheon+SH0ES sample. The present value obtained of the EoS parameter is quite consistent with recent cosmological investigations.  
The statefinder diagnostic provides a geometric method for differentiating various DE models. It is based on the statefinder parameters, denoted as $r$ and $s$. These parameters are  constructed solely from the scale factor and its time derivatives. The statefinder parameters are defined as follows,
 \begin{equation}\label{5a}
  r =\frac{\dddot{a}}{aH^3} \:\: \text{and} \:\: s=\frac{(r-1)}{3(q-\frac{1}{2})} \text{.}
\end{equation}

 The values $(r < 1, s > 0)$ represents the quintessence DE, whereas the region $(r > 1, s < 0)$ represents the phantom scenario and the values $(r = 1, s = 0)$ represents the standard $\Lambda$CDM model.

\begin{figure}[H]
\centering
\includegraphics[width=10cm,height=10cm]{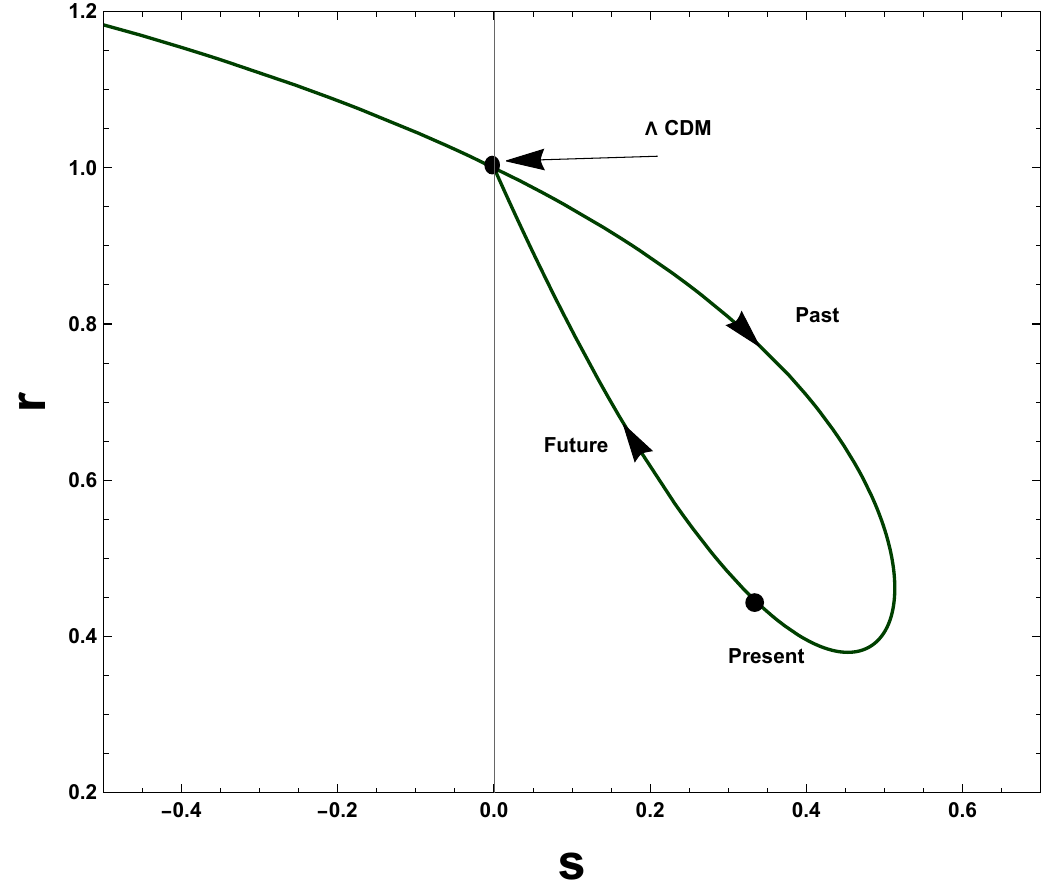}
\caption{The figure shows the behavior of the statefinder parameters in the $r-s$ plane.}
\label{f8a}
\end{figure}

The behavior of statefinder parameters in the $r-s$ plane is presented in Figure \eqref{f8a}. The present time values of the statefinder parameters for our model are $(r_0,s_0) = (0.44,0.33)$, corresponding to the combined CC+Pantheon + SH0ES sample. Hence, the present time values of the statefinder parameters of the viscous model considered favor quintessence-type behavior.

The Om diagnostic serves as a simple testing method that depends only on the first-order derivative of the cosmic scale factor. For a spatially flat Universe, its expression is as follows,
\begin{equation}\label{5b}
 Om(z)= \frac{\big(\frac{H(z)}{H_0}\big)^2-1}{(1+z)^3-1}\text{.}
 \end{equation}

The slope of the $Om(z)$ curve can tell us about the behavior of the assumed viscous model. If the slope of the $Om(z)$ curve is descending throughout the domain, it indicates quintessence-like behavior of the model, while the ascending slope of the $Om(z)$ curve indicates the phantom behavior of the model. On the other hand, a constant $Om(z)$ represents the $\Lambda$CDM model.

\begin{figure}[H]
\centering
\includegraphics[width=11cm,height=6cm]{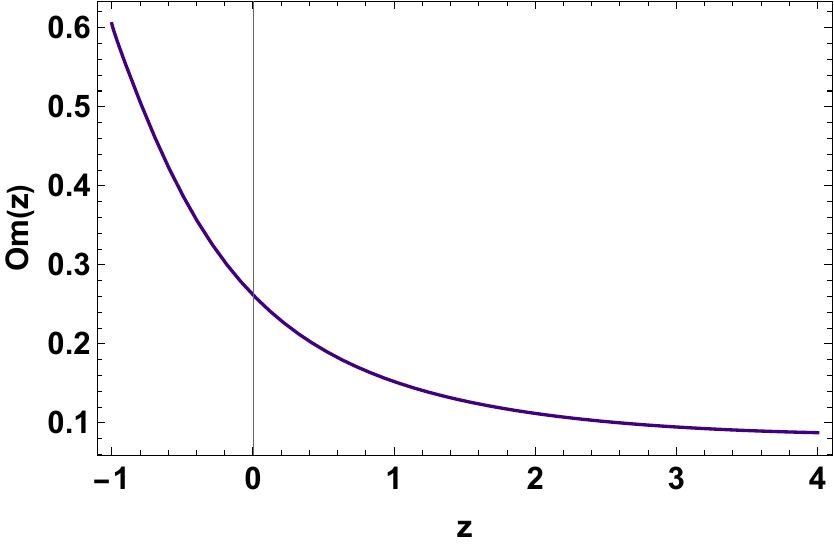}
\caption{The figure displays the behavior of the Om diagnostic parameter}
\label{f9a}
\end{figure}

The behavior of the $Om(z)$ parameter for the assumed viscus $f(Q)$ model is represented in Figure \eqref{f9a}. The slope of the $Om(z)$ curve descends throughout the domain for the estimated values of the parameters. Therefore, on the basis of Om diagnostic test, we can say that our viscous fluid cosmological model shows quintessence-like behavior.
 
\section{Conclusions}\label{cha2sec5}

\justifying

 The study of $f(Q)$ gravity has gained substantial attention from cosmologists, encompassing a range of topics, including wormholes, black holes, late-time acceleration, inflation, etc. Meanwhile, cosmological models involving viscous fluids have gained significant interest, particularly for their description of the Universe's early stages and offering insights into late-time expansion. Considering hydrodynamics, incorporating the influence of viscosity in the cosmic fluid is a reasonable assumption given that a perfect fluid ultimately is an idealization. In this study, we explored the significance of viscosity coefficients in explaining the observed cosmic acceleration in the $f(Q)$ gravity background.

We begin with the power law model $f(Q)=\alpha Q^n $, where $\alpha$ and $n$ are arbitrary constants, along with the fluid part incorporating the bulk viscosity coefficient $\zeta= \zeta_0 \rho H^{-1} + \zeta_1 H $. The analytical solutions of the corresponding field equations for a flat FLRW environment are presented in equation \eqref{3oa}. The free parameters of the obtained solutions have been constrained using the CC+Pantheon+SH0ES sample.
We performed Bayesian statistical analysis to estimate the posterior probability utilizing the likelihood function and the MCMC random sampling technique. We have obtained estimated values of the model parameter and viscous parameters by using CC+Pantheon+ sample. In addition, the contour plots for the free parameters $H_0, \alpha, \zeta_0$, $\zeta_1$ and $n$ within the $1\sigma-3\sigma$ confidence interval are presented in Figure \eqref{cha2f1}. Moreover, we performed a statistical analysis using the AIC and BIC to assess the robustness of our MCMC analysis. We obtain $\Delta AIC=0.515 $ and $\Delta BIC=15.856 $.
Thus, the value $\Delta AIC$ clearly indicates strong support for the proposed theoretical $f(Q)$ model, whereas a high value $\Delta BIC$ can be offset by a large number of parameters due to the presence of a factor $d$ in multiplication of $ln(N)$. 
We also examined the evolutionary behavior of some prominent cosmological parameters. The effective EoS parameter predicts the accelerating behavior of the expansion phase of the Universe (see Figure \eqref{f7a}). The value of the EoS parameter in the present redshift $(z=0)$ is $\omega_0 \approx -0.7378 $ corresponding to the combined CC+Pantheon + SH0ES sample.
 
Furthermore, Figure \eqref{f8a} illustrates the behavior of the assumed viscous $f(Q)$ model in the $r-s$ plane. The corresponding present time value of the statefinder parameters for our model is $(r_0,s_0) = (0.44, 0.33)$ corresponding to the model parameter constraints obtained by the combined CC+Pantheon+SH0ES sample. Hence, the statefinder parameters  favor quintessence-type behavior. Lastly, in Figure \eqref{f9a}, we illustrate the behavior of the $Om(z)$ curve, which represents a consistent negative slope across the entire domain for our assumed model. Thus, it can be inferred that our assumed viscous fluid $f(Q)$ models embodies quintessence-like behavior and can successfully describe the late-time scenario. In this section, we update the constraints on the bulk viscous and model parameters utilizing Pantheon+SH0ES samples along with the CC sample, while the existing literature \cite{ADD3} considered only Hubble and BAO samples but not the supernovae samples, while the references \cite{ADD6} utilized old Pantheon samples. Moreover, the $H_0$ constraint corresponding to the Hubble, Pantheon, and joint samples obtained in references \cite{ADD5,ADD6} is almost identical (nearly $H_0 \approx 67 \: km/s/Mpc$) which is not consistent. Since the discrepancy in the value $H_0$ for the Hubble and Pantheon sample is well known in the literature, known as the $H_0$ tension. In the present chapter, we obtained $H_0 \approx 72 \: km/s/Mpc$ for the joint analysis that is less tension. For future work, one can thoroughly investigate the $\sigma_8$ and $S_8$ tension. \\



 \chapter{Asymptotic behavior and observational bounds of $f(Q)$ gravity with $\Omega$-dependent viscosity} 

 \label{Chapter3} 

 \lhead{Chapter 3. \emph{Asymptotic behavior and observational bounds of $f(Q)$ gravity with $\Omega$-dependent viscosity}} 


 \vspace{10 cm}
 * The work, in this chapter, is covered by the following publication:

 \textit{Power law $f(Q)$
  cosmology with bulk viscous fluid}, Universe \textbf{536}, 2400072 (2024).
 \clearpage
In this chapter, we explore a power law model $f(Q)$, specifically $f(Q)= \alpha Q^n$, together with a viscous matter fluid that has a transport coefficient $\zeta = \zeta_0 \sqrt{\Omega} + \zeta_1 \Omega H $. The corresponding analytical solution has been derived and then confronted with recent cosmic data. We utilized the MCMC sampling technique to estimate the mean value of arbitrary parameters by incorporating CC and the recently published Pantheon+ dataset. In addition, we reconstructed some cosmological parameters by resampling the chains obtained by emcee, incorporating $6000$ samples. We find that the matter-energy density depicts the expected positive behavior, whereas the effective pressure indicates the negative behavior that is leading to the accelerating expansion, which is further predicted in the effective EoS parameter. In addition, we investigate the asymptotic nature of the assumed model by invoking a phase-space analysis. We conclude that the assumed viscous $f(Q)$ model successfully predicts an evolution of the Universe from a decelerated epoch to a stable accelerated de-Sitter epoch.

\section{Introduction}\label{sec1}

Over the past two decades, the standard $\Lambda$CDM model has been widely accepted as an accurate representation of the observed cosmological scenario, with GR serving as the background geometry. However, the model has been struggling to reconcile the value of the cosmological constant $\Lambda$, indicating that the standard GR may not be sufficient to address issues related to dark matter and DE. As a result, there has been a growing interest in exploring alternative theoretical models of cosmology that can describe cosmological observations with high precision.
The theory of gravitation based on curvature, specifically, GR and its various extensions, including $f(R)$, $f(R, G)$, and $f(R,T)$ gravity, has been widely explored \cite{cantata,modifiedgrav}. Another promising approach to modify the geometry of the background is by formulating it on a flat spacetime, and the central object manifesting the gravitational interactions is either the torsion or the non-metricity. The former approach is called metric teleparallel theories, while the second one is widely known as symmetric teleparallel theories. The symmetric teleparallel theories introduced by Nester \cite{R50} was further developed by Jimenez et al. to formulate the $f(Q)$ gravity. The intrinsic presence of ghosts in the symmetric teleparallel framework along with the number of propagating degrees of freedom in these theories is discussed in \cite{ADDP1}. In the context of $f(Q)$ cosmology, the study \cite{ADDP2} reveals that, for a particular background connection, where all seven modes propagate, there is at least one ghost degree of freedom. For all other choices of the connection, the ghost can be avoided at the cost of a strong coupling problem, where only four degrees of freedom propagate. Hence, all of the cosmologies within the symmetric teleparallel background suffer either from strong coupling or from ghost instabilities. However, the $f(Q)$ theory had been successfully investigated as a toy model. Recently, a significant amount of research has been conducted on $f(Q)$ theories  \cite{jimenez2,lcdm,accfQ3, fQfT3, gde, ad/ec, ad/bianchi,ad/ec, signa,perturb}.

The present studies explore a viscous cosmological $f(Q)$ model with recent observational constraints to probe the accelerating expansion of the cosmos. The present chapter is organized as follows. In sec \ref{cha3sec2}, we present the analytic solution of the  power law $f(Q)$ model along with the cosmic matter fluid that has viscous effects. In sec \ref{cha3sec3}, we estimate the mean value of arbitrary parameters of our $f(Q)$ cosmological model, utilizing the CC and recent Pantheon+ dataset. In sec \ref{cha3sec4},  we investigate the asymptotic nature of the assumed model by invoking phase-space analysis. Lastly, in sec \ref{cha3sec5}, we conclude our findings.

\section{The cosmological $f(Q)$ model in FLRW background with viscous matter}\label{cha3sec2}

 We begin with the following homogeneous and isotropic flat FLRW line element \eqref{FLRW} given in cartesian coordinates. The gauge considered in the line element \eqref{FLRW} is a coincident gauge coordinate, and thus it follows that the metric is only a fundamental variable.

We consider a cosmic fluid with bulk viscosity, where the dissipative phenomenon is driven by transport coefficients. It turns out that the viscosity of the matter content of the Universe provides a fraction of the kinetic energy of the cosmic expansion. The transport coefficient $\zeta$ emerging as a viscous effect in late-time acceleration is given as  $\zeta=\zeta_0 + \zeta_1 H$ \cite{IB-3}. Another widely discussed transport coefficient is $\zeta \approx \rho^s$ \cite{vs2}. Moreover, for the matter era, it turns out to be $s=\frac{1}{2}$. However, a growing Hubble parameter corresponds to the early phase of the cosmos. Therefore, we consider the following viscosity coefficient that incorporates both early and late time phenomena introduced in \cite{vsform},
\begin{equation}\label{3cc}
\zeta = \zeta_0 \sqrt{\Omega} + \zeta_1 \Omega H   \text{.}
\end{equation}
Note that the viscosity coefficient $\zeta_0$ has dimension $[ML^{-1}T^{-1}]$, while $\zeta_1$ has dimension $[ML^{-1}]$. One can observe that equation \eqref{3cc} is a simple linear combination of the two transport coefficients mentioned above and preserves property $\zeta \approx \rho^s$ as well as reduces to the form $\zeta=\zeta_0 + \zeta_1 H$. Now, for the matter dominated era $\zeta$ would give a contribution like $\rho^{\frac{1}{2}}$ and for the late time acceleration $\zeta$ takes the form  $\zeta = \zeta_0  + \zeta_1 H  $. Hence such a choice of viscous coefficient has advantage of explaining both matter dominated as well as late time acceleration of Universe.

Now, the Friedmann equations in GR like form for an arbitrary $f(Q)$ function incorporating aforementioned viscous effects in the cosmic fluid are given as in \eqref{fqfe1} and the second field equation \eqref{fqfe2} can be written as follows,
\begin{equation}\label{3f}
2\dot{H}+3H^{2} = -p_{eff} = \frac{1}{2f_{Q}} \left( 2p_v + f+Q f_Q - 4\dot{f_{Q}} H \right) \text{.}
\end{equation}

\justify We assume the following power law $f(Q)$ model \cite{RSU},
\begin{equation}\label{3ha}
f(Q)= \alpha Q^n  \text{.}
\end{equation}
Here, $n$ and $\alpha$ are arbitrary parameters. The motivation behind the considered power law $f(Q)$ model is the dynamical investigations performed in \cite{ADDP5,ADDP6,ADDP7} utilizing different connection settings that successfully describe various early Universe epochs along with the de-Sitter type expansion phase. However, in this work, we restrict our analysis to the coincident gauge connection in order to perform observational analysis of the obtained solution, since the other two connections involve highly non-linear complex equations that make it impossible to find analytical solution. Then the Friedmann equations correspond to the power law function \eqref{3ha} for a pressureless Universe becomes,
\begin{equation}\label{3iii}
2\rho=\left(1-2n \right) \alpha 6^n H^{2n}  \text{,}
\end{equation}
and
\begin{equation}\label{3jjj}
\dot{H} + \frac{3}{2n} H^2 = \frac{p_v}{2n(2n-1)\alpha 6^{n-1}H^{2n-2}}     \text{.}
\end{equation}
 Now, in evaluating \eqref{3iii} in the present $z=0$, we obtain the following,
 \begin{equation}\label{3kkk}
 \alpha= \frac{\Omega_0}{(1-2n) 6^{n-1}H_0^{2n-2}}    \text{.}
 \end{equation}
Thus, solving \eqref{3iii} and \eqref{3jjj} with \eqref{3kkk}, we have
\begin{equation}\label{3ll}
\rho=3\Omega_0 H_0^2 \left(\frac{H}{H_0}\right)^{2n} \text{,}
\end{equation}
and
\begin{equation}\label{3mm}
\dot{H}+\frac{(3-\bar{\zeta_1})}{2n} H^2=\frac{\bar{\zeta_0} {H_0}^n}{2n \sqrt{\Omega_0}H^{n-2}} \text{.}
\end{equation}
Here, $\bar{\zeta_0}=\frac{3\zeta_0}{H_0}$, $\bar{\zeta_1}=3\zeta_1$, and $\Omega_0=\frac{\rho_0}{3H^2}$ are, respectively, dimensionless viscosity and matter density parameter. In further solving, we have the following,
\begin{equation}\label{3nn}
\frac{dH}{dlna}+\frac{(3-\bar{\zeta_1})}{2n} H=\frac{\bar{\zeta_0} {H_0}^n}{2n \sqrt{\Omega_0}H^{n-1}} \text{.}
\end{equation}
and the corresponding solution,
\begin{equation}\label{3oo}
H(z)=H_0 \Biggl\{(1+z)^{\frac{3-\bar{\zeta_1}}{2}} +\left(\frac{\bar{\zeta_0}}{3-\bar{\zeta_1}}\right) \frac{1}{\sqrt{\Omega_0}} \left[1-(1+z)^{\frac{3-\bar{\zeta_1}}{2}}\right]    \Biggl\}^{\frac{1}{n}} \text{.}
\end{equation}
In particular, one can obtain the usual GR solution for the parameter choices $n=1$ and $\bar{\zeta_0}=\bar{\zeta_1}=0$. Moreover, the standard $\Lambda$CDM solution can be obtained corresponding to the parameter choices $n=2$, $\bar{\zeta_0}= 6\sqrt{\Omega_0} (1-\Omega_0)$, and $\bar{\zeta_1}=-3$. 

\section{Model parameters estimation}\label{cha3sec3}

In this section, we intend to estimate the mean value of arbitrary parameters of our $f(Q)$ cosmological model. We utilize the MCMC sampling technique along with Bayesian analysis and Scipy optimization method in Python package emcee \cite{Mackey/2013}.

\subsubsection{Cosmic chronometers}
\justify
The CC datasets incorporate the $H(z)$ measurements of passively evolving massive galaxies covering the redshift range $0.07 \leq z \leq 2.41$ and are estimated using the differential age method \cite{CYU}. The list of $H(z)$ measurements is available in the reference \cite{RS}. The $\chi^2$ function for CC dataset is given in equation \eqref{4acc}. 


\subsubsection{Pantheon+ dataset}
\justify
We utilized recently published Pantheon+ dataset that incorporate measurements of 1701 light curves acquired from the 1550 distinct samples of SN Ia covering the redshift range $0.001 \leq z \leq 2.27$ \cite{Brout}. The corresponding $\chi^2$ function is given as
\begin{equation}\label{4b}
\chi^2_{SN}=\sum_{i,j=1}^{1701}\bigtriangledown\mu_{i}\left(C^{-1}_{SN}\right)_{ij}\bigtriangledown\mu_{j}  \text{.}
\end{equation}
Here, $C_{SN}$ is the covariance matrix \cite{Brout} and
\begin{align*}\label{4c}
\quad \bigtriangledown\mu_{i}=\mu^{th}(z_i,\theta)-\mu_i^{obs}.
\end{align*} 
We estimate the distance modulus for the $f(Q)$ model considered using the following relation
\begin{equation}\label{4d}
\mu(z)= 5log_{10} \left[ \frac{D_{L}(z)}{1 Mpc}  \right]+25  \text{.} 
\end{equation}
where $D_{L}(z)$ represents the luminosity distance and can be calculated as,
\begin{equation}\label{4e}
D_{L}(z)= c(1+z) \int_{0}^{z} \frac{ dx}{H(x,\theta)} \text{.}
\end{equation}
Here, $\theta$ is the usual parameter space.

\begin{figure}[H]
\centering
\includegraphics[width=17cm,height=18cm]{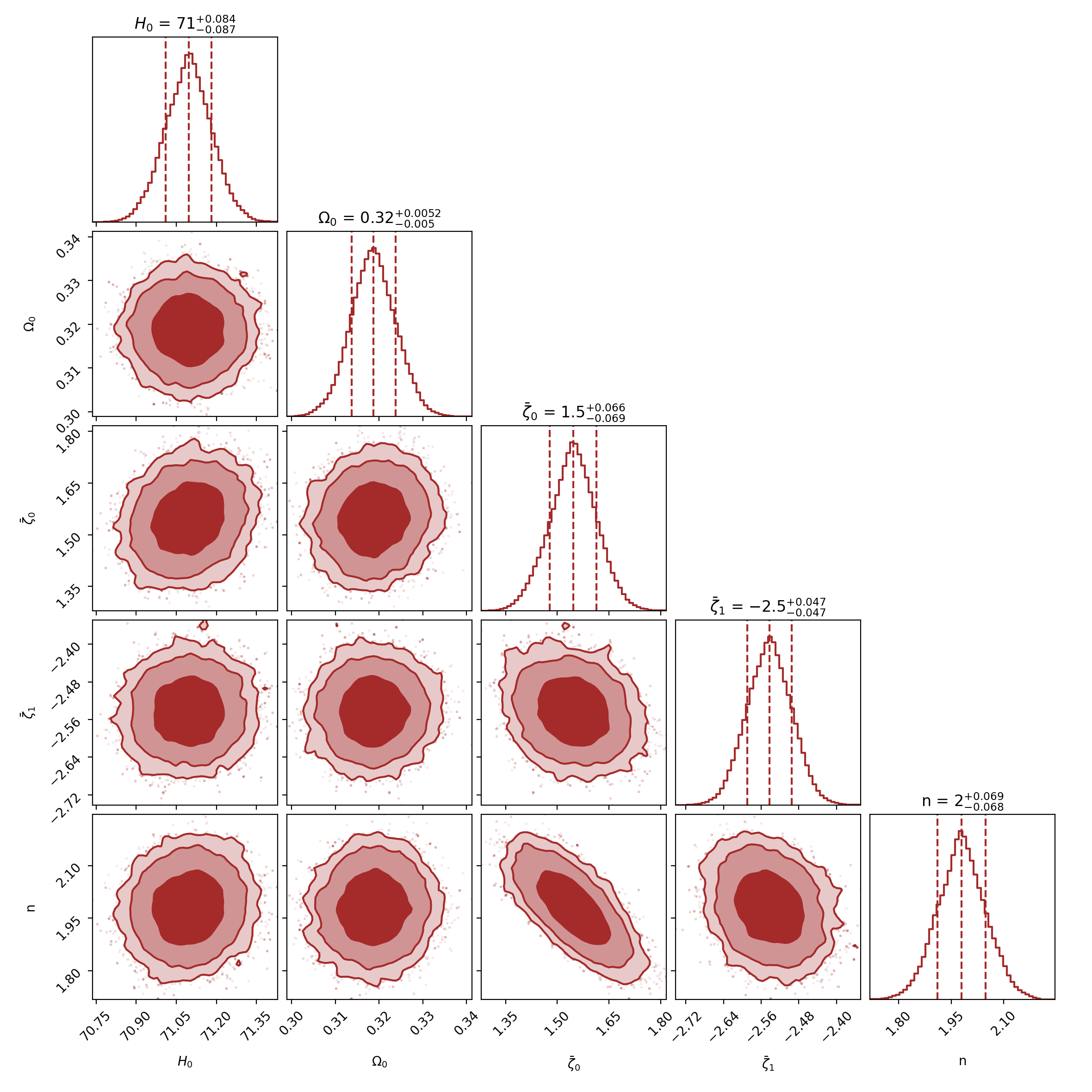}
\caption{The $1-\sigma$ to $3-\sigma$ contour for the considered viscous cosmological $f(Q)$ model using CC+Pantheon+ dataset.}
\label{cha3f1}
\end{figure}

\justify Now, the total $\chi^{2}$ function corresponding to CC+Pantheon+ dataest reads as
\begin{equation}\label{4f}
\chi^{2}= \chi_{CC}^{2} + \chi^2_{SN}  \text{.}
\end{equation}
We minimize the total $\chi^{2}$ function that is equivalent to maximize the total likelihood function, usually given as $\mathcal{L}= \exp(-\chi^{2}/2)$. The $1-\sigma$ to $3-\sigma$ contour for the model parameters $\bar{\zeta_0}$, $\bar{\zeta_1}$, $n$, $\Omega_0$, and $H_0$ for the CC+Pantheon+ dataset is presented in Figure \eqref{cha3f1}.

\justify We obtain the mean value of the parameters as $H_0 = 71^{+0.084}_{-0.087} \: kms^{-1}Mpc^{-1}$, $\Omega_0 = 0.32 \pm 0.005$, $\bar{\zeta_0} = 1.5^{+0.066}_{-0.069}$, $\bar{\zeta_1} = -2.5 \pm 0.047$, and $n=2^{+0.069}_{-0.068}$. Now we present the profiles of matter-energy density, effective pressure, and the EoS parameter along with the dimensionless Hubble parameter.

\begin{figure}[H]
\centering
\includegraphics[width=13cm,height=8cm]{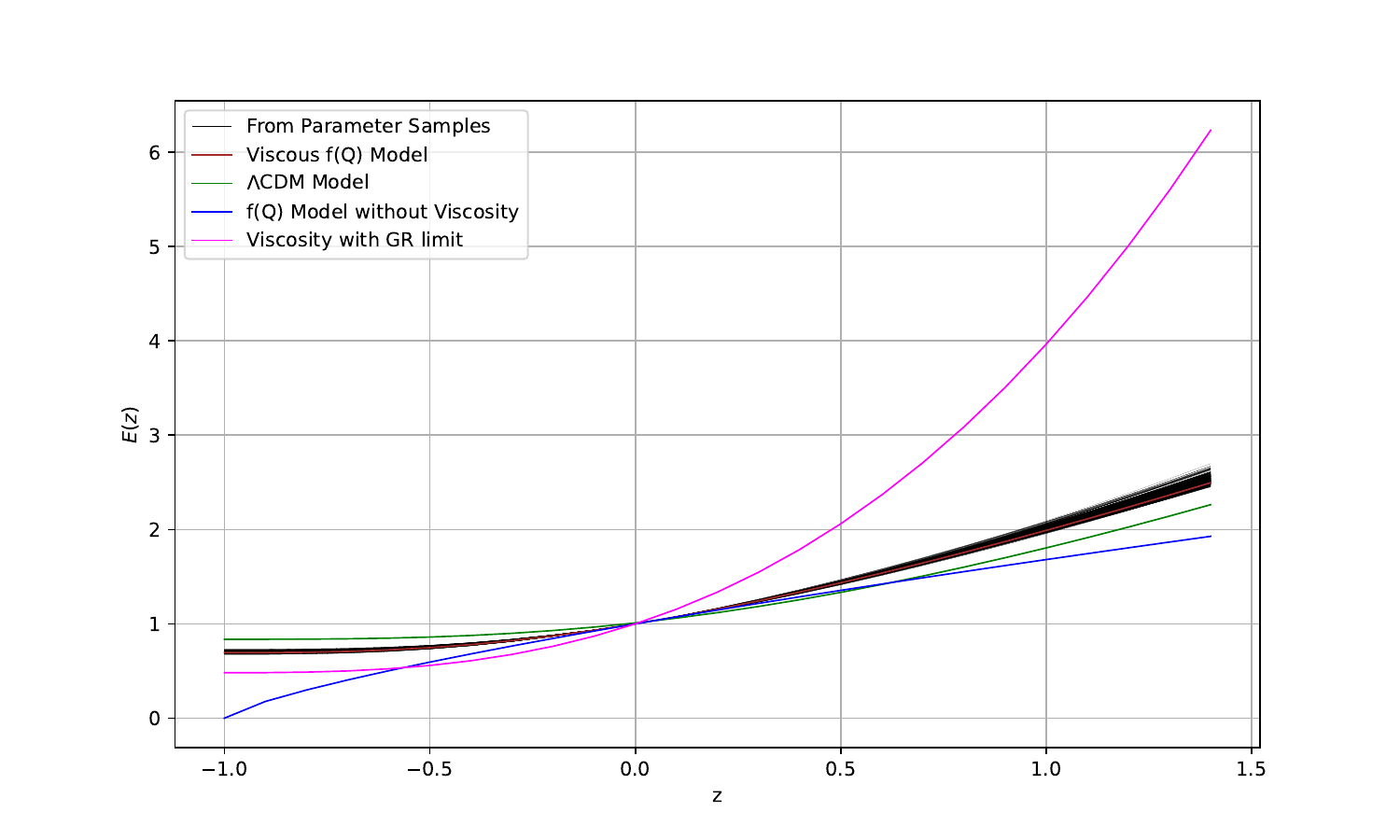}
\caption{Behavior of the dimensionless Hubble parameter produced by resampling the chains in emcee, utilizing 6000 samples (black curves) with mean values of parameters (brown curve). The green curve represents the $\Lambda$CDM predictions for the values $H_0=69 \: kms^{-1}Mpc^{-1}$ and $\Omega_0=0.3$. The blue curve depicts the prediction of considered $f(Q)$ gravity model in the absence of viscosity, whereas the magenta curve represents the prediction of model in the GR limit along with viscosity coefficients.}
\label{cha3f2}
\end{figure}

\begin{figure}[H]
\centering
\includegraphics[width=13cm,height=8cm]{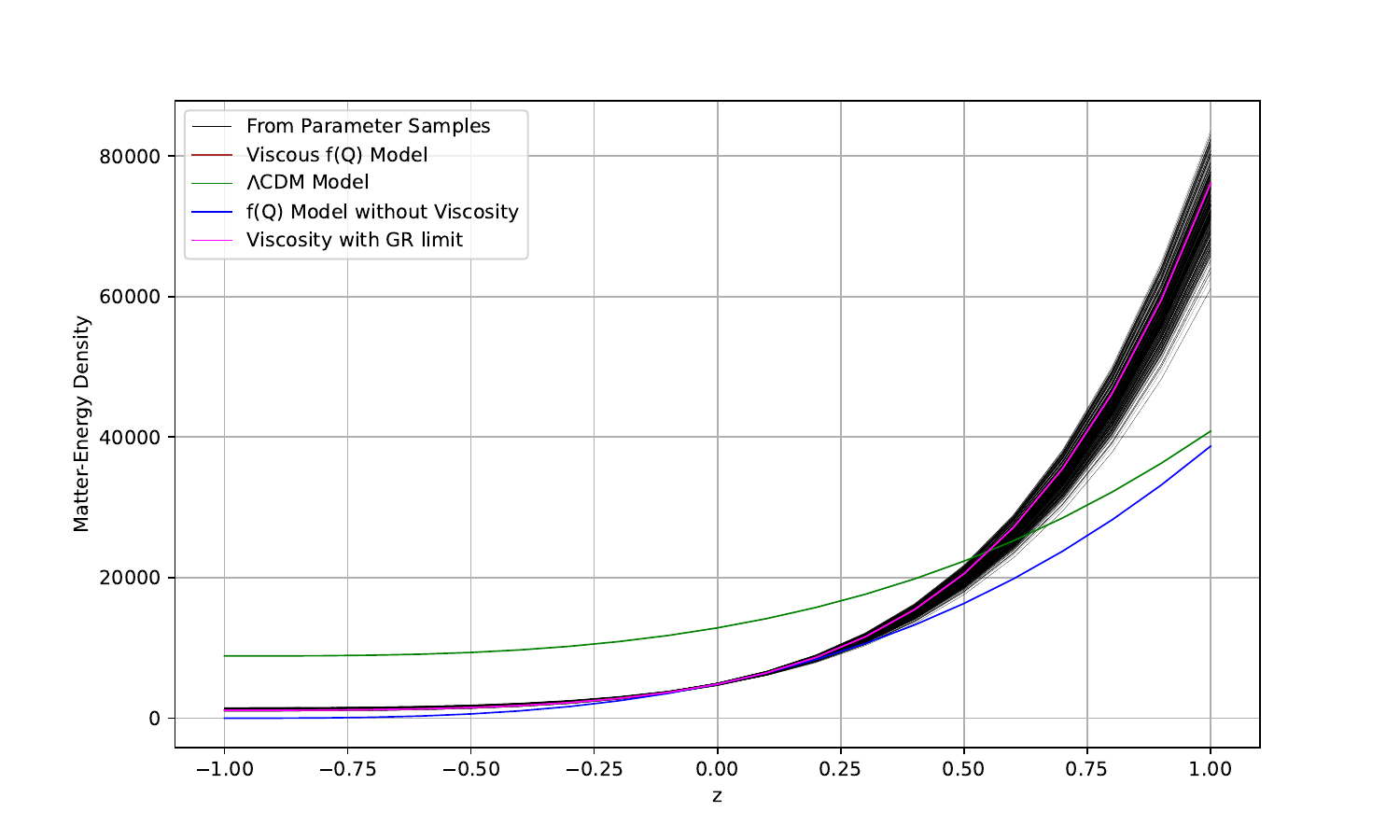}
\caption{Behavior of the matter-energy density \textbf{(in $kg m^{-3}$)} parameter produced by resampling the chains in emcee, utilizing 6000 samples (black curves) with mean values of parameters (brown curve). The green curve represents the $\Lambda$CDM predictions for the values $H_0=69 \: kms^{-1}Mpc^{-1}$ and $\Omega_0=0.3$. The blue curve depicts the prediction of considered $f(Q)$ gravity model in the absence of viscosity, whereas the magenta curve represents the prediction of model in the GR limit along with viscosity coefficients.}
\label{cha3f3}
\end{figure}

\begin{figure}[H]
\centering
\includegraphics[width=13cm,height=8cm]{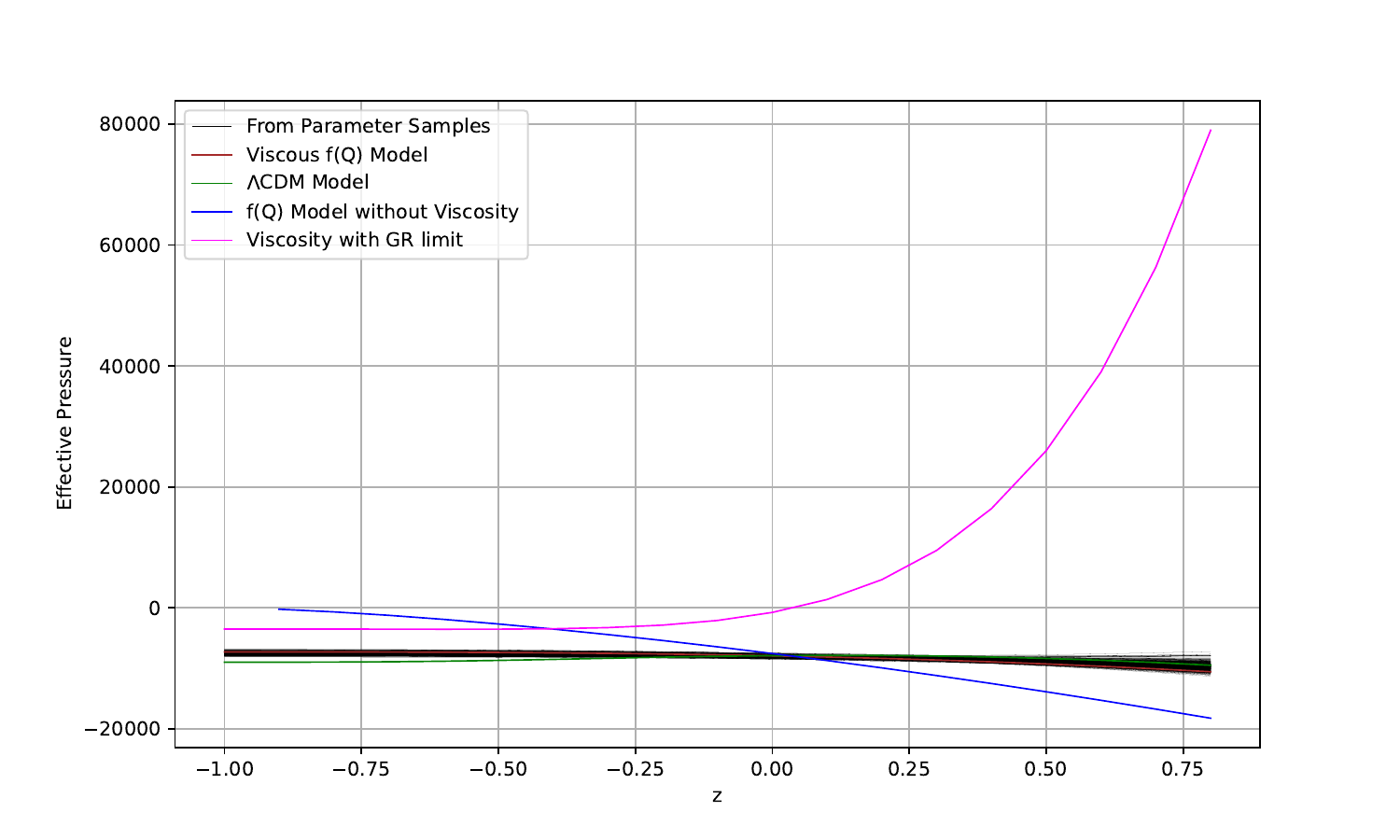}
\caption{Behavior of the effective pressure \textbf{(in Pascals)} produced by resampling the chains in emcee, utilizing 6000 samples (black curves) with mean values of parameters (brown curve). The green curve represents the $\Lambda$CDM predictions for the values $H_0=69 \: kms^{-1}Mpc^{-1}$ and $\Omega_0=0.3$. The blue curve depicts the prediction of the considered $f(Q)$ gravity model in the absence of viscosity, whereas the magenta curve represents the prediction of model in the GR limit along with viscosity coefficients.}
\label{cha3f4}
\end{figure}

\begin{figure}[H]
\centering
\includegraphics[width=13cm,height=8cm]{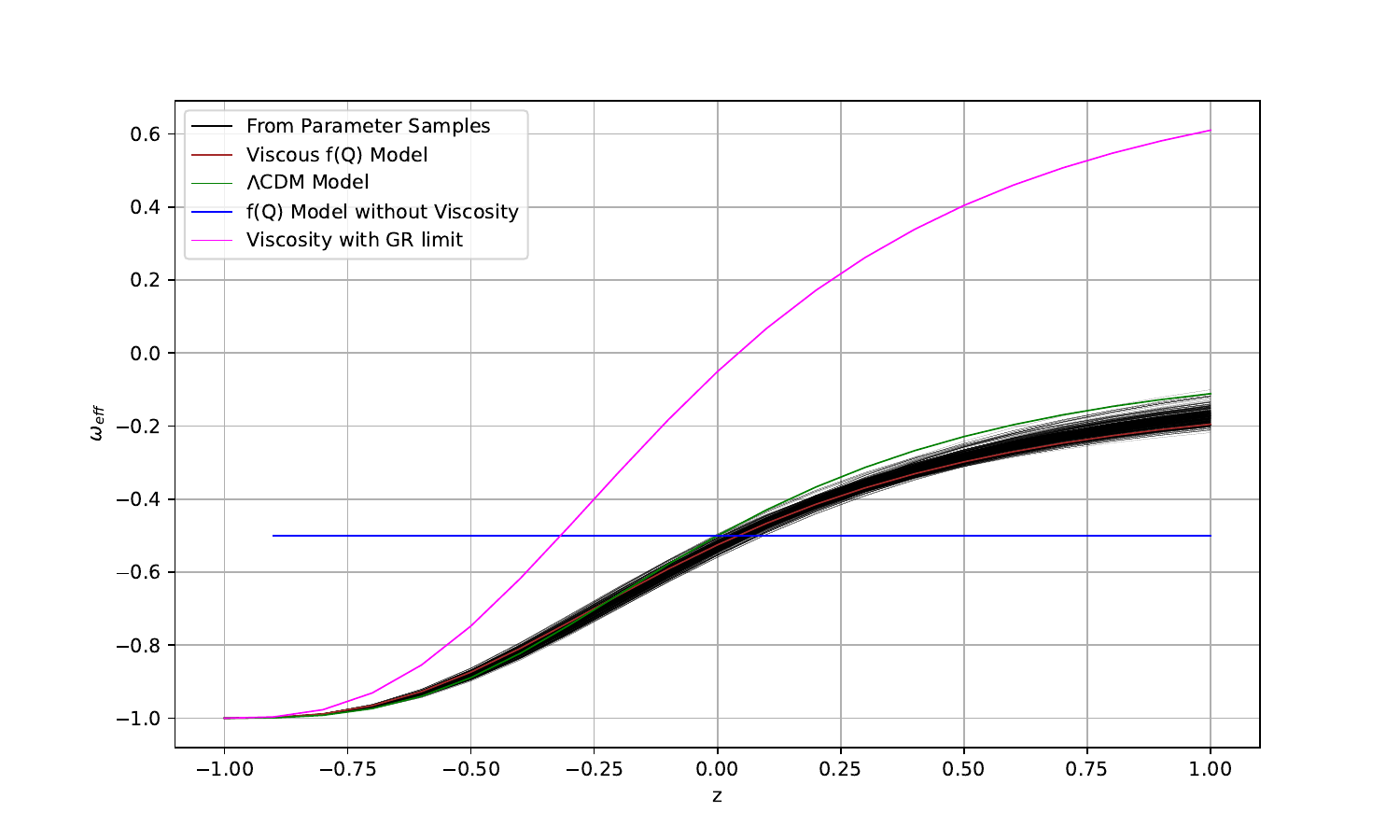}
\caption{Behavior of the effective EoS parameter produced by resampling the chains in emcee, utilizing 6000 samples (black curves) with mean values of parameters (brown curve). The green curve represents the $\Lambda$CDM predictions for the values $H_0=69 \: kms^{-1}Mpc^{-1}$ and $\Omega_0=0.3$. The blue curve depicts the prediction of considered $f(Q)$ gravity model in the absence of viscosity, whereas the magenta curve represents the prediction of model in the GR limit along with viscosity coefficients.}
\label{cha3f5}
\end{figure}

\justify We have resampled the chains by incorporating 6000 samples and plotted all corresponding curves (black curves) along with the mean value of the parameters obtained (brown curve). The reconstructed cosmological parameters obtained are presented in Figures \eqref{cha3f2}-\eqref{cha3f5}. Figure \eqref{cha3f2} depicts the behavior of the dimensionless Hubble parameter given by $E(z)= H(z)/H_0$. Note that the curve of the dimensionless Hubble parameter corresponding to the viscous model in the GR limit overestimates the predictions of the established $\Lambda$CDM model at high redshift, whereas the curve corresponding to the $f(Q)$ model when we switch off the viscosity terms underestimates its value in future regime. Thus, we found that the predictions of the combined viscous $f(Q)$ model considered represented by the brown curve efficiently mimic the standard $\Lambda$CDM model. Figure \eqref{cha3f3} depicts the expected positive behavior of matter-energy density and that will further vanish in the far future with the expansion of the Universe corresponding to each case. Moreover, Figure \eqref{cha3f4} indicates the negative behavior of the effective pressure that leads to the accelerated expansion predicted in the effective EoS parameter of our viscous $f(Q)$ model, presented in Figure \eqref{cha3f5}. Note that the effective pressure corresponding to the viscous model in the GR limit possesses a high positive value that is not suitable for the accelerating phase of the expansion. In Figure \eqref{cha3f5}, we further found that the $f(Q)$ model in the absence of viscosity can describe the accelerating phase of the expansion, but it fails to explain the early deceleration epoch, whereas the viscous model in the GR limit cannot describe the proper transition from the deceleration epoch to the acceleration epoch as it predicts the present state of the Universe to be decelerating (since $\omega_{eff} > -\frac{1}{3}$). The combined model, i.e., the $f(Q)$ gravity model in the presence of a viscous fluid, can successfully describe the present accelerating epoch (since $\omega_{eff} < -\frac{1}{3}$ at the present redshift $z=0$) and the early deceleration epoch (since $\omega_{eff} > -\frac{1}{3}$ at the redshift $z > 0.48$). Thus, the $f(Q)$ model alone or the viscous fluid in the GR limit cannot describe the late-time phenomenological evolution of the Universe, and hence the $f(Q)$ model considered together with the viscous fluid is found to be more physically viable.

\section{Asymptotic behavior of the model}\label{cha3sec4}

In this section, we explore the asymptotic nature of the assumed viscous cosmological $f(Q)$ model by invoking phase-space analysis. In order to do so, we transform the obtained Friedmann like equations into an autonomous form by utilizing phase-space variables. We define the following dimensionless variables as the governing phase-space variables,

\begin{equation}\label{5a}
x= \frac{\rho}{3\Omega_0 H_0^2 E(z)^{2n}}  \:\: \text{and} \:\:   y=\frac{E(z)}{E(z)+1} \text{.}
\end{equation}
Here, $E(z)= H(z)/H_0$ is the dimensionless Hubble parameter. Using equation \eqref{3ll} and the definition of $y$, we have constraints $x=1$ and $0 \leq y \leq 1$. Then using equations \eqref{3ll} and \eqref{3mm} with the matter-energy conservation equation, we obtained the following autonomous form corresponding to the variable $N=ln(a)$, 

\begin{equation}\label{5bbb}
x'=\frac{dx}{dN}=\frac{\bar{\zeta_0}(1-x)}{\sqrt{\Omega_0}} \left( \frac{1-y}{y} \right)^n = F(x,y),
\end{equation}
and
\begin{equation}\label{5ccc}
y'=\frac{dy}{dN}=\frac{(1-y)y}{2n} \left[ \bar{\zeta}_1 -3 + \frac{\bar{\zeta_0}}{\sqrt{\Omega_0}} \left( \frac{1-y}{y} \right)^n \right] = G(x,y)\text{.}
\end{equation}

\begin{figure}[H]
\centering
\includegraphics[width=10cm,height=10cm]{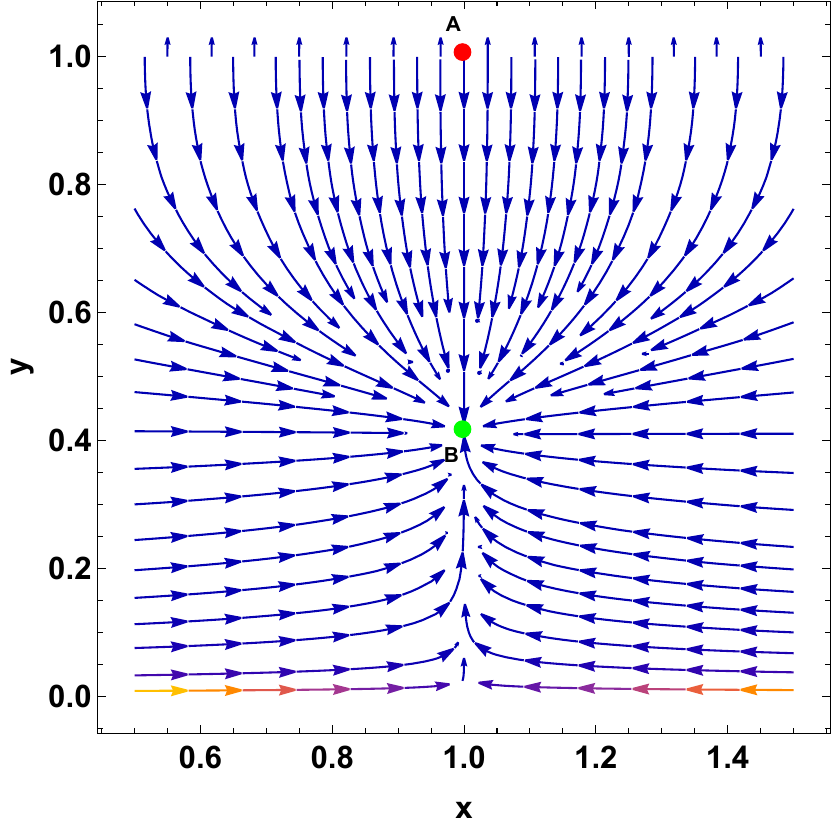}
\caption{Phase plot corresponding to viscous $f(Q)$ model with red dot represents the unstable critical point, whereas the green dot denotes the stable one.}
\label{cha3f6}
\end{figure}

\justify In addition, by the definition of EoS and the deceleration parameter, we have the following
\begin{equation}\label{5d}
q = -1 + \frac{1}{2n} \left[ 3-\bar{\zeta}_1 - \frac{\bar{\zeta_0}}{\sqrt{\Omega_0}} \left( \frac{1-y}{y} \right)^n \right] \text{,}
\end{equation}
and
\begin{equation}\label{5e}
\omega_{eff} = -1 + \frac{1}{3n} \left[ 3-\bar{\zeta}_1 - \frac{\bar{\zeta_0}}{\sqrt{\Omega_0}} \left( \frac{1-y}{y} \right)^n \right] \text{.}
\end{equation}
We obtain the following critical points $(x_c,y_c)$ by solving equations $x'=0$ and $y'=0$,
\begin{equation}\label{5fff}
(x_c,y_c)=(1,1) \:\: \text{and} \:\: (x_c,y_c)=\left( 1,\frac{1}{\left(-\frac{(\bar{\zeta_1}-3) \sqrt{\Omega_0}}{\bar{\zeta_0}} \right)^{1/n}+1} \right)\text{.}
\end{equation}
We linearize the autonomous system \eqref{5bbb} and \eqref{5ccc} by taking small perturbations near the critical points $(x,y)\longrightarrow (x_c+\delta x,y_c+\delta y)$ that satisfy the two indices
\begin{equation*}\label{5g}
\left[
\begin{array}{c}
\delta x'  \\ 
\delta y'  \\ 
\end{array}
\right] \,=   \left[
\begin{array}{cc}
\left( \frac{\partial F}{\partial x} \right)_{0} & \left( \frac{\partial F}{\partial y}  \right)_{0}  \\ 
\left( \frac{\partial G}{\partial x}  \right)_{0} & \left( \frac{\partial G}{\partial x}  \right)_{0} \\ 
\end{array}
\right]  \left[
\begin{array}{c}
\delta x \\ 
\delta y  \\ 
\end{array}
\right] \text{.}
\end{equation*}
Here, the suffix 0 denotes the critical points $(x_c,y_c)$ at which the above Jacobian matrix is estimated and that is given as,
\begin{small}
\begin{equation}\label{5h}
J= \left[
\begin{array}{cc}
\left( -\frac{\bar{\zeta_0} \left(\frac{1}{y}-1\right)^n}{\sqrt{\Omega_0}} \right)_{0} & \left( \frac{\bar{\zeta_0} n (x-1) (1-y)^{n-1} \left(\frac{1}{y}\right)^{n+1}}{\sqrt{\Omega_0}} \right)_{0}  \\ 
0 & \left( \frac{-\frac{\bar{\zeta_0} (n+2 y-1) \left(\frac{1}{y}-1\right)^n}{\sqrt{\Omega_0}}-2 \bar{\zeta_1} y+\bar{\zeta_1}+6 y-3}{2 n} \right)_{0} \\ 
\end{array}
\right] \text{.}
\end{equation}
\end{small}
\justify The respective eigenvalues corresponding to the critical points in equation \eqref{5fff} estimated using the above Jacobian $J$ are,
\begin{equation}\label{5i}
(\lambda_1,\lambda_2)= \left(0,\frac{3-\bar{\zeta_1}}{2} \right) \:\: \text{and} \:\: (\lambda_1,\lambda_2)= \left(\bar{\zeta_1}-3,\frac{\bar{\zeta_1}-3}{2} \right) \text{.}
\end{equation}
Incorporating the mean values of the parameters, we obtained the following behavior of the critical points corresponding to our viscous $f(Q)$ model, presented in Table \eqref{Table-1a} and Figure \eqref{cha3f6}.

\begin{table}[H]
\begin{center}\caption{Table shows the critical points and their behavior for the assumed viscous $f(Q)$ model.}
\begin{tabular}{|c|c|c|c|c|}
\hline
Critical Points $(x_c,y_c)$ & Eigenvalues $\lambda_1$ and $\lambda_2$ & Nature of critical point  & $q$ & $\omega_{eff}$ \\
\hline 
$A(1,1)$ & $ 0\:\: \text{and} \:\: 2.75$ & Unstable & $0.375$ & $-0.083$ \\
\hline
$B(1,0.403)$ & $-5.5 \:\: \text{and} \:\: -2.75$ & Stable & $-1$ & $-1$ \\
\hline
\end{tabular}\label{Table-1a}
\end{center}
\end{table}

\begin{figure}[H]
\centering
\includegraphics[width=13cm,height=8cm]{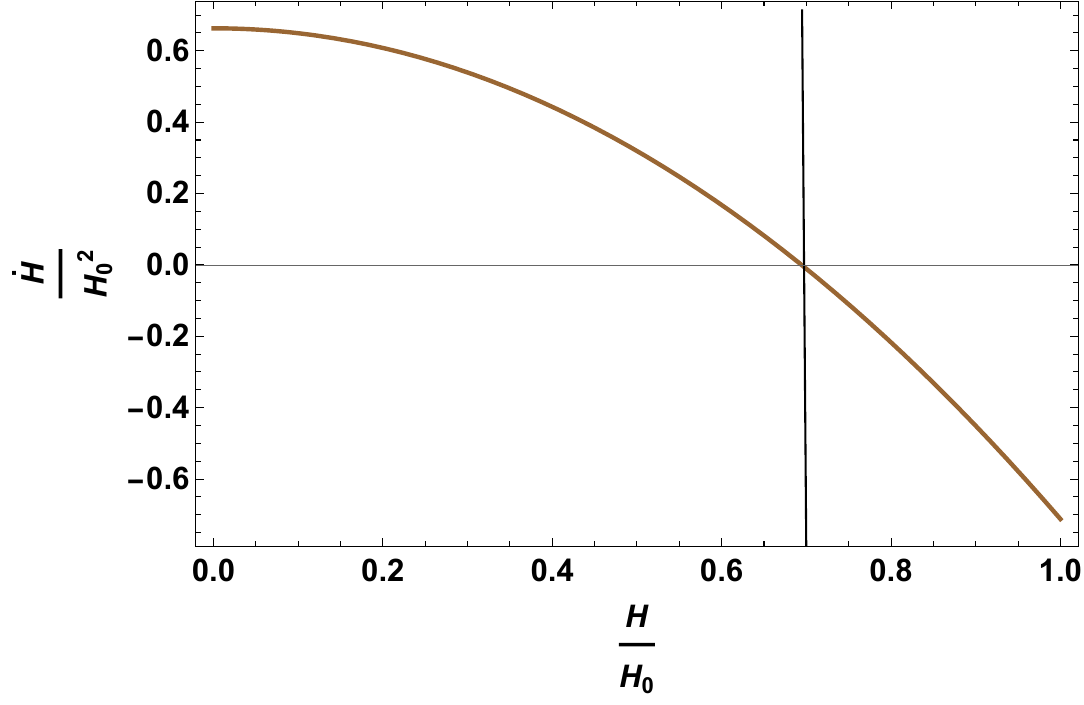}
\caption{$\dot{H}$ as a function of $\frac{H}{H_0}$ from equation \eqref{3mm} with obtained mean values of parameters, where vertical black lines indicating the fixed point that corresponds to $\dot{H}=0$.}
\label{cha3f7}
\end{figure}

From Table \eqref{Table-1a} and Figure \eqref{cha3f6}, it is evident that the critical point $A(1,1)$ indicates an unstable past attractor, while the point $B(1,0.403)$ is stable. Thus, the assumed viscous $f(Q)$ model with the obtained best-fit values of parameters (i.e. $H_0 = 71^{+0.084}_{-0.087} \: kms^{-1}Mpc^{-1}$, $\Omega_0 = 0.32 \pm 0.005$, $\bar{\zeta_0} = 1.5^{+0.066}_{-0.069}$, $\bar{\zeta_1} = -2.5 \pm 0.047$, and $n=2^{+0.069}_{-0.068}$), successfully represents an evolution of the Universe from a decelerated epoch to a stable accelerated de-Sitter epoch. Moreover, as presented in Figure \eqref{cha3f7}, we obtain $H \approx 0.69 H_0$ corresponding to the de-Sitter type fixed point. 

\section{Conclusions}\label{cha3sec5}

In standard cosmological models, such as the $\Lambda$CDM model, the cosmic fluid is often assumed to be a perfect fluid with no viscosity. The inclusion of bulk viscosity is relevant in the late-time Universe, especially when we are trying to understand the nature of DE. In some modified gravity theories, where the standard Einstein-Hilbert action is modified, the resulting field equations may involve additional terms representing viscosity-like effects. Considering viscous fluids in these theories allows for a more comprehensive exploration of alternative gravitational theories and their cosmological implications. For instance, in \cite{CPCP1} the author analyzed the Friedmann model with viscous cosmology in modified $f(R, T)$ gravity theory and shown that the bulk viscosity plays a major role in the expansion of the Universe by comparing the viscous models with the non-viscous one. In addition, in \cite{vsform} the author presented bulk viscosity as practical corrections to the perfect fluid models of baryonic and dark matter, since the material fluids in the real world have viscosity due to thermodynamics. In addition, in \cite{CPCP2,IB-1} the authors investigated the importance of viscosity in modified gravity.
The manuscript investigates late time acceleration under the $f(Q)$ gravity background in order to explore the cosmological phenomenon beyond the standard $\Lambda$CDM model and search an alternate description of controversial DE component $\Lambda$. For more on $f(Q)$ gravity, we must check the review in \cite{CPCP3}. We consider the power law $f(Q)$ model, specifically $f(Q)= \alpha Q^n$, along with the cosmic matter fluid having viscous effects with transport coefficient $\zeta = \zeta_0 \sqrt{\Omega} + \zeta_1 \Omega H $. The corresponding analytic solution is presented in equation \eqref{3oo} and is further encountered with recent cosmic data. We utilize the MCMC sampling technique along with Scipy optimization method in Python package emcee. The estimated mean value of the parameters for the combined CC and Pantheon+ samples are $H_0 = 71^{+0.084}_{-0.087} \: kms^{-1}Mpc^{-1}$, $\Omega_0 = 0.32 \pm 0.005$, $\bar{\zeta_0} = 1.5^{+0.066}_{-0.069}$, $\bar{\zeta_1} = -2.5 \pm 0.047$, and $n=2^{+0.069}_{-0.068}$. The experimental prediction of the Planck 2018 results for a flat $\Lambda$CDM model suggests the Hubble constant value $H_0= 67.27 \pm 0.60 \: kms^{-1}Mpc^{-1}$ \cite{O1}, while the Planck 2018+CMB lensing suggests the value $H_0 = 67.36 \pm 0.54 \: kms^{-1}Mpc^{-1}$ \cite{O1}. The nine-year data released for Wilkinson Microwave Anisotropy Probe experiments for the $\Lambda$CDM model presented a value of Hubble constant $H_0 = 70 \pm 2.2 \: kms^{-1}Mpc^{-1}$ \cite{O2}. Apart from the results obtained from the CMB data analysis, various results were also presented for BAO and its combined analysis with CMB considering different cosmological scenarios. For example, BOSS Data Release 12 provides $H_0= 67.9 \pm 1.1 \: kms^{-1}Mpc^{-1}$ \cite{O3} and $H_0= 68.36^{+0.53}_{-0.52} \: kms^{-1}Mpc^{-1}$ for WAMP+BAO \cite{O4}. Hence, one can observe the difference in the $H_0$ values obtained from different surveys that probe the high redshift results and the estimated values in our analysis using Pantheon+ data that probe the low redshift results. For more details on $H_0$ tension, one can check the references \cite{O5,O6,O7}.

It is interesting to note that, a model independent reconstruction of the $f(Q)$ theory given in the reference \cite{SCAPE} suggests that the best approximation for describing the accelerated expansion of the Universe is represented by a scenario $f(Q) = \alpha + \beta Q^n $ with the parameter values $ (\alpha,\beta,n)=(2.492, 0.757, 1.118) $. It indicates deviations from the $\Lambda$CDM model as the redshift increases. Moreover, the constraints obtained by propagation of gravitational waves from inspiraling of binary systems in the reference \cite{RCN} supports the choice of our power law $f(Q)$ function, but with small deviations from the $\Lambda$CDM one since the parameter $n \approx 0.99$. We reproduced the cosmological parameters presented in Figures \eqref{cha3f2}-\eqref{cha3f5}, resampling the chains obtained by the emcee, incorporating 6000 samples. We found that the matter-energy density depicts the expected positive behavior, whereas the effective pressure indicates the negative behavior that is leading to the accelerating expansion, which is further predicted in the effective EoS parameter. In addition, we investigate the asymptotic nature of the assumed viscous cosmological $f(Q)$ model by invoking phase-space analysis. The results obtained are presented in Table \eqref{Table-1a} and Figures \eqref{cha3f6}-\eqref{cha3f7}. We conclude that the assumed viscous $f(Q)$ model with the obtained constraints on parameters successfully predicts an evolution of the Universe from matter dominated decelerated epoch to stable accelerated de-Sitter epoch. Thus, the considered cosmological scenario can be a good alternative to describe the late-time phenomenon of the Universe.

 \chapter{Asymptotic and observational analysis of linear $f(Q)$ model and asymptotic behavior of quadratic $f(Q)$ model under generalized bulk viscosity} 

 \label{Chapter4} 

 \lhead{Chapter 4. \emph{Asymptotic and observational analysis of linear $f(Q)$ model and asymptotic behavior of quadratic $f(Q)$ Model under generalized bulk viscosity}} 

 \vspace{8 cm}
 * The work, in this chapter, is covered by the following publications: \\
 
 \textit{Viscous fluid cosmology in symmetric teleparallel gravity}, Fortschritte der Physik {\bf 71}, 2200202 (2023).

 \clearpage

In this chapter, we analyze the viscous fluid cosmological model in the framework of $f(Q)$ gravity by assuming three different forms of bulk viscosity coefficients, specifically $(i)\zeta =\zeta_{0}+\zeta_{1}\left( \frac{\dot{a}}{a}\right) +\zeta_{2}\left( \frac{{\ddot{a}}}{\dot{a}}\right) $, $(ii)\zeta =\zeta_{0}+\zeta_{1}\left( \frac{\dot{a}}{a}\right)$, and $(iii)\zeta =\zeta_{0}$ and a linear $f(Q)$ model, particularly $f(Q)=\alpha Q$ where $\alpha \neq 0$ is the free model parameter. We estimate the bulk viscosity coefficients and the values of the model parameters using the combined dataset $H(z)$+Pantheon+BAO. We study the asymptotic behavior of our cosmological bulk viscous model by utilizing the phase space method. We find that corresponding to all three cases, our model depicts the evolution of the Universe from a matter-dominated decelerated epoch (a past attractor) to a stable de-sitter accelerated epoch (a future attractor). In addition, we study the physical behavior of the effective pressure, the effective EoS, and the statefinder parameters. We find that the pressure component in the presence of bulk viscosity shows negative behavior and the effective EoS parameter predicts the accelerated expansion phase of the Universe for all three cases. Moreover, we see that the trajectories of our model lie in the quintessence region and it converges to the $\Lambda$CDM fixed point in the distant future. We find that the accelerated de-Sitter type phase comes purely from the $\bar{\zeta}_0$ case without any geometrical modification to GR. Moreover, we find that the late-time behavior of all three cases of viscosity coefficients is identical. Further, we consider a non-linear $f(Q)$ model, specifically $f(Q)=-Q+\beta Q^2$, and then we analyze the behavior of model using dynamical approach. We find that the late-time behavior of the considered non-linear model $f(Q)=-Q+\beta Q^2$ with $\beta \leq 0$ is similar to the linear case, whereas for the case $\beta > 0$ the results are quite different.

\section{Introduction}\label{sec1z}
The accelerated expansion of the Universe is supported by several observations, including Supernovae \cite{Riess,Perlmutter}, BAO \cite{D.J.,W.J.}, WMAP \cite{C.L.,D.N.}, and CMB measurements \cite{R.R.,Z.Y.}. Although the $\Lambda$CDM model successfully explains these observations through a DE component, it suffers from fundamental issues such as cosmic coincidence and the cosmological constant \cite{COP}. An alternative approach to address these challenges is to modify the geometric framework of GR, leading to the development of modified gravity theories. This approach can be seen in references \cite{L.A.,SA,R44}.
In this work, we analyze cosmological scenarios within the framework of $f(Q)$ gravity \cite{R49}, a natural extension of symmetric teleparallel gravity based on non-metricity. The $f(Q)$ gravity leads to second-order field equations and automatically satisfies Bianchi identities, allowing greater freedom in model construction. Recently, several interesting cosmological and astrophysical implications of $f(Q)$ gravity has been appeared, for instance, cosmography \cite{SM1}, energy conditions \cite{SM2}, gravitational waves \cite{Hohmann1,e4,n1}, black hole solution \cite{e7},  quantum cosmology \cite{ND}, general covariant symmetric teleparallel cosmology \cite{Hohmann2}, wormhole solution \cite{e8}, evidence that non-metricity $f(Q)$ gravity can challenge $\Lambda$CDM \cite{e1}, bouncing cosmology \cite{bb1}, Hamiltonian analysis and ADM formulation \cite{hh1}, statefinder analysis of $f(Q)$ cosmology \cite{rs1}, and the cosmological perturbations \cite{R49}.

 This chapter is organized as follows.  In sec \ref{cha4sec2}, we present the flat FLRW Universe in symmetric teleparallel cosmology with bulk viscous matter. In sec \ref{cha4sec3}, we derive the expressions for the Hubble parameter by assuming three different bulk viscosity coefficients and a linear $f(Q)=\alpha Q$ model with $\alpha \neq 0$. In sec \ref{cha4sec4} We present the best fit values of the viscosity coefficients and the model parameter by using the combined $H(z)$+Pantheon+BAO dataset. Sec \ref{cha4sec5} presents an analysis of the asymptotic behavior of our model using the dynamical system approach. In addition, in sec \ref{cha4sec6} we investigate the behavior of effective pressure, effective EoS parameter, and the statefinder parameters. In sec \ref{cha4sec7}, we consider a non-linear $f(Q)$ model, specifically $f(Q)=Q+\beta Q^2$ and then investigate its dynamical behavior. In the last sec \ref{cha4sec8}, we present the results of our investigation.

\section{Flat FLRW Universe in symmetric teleparallel cosmology with bulk viscous matter}\label{cha4sec2}

We begin with following spatially flat FLRW line element \eqref{FLRW} in
Cartesian coordinates, which is, as a matter of fact, also a coincident gauge coordinates, therefore from now connection becomes trivial and metric is only a fundamental variable,

\section{Linear $f(Q)$ model }\label{cha4sec3}
\justify
We consider the following $f(Q)$ function for our analysis \cite{RS,ZHH},  
\begin{equation}\label{3g}
f(Q)=\alpha Q,\ \ \ \alpha \neq 0  \text{.}
\end{equation}
Then the Friedmann equations for this specific $f(Q)$ function become 
\begin{equation}\label{3h}
\rho =-3\alpha H^{2}  \text{,}
\end{equation}
and 
\begin{equation}\label{3ii}
p_v=2\alpha \dot{H}+3\alpha H^{2} \text{.}
\end{equation}
In particular, for the case $\alpha=-1$, one can retrieve the usual Friedmann equations of GR.

In fluid mechanics, it is well known that the coefficient of bulk viscosity is associated with the rate of expansion or compression of the fluid \cite{DZ1}. In the context of the cosmological model, the cosmic fluid is comoving with the expanding Universe, therefore, the velocity $\dot{a}$ and acceleration $\ddot{a}$ of the expanding Universe and that of the cosmic fluid coincide. Thus, it is evident that the bulk viscosity coefficient $\zeta$ is proportional to the velocity and acceleration term. We consider the following bulk viscosity form, which is nothing more than a linear combination of velocity and acceleration terms with a constant \cite{DZ2}, 
\begin{equation}\label{3jj}
\zeta =\zeta_{0}+\zeta_{1}\left( \frac{\dot{a}}{a}\right) +\zeta_{2}\left( \frac{{
\ddot{a}}}{\dot{a}}\right) =\zeta_{0}+\zeta_{1}H+\zeta_{2}\left( \frac{\dot{H}}{
H}+H\right)  \text{.}
\end{equation}
Now we set $\bar{\zeta_{0}}$, $\bar{\zeta_{1}}$, and $\bar{\zeta_{2}}$ as
\begin{equation}\label{3kk}
\frac{3\zeta _{0}}{H_{0}}=\bar{\zeta_{0}}, \ \ 3\zeta_{1}=\bar{\zeta_{1}}, \ \ 3\zeta_{2}=
\bar{\zeta_{2}} \text{.}
\end{equation}
where $H_{0}$ is the Hubble parameter value at present $z=0$ and these parameters are known as dimensionless bulk viscous parameters.
Then using \eqref{3jj} and \eqref{3kk} in equation \eqref{3ii} and the fact that $\frac{d}{dt}=H\frac{d}{dln(a)}$, we have 
\begin{equation} \label{3l}
\frac{dH}{dln(a)}+\left( \frac{3\alpha +\bar{\zeta_{1}}+ \bar{\zeta_{2}}}{2\alpha +\bar{\zeta_{2}}}\right) H+\left( \frac{\bar{\zeta_{0}}}{2\alpha +\bar{\zeta_{2}}}\right)H_{0}=0 \text{.}
\end{equation}
Now by integrating the above equation, we obtained the following expression of Hubble parameter in terms of redshift corresponding to the bulk viscous non-relativistic matter dominated Universe 

\begin{equation}\label{3m}
H(z)=H_{0}\left[ (1+z)^{\left( \frac{3\alpha +\bar{\zeta_{1}}+ \bar{\zeta_{2}}}{2\alpha +\bar{
\zeta_{2}}}\right) }\left( 1+\frac{\bar{\zeta_{0}}}{3\alpha +\bar{\zeta_{1}}+ \bar{\zeta_{2}}}
\right) -\frac{\bar{\zeta_{0}}}{3\alpha +\bar{\zeta_{1}}+ \bar{\zeta_{2}}}\right]  \text{.}
\end{equation}

Now we consider the following three different cases on dimensionless viscous parameters which is well known in the literature. 
\justify Case I: When the viscosity coefficient depends on both the velocity and acceleration, i.e. $\bar{\zeta_{0}}$, $\bar{\zeta_{1}}$, and $\bar{\zeta_{2}}$ all are non-zero.\\
Case II: When viscosity coefficient depends on velocity but not on acceleration, i.e. $\bar{\zeta_{0}}$, $\bar{\zeta_{1}}$ are non-zero, while $\bar{\zeta_{2}}=0$. In this case, the expression for the Hubble parameter becomes
\begin{equation}\label{3n}
H(z)=H_{0}\left[ (1+z)^{\left( \frac{3\alpha +\bar{\zeta_{1}}}{2\alpha} \right) }\left( 1+\frac{\bar{\zeta_{0}}}{3\alpha +\bar{\zeta_{1}}} \right) -\frac{\bar{\zeta_{0}}}{3\alpha +\bar{\zeta_{1}}} \right]  \text{.}
\end{equation}
Case III: When the viscosity coefficient does not depend on both the velocity and acceleration, i.e.  $\bar{\zeta_{1}}=0$, $\bar{\zeta_{2}}=0$ whereas $\bar{\zeta_{0}}$ is non-zero. In this case, the expression for the Hubble parameter becomes
\begin{equation}\label{3o}
H(z)=H_{0}\left[ (1+z)^{\frac{3}{2}} \left( 1+\frac{\bar{\zeta_{0}}}{3\alpha} \right) -\frac{\bar{\zeta_{0}}}{3\alpha } \right]  \text{.}
\end{equation}
In particular, when $\bar{\zeta_{0}}=\bar{\zeta_{1}}=\bar{\zeta_{2}}=0$, then $H(z)$ reduces to $H(z)=H_{0}(1+z)^{\frac{3}{2}}$, which corresponds to the non-viscous matter dominated Universe.

\section{Parameter estimation using observational data}\label{cha4sec4}

In this section, we consider updated $H(z)$ data, Pantheon Supernovae data and BAO data to estimate the parameter values corresponding to all three different cases. We apply the MCMC approach along with Bayesian analysis to explore the parameter space of our cosmological bulk viscous model by utilizing \texttt{emcee} python library \cite{Mackey/2013}.

\subsubsection{$H(z)$ datasets}
\justify
We are familiar with the fact that the Hubble parameter can directly predict the rate of cosmic expansion. In general, there are two very popular approaches to extracting the Hubble parameter value at a definite redshift, namely, the differential age and line of sight BAO technique. In this chapter, we work with an updated list of $H(z)$ data points. The reference for the complete set of data points can be checked \cite{RS}. Furthermore, we have taken $H_0=67.9$ Km/s/Mpc \cite{Planck} for our investigation. 
The $\chi^2$ function corresponding to the $H(z)$ data points is read as,
\begin{equation}\label{4a}
\chi_{H}^{2}=\sum\limits_{k=1}^{57}
\frac{[H_{th}(z_{k},\theta)-H_{obs}(z_{k})]^{2}}{
\sigma _{H(z_{k})}^{2}}.  
\end{equation}
Here, $H_{obs}$ is the Hubble parameter value extracted from cosmic observations, while $H_{th}$ represents its theoretical value calculated at $z_{k}$ with parameter space $\theta$, and $\sigma_{H(z_{k})}$ denotes the corresponding error.

\subsubsection{Pantheon datasets}
\justify
In the present manuscript, we work with a recently published Pantheon supernova dataset that contains 1048 supernova samples with their distance moduli $\mu^{obs}$ in the redshift range $z \in [0.01, 2.3]$ \cite{Scolnic/2018}. The $\chi^2$ function corresponding to the Pantheon data points reads as
\begin{equation}\label{4b}
\chi^2_{SN}=\sum_{i,j=1}^{1048}\bigtriangledown\mu_{i}\left(C^{-1}_{SN}\right)_{ij}\bigtriangledown\mu_{j} \text{,}
\end{equation}
Here, $C_{SN}$ is the covariance matrix \cite{Scolnic/2018} and
\begin{align*}\label{4c}
\quad \bigtriangledown\mu_{i}=\mu^{th}(z_i,\theta)-\mu_i^{obs},
\end{align*} 
is the difference between the observed value of the distance modulus extracted from cosmic observations and its theoretical values calculated from the model with a given parameter space $\theta$. We define the distance modulus by $\mu=m_B-M_B$, where $m_B$ and $M_B$ denote, respectively, the apparent magnitude observed and the absolute magnitude at a given redshift (Retrieving the nuisance parameter following the recent approach called BEAMS with Bias Correction (BBC) \cite{BMS}. Furthermore, its theoretical value is given by
\begin{equation}\label{4d}
\mu(z)= 5log_{10} \left[ \frac{D_{L}(z)}{1 Mpc}  \right]+25  \text{.} 
\end{equation}
where, 
\begin{equation}\label{4e}
D_{L}(z)= c(1+z) \int_{0}^{z} \frac{ dx}{H(x,\theta)} \text{.}
\end{equation}

\subsubsection{BAO datasets}
\justify
BAO investigates oscillations produced in the early phase of the Universe due to cosmological perturbations in the fluid consisting of photons, baryons, and dark matter, which is tightly coupled through Thompson scattering. The BAO measurements consist of the Sloan Digital Sky Survey (SDSS), Six Degree Field Galaxy Survey (6dFGS), and the Baryon Oscillation Spectroscopic Survey (BOSS) \cite{W.J.,BAO1}. The relations used in BAO measurements are
\begin{equation}\label{4f}
d_{A}(z)=\int_{0}^{z}\frac{dz^{\prime }}{H(z^{\prime })},
\end{equation}
\begin{equation}\label{4g}
D_{V}(z)=\left( d_{A}(z)^{2}z/H(z) \right)^{1/3},
\end{equation}
and
\begin{equation}\label{4h}
\chi _{BAO}^{2}=X^{T}C^{-1}X  .
\end{equation}
Here, $C$ is the covariance matrix \cite{BAO6}, $d_{A}(z)$ denotes the angular diameter distance while $D_{V}(z)$ represents the dilation scale.

\subsubsection{Observational results}
\justify
We obtained the constraints on the parameters of our cosmological model with bulk viscous matter corresponding to all three different cases for the combined $H(z)$+Pantheon+BAO dataset by minimizing the total chi-squared function $\chi_{H}^{2}+\chi_{SN}^{2}+\chi_{BAO}^{2}$.

\begin{center}
\begin{figure}[H]
\includegraphics[scale=0.90]{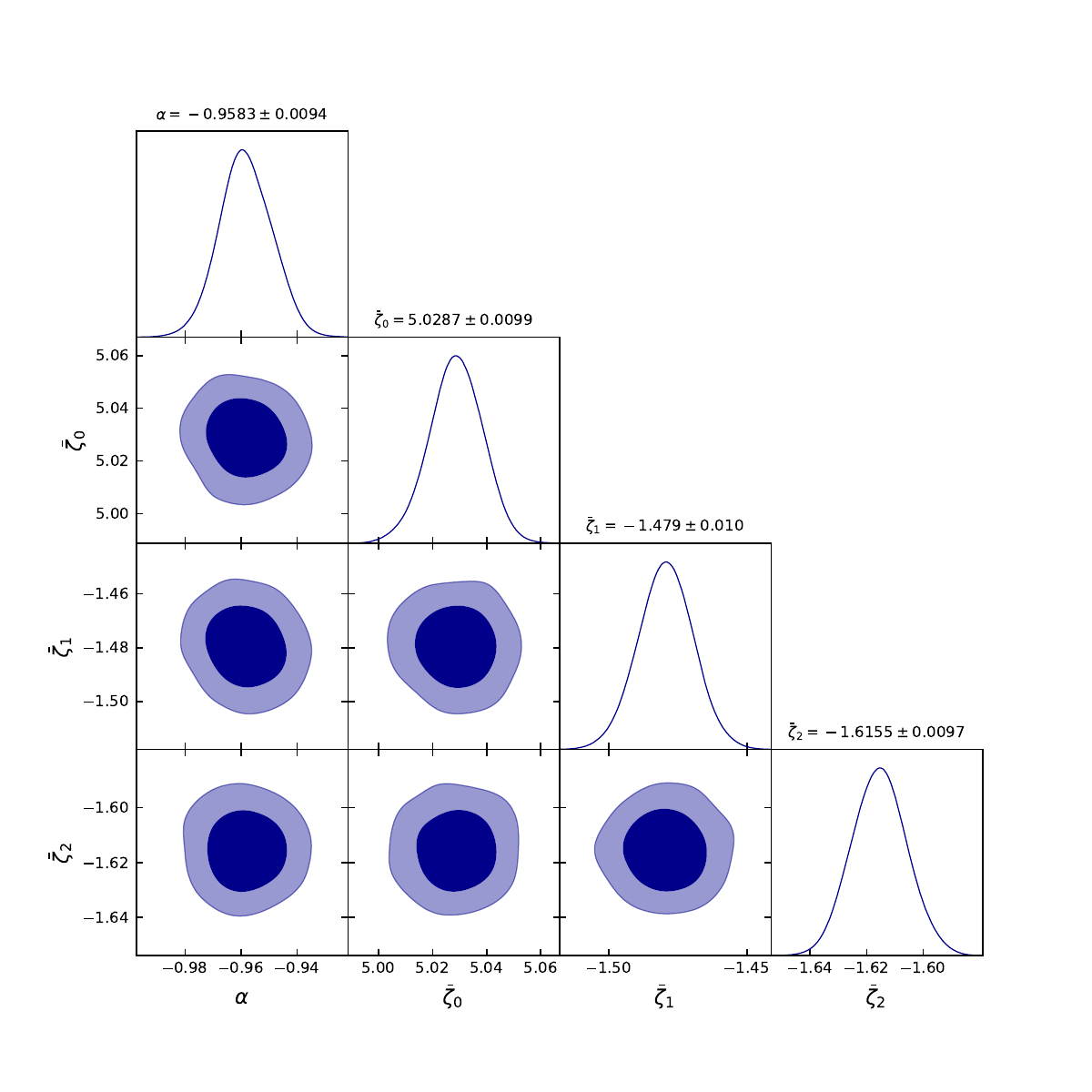}
\caption{Constraints on the model and bulk viscous parameters corresponding to case I at $1-\sigma$ and $2-\sigma$ confidence interval using the combined $H(z)$+Pantheon+BAO dataset.}\label{cha4f1}
\end{figure}
\end{center}

\begin{figure}[h]
\centering
\includegraphics[width=14cm,height=14cm]{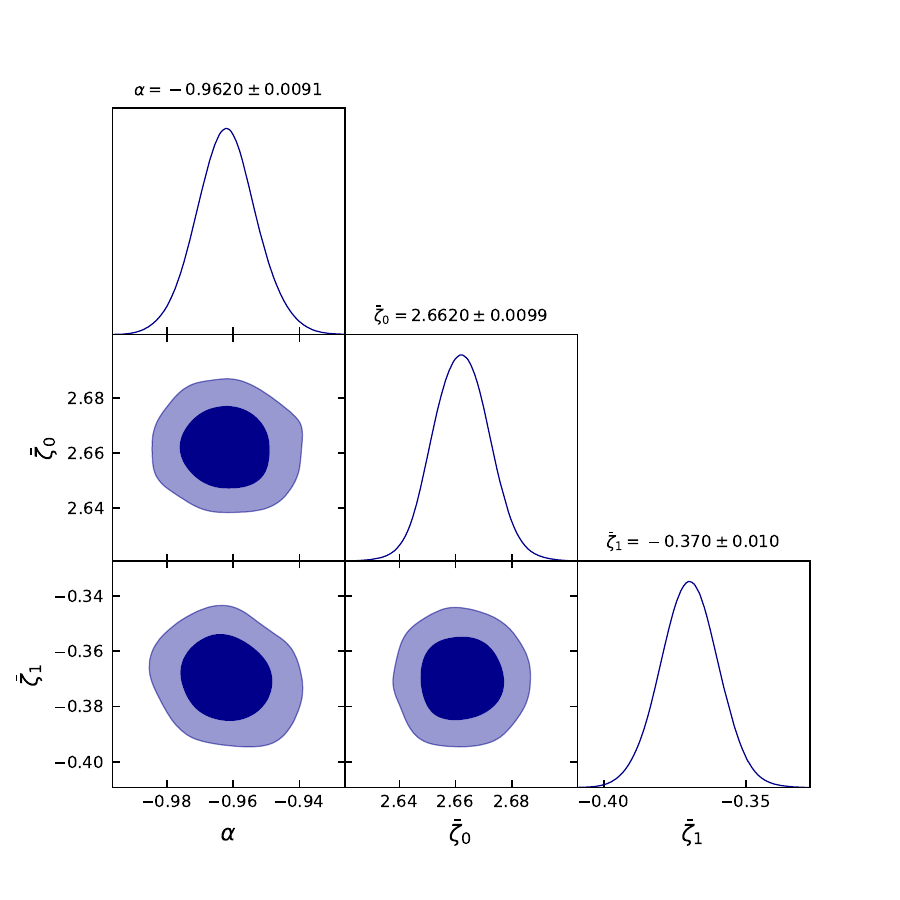}
\caption{Constraints on the model and bulk viscous parameters corresponding to case II at $1-\sigma$ and $2-\sigma$ confidence interval using the combined $H(z)$+Pantheon+BAO dataset.}
\label{cha4f2}
\end{figure}

\begin{figure}[h]
\centering
\includegraphics[width=13cm,height=12cm]{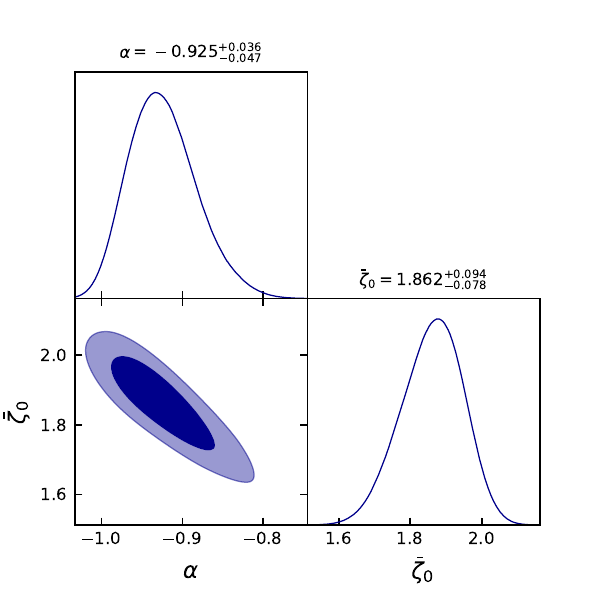}
\caption{Constraints on the model and bulk viscous parameters corresponding to case III at $1-\sigma$ and $2-\sigma$ confidence interval using the combined $H(z)$+Pantheon+BAO dataset.}
\label{cha4f3}
\end{figure}

The constraints obtained on the model and the bulk viscous parameters correspond to all three cases for the combined $H(z)$+Pantheon+BAO dataset presented in Table \ref{Table-1aa}.

\begin{table}[H]
\begin{center}\caption{Table shows the constraints on the model and bulk viscous parameters corresponds to all three cases for the combined $H(z)$+Pantheon+BAO dataset.}
\begin{tabular}{|c|c|c|c|c|}
\hline
Cases & $\alpha$ & $\bar{\zeta}_0$  & $\bar{\zeta}_1$ & $\bar{\zeta}_2$ \\
\hline 
Case I & $-0.9583 \pm 0.0094$ & $5.0287 \pm 0.0099$ & $-1.479 \pm 0.010$ & $-1.6155 \pm 0.0097$ \\
\hline
Case II & $-0.9620 \pm 0.0091$ & $2.6620 \pm 0.0099$ & $-0.370 \pm 0.010$ & $0$ \\
\hline
Case III & $-0.925^{+0.036}_{-0.047} $ & $1.862^{+0.094}_{-0.078} $ & $0$ & $0$ \\
\hline
\end{tabular}\label{Table-1aa}
\end{center}
\end{table}

\section{Behavior of cosmological parameters}\label{cha4sec5}

In this section, we present physical behavior of well known cosmological parameters such as the pressure component in the presence of viscosity, the effective EoS and the $r-s$ parameter \cite{V.S.}.

\begin{figure}[H]
\centering
\includegraphics[width=12cm,height=7cm]{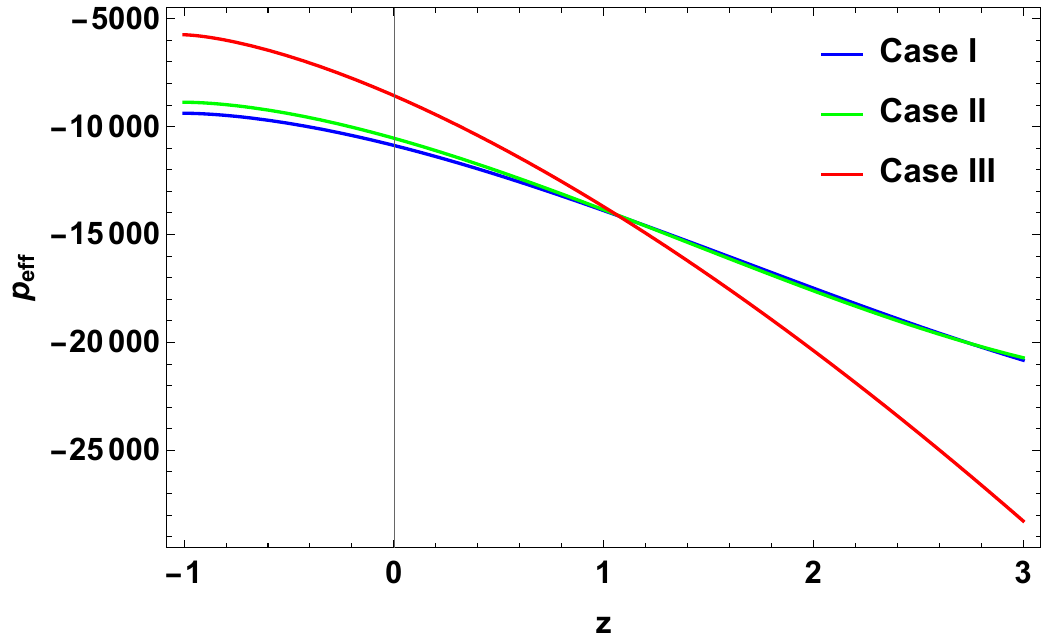}
\caption{Plot for pressure component in the presence of bulk viscosity corresponding to the observational constraints obtained using combined $H(z)$+Pantheon+BAO dataset.}
\label{cha4f7}
\end{figure}

\begin{figure}[H]
\centering
\includegraphics[width=12cm,height=7cm]{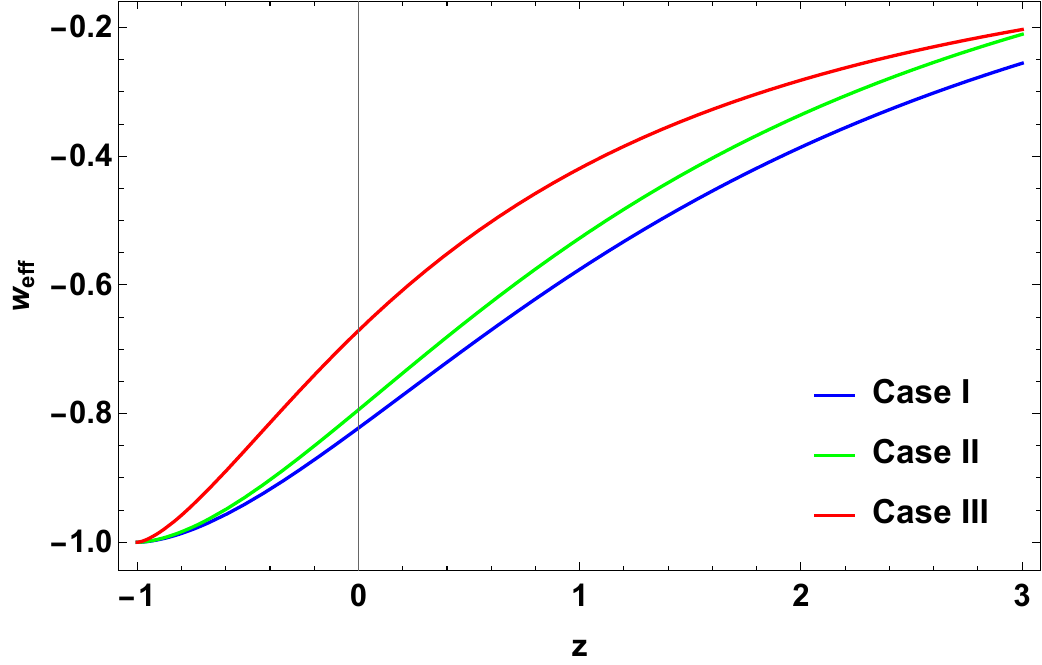}
\caption{Plot for effective EoS parameter corresponding to the observational constraints obtained using combined $H(z)$+Pantheon+BAO dataset.}
\label{cha4f8}
\end{figure}

\begin{figure}[H]
\centering
\includegraphics[width=10cm,height=12cm]{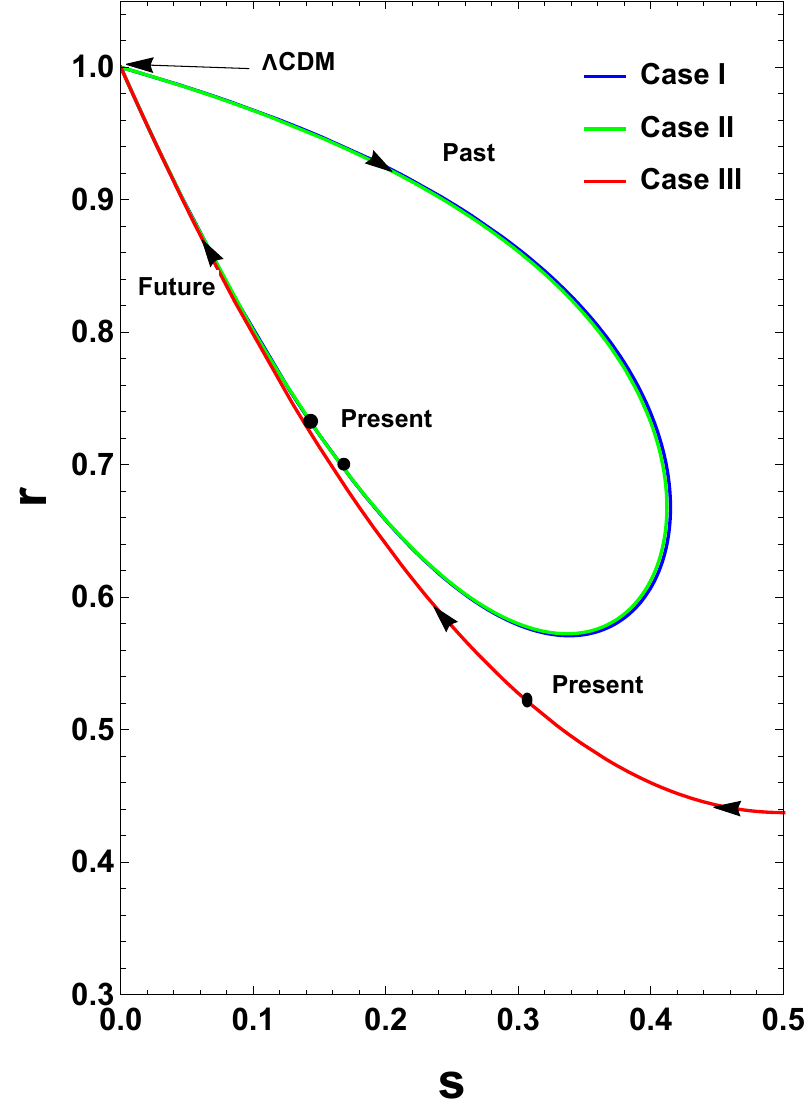}
\caption{Plot for $r-s$ parameter corresponding to the observational constraints obtained using combined $H(z)$+Pantheon+BAO dataset.}
\label{cha4f9}
\end{figure}

\justify Figure \eqref{cha4f7} indicates that the pressure component in the presence of bulk viscosity shows a negative behavior for all three cases. The negative behavior of the pressure component confirms the existence of a mysterious DE component leading to acceleration. The plot of the effective EoS parameter in Figure \eqref{cha4f8} supports the recently observed acceleration, and it will converge to the $\Lambda$CDM EoS in the far future. In Figure \eqref{cha4f9}, we present the evolutionary trajectory of the given viscous fluid model in the $r-s$ plane corresponding to all three cases. Since all three trajectories lie in the region $r<1$ and $s>0$, our bulk viscous model follows the quintessence scenario. Furthermore, we observed that, corresponding to all three cases, the trajectories of our model converge to the $\Lambda$CDM fixed point. Thus, from the $r-s$ diagram we conclude that our viscous fluid cosmological model represents a stable de-Sitter phase of the Universe in the far future.

\section{Phase space analysis}\label{cha4sec6}
In this section, we investigate the asymptotic behavior of cosmological bulk viscous model by utilizing the dynamical system approach. First, we convert the cosmological equations of our model into a set of autonomous differential equations and then analyze the corresponding phase space. Now, we define the following dimensionless variables,
\begin{equation}\label{5a}
x= \frac{-\rho}{3\alpha H^2}  \:\: \text{and} \:\:   y=\frac{1}{\frac{H_0}{H}+1} \text{.}
\end{equation}
The variables $x$ and $y$ are called phase space variables and are in the range $0 \leqslant y \leqslant 1$, whereas the first Friedmann equation implies $ x = 1$.
\justify \textbf{Case I: When the viscosity coefficient depends both on the velocity and the acceleration, i.e., $\zeta =\zeta_{0}+\zeta_{1}\left( \frac{\dot{a}}{a}\right) +\zeta_{2}\left( \frac{{\ddot{a}}}{\dot{a}}\right) =\zeta_{0}+\zeta_{1}H+\zeta_{2}\left( \frac{\dot{H}}{H}+H\right)$}.

\justify We start with defining the variable $N=ln(a)$. Then by solving the Friedmann equations with the conservation equation for the bulk viscous matter, we obtained the following autonomous differential equations for the variables $x$ and $y$, 

\begin{equation}\label{5bb}
x'=\frac{dx}{dN}=\frac{(x-1)}{(2\alpha+\bar{\zeta}_2)y} \left[ 2\bar{\zeta}_0 (1-y)+(2\bar{\zeta}_1-\bar{\zeta}_2)y \right] = F_1(x,y) \text{,}
\end{equation}
and
\begin{equation}\label{5cc}
y'=\frac{dy}{dN}=\frac{(y-1)}{(2\alpha+\bar{\zeta}_2)} \left[ \bar{\zeta}_0 (1-y)+(\bar{\zeta}_1+\bar{\zeta}_2+3\alpha)y \right] = F_2(x,y) \text{.}
\end{equation}

Further, by using the definition of the EoS and deceleration parameter, we acquired
\begin{equation}\label{5d}
q=\frac{1}{(2\alpha+\bar{\zeta}_2)} \left[ \alpha+ \bar{\zeta}_1+ \frac{\bar{\zeta}_0 (1-y)}{y} \right]     \text{,}
\end{equation}
and
\begin{equation}\label{5e}
\omega= \frac{1}{3(2\alpha+\bar{\zeta}_2)} \left[ 2\bar{\zeta}_1 - \bar{\zeta}_2 + \frac{2\bar{\zeta}_0 (1-y)}{y} \right]  \text{.}
\end{equation}
Now, by solving equations $x'=0$ and $y'=0$, we obtain the coordinates of the critical points $(x_c,y_c)$ corresponding to the autonomous equations \eqref{5bb} and \eqref{5cc} as
\begin{equation}\label{5ff}
(x_c,y_c)=(1,1) \:\: \text{and} \:\: (x_c,y_c)=(1,\frac{\bar{\zeta}_0}{\bar{\zeta}_0-\bar{\zeta}_1-\bar{\zeta}_2-3\alpha}) \text{.}
\end{equation}
We investigate the stability of the given autonomous system in the neighborhood of the critical points. First, we linearize the given autonomous dynamical system by assuming small perturbations near the critical points $(x,y)\longrightarrow (x_c+\delta x,y_c+\delta y)$ satisfying

\begin{equation}\label{5g}
\left[
\begin{array}{c}
\delta x'  \\ 
\delta y'  \\ 
\end{array}
\right] \,=   \left[
\begin{array}{cc}
\left( \frac{\partial F_1}{\partial x} \right)_{0} & \left( \frac{\partial F_1}{\partial y}  \right)_{0}  \\ 
\left( \frac{\partial F_2}{\partial x}  \right)_{0} & \left( \frac{\partial F_2}{\partial x}  \right)_{0} \\ 
\end{array}
\right]  \left[
\begin{array}{c}
\delta x \\ 
\delta y  \\ 
\end{array}
\right]  \text{.}
\end{equation}
Here, suffix 0 implies that the above Jacobian matrix was calculated at the critical points $(x_c,y_c)$ and is given as

\begin{equation}\label{5h}
J= \left[
\begin{array}{cc}
\left( \frac{1}{(2\alpha+\bar{\zeta}_2)y} \left[ 2\bar{\zeta}_0 (1-y)+(2\bar{\zeta}_1-\bar{\zeta}_2)y \right] \right)_{0} & \left( \frac{-2\bar{\zeta}_0 (x-1)}{(2\alpha+\bar{\zeta}_2)y^2}  \right)_{0}  \\ 
0 & \left( \frac{1}{(2\alpha+\bar{\zeta}_2)} \left[ -2\bar{\zeta}_0 (y-1)+(\bar{\zeta}_1+\bar{\zeta}_2+3\alpha)(2y-1) \right] \right)_{0} \\ 
\end{array}
\right]  \text{.}
\end{equation}

Now we analyze the behavior of each of the critical points obtained in equation \eqref{5ff}.
\justify \textbf{(i)} $(x_c,y_c)=(1,1)$ : In this case, the eigenvalues obtained for the Jacobian matrix $J$ are 
\begin{equation}\label{5i}
\lambda_1= \frac{(2\bar{\zeta}_1-\bar{\zeta}_2)}{(2\alpha+\bar{\zeta}_2)}  \:\: \text{and} \:\: \lambda_2= \frac{(\bar{\zeta}_1+\bar{\zeta}_2+3\alpha)}{(2\alpha+\bar{\zeta}_2)} \text{.}
\end{equation}
Using the constrained values of the parameters, one can have the following,
\begin{equation}\label{5j}
\lambda_1= 0.38 \:\: \text{and} \:\: \lambda_2= 1.69 \text{.}
\end{equation}
Since, $\lambda_1>0$ and $\lambda_2>0$, therefore, the critical point $(1,1)$ is unstable. Furthermore, $y_c=1$ implies $\frac{1}{\frac{H_0}{H}+1}=1$ i.e., $\frac{H_0}{H}=0$ which indicates either $H_0=0$ or $H \rightarrow \infty $. Since $H_0 \neq 0$, therefore the critical point $(x_c,y_c)=(1,1)$ represents the initial singularity characterized by $H \rightarrow \infty $, i.e. a past attractor. In addition, from equations \eqref{5d} and \eqref{5e}, we obtain $q \sim 0.69$ and $\omega \sim 0.12$.

\justify \textbf{(ii)} $(x_c,y_c)=(1,\frac{\bar{\zeta}_0}{\bar{\zeta}_0-\bar{\zeta}_1-\bar{\zeta}_2-3\alpha})=(1,0.4572)$ : In this case, the eigenvalues obtained for the Jacobian matrix $J$ are, 
\begin{equation}\label{5k}
\lambda_1= -3  \:\: \text{and} \:\: \lambda_2= -\frac{(\bar{\zeta}_1+\bar{\zeta}_2+3\alpha)}{(2\alpha+\bar{\zeta}_2)} \text{.}
\end{equation}
Again, by using the constrained values of the parameters, we can have
\begin{equation}\label{5l}
\lambda_1= -3 \:\: \text{and} \:\: \lambda_2= -1.69 \text{.}
\end{equation}
Since, $\lambda_1<0$ and $\lambda_2<0$, therefore, the critical point $(1,0.4572)$ is stable. In addition, from equations \eqref{5d} and \eqref{5e}, we found $q \sim -1$ and $\omega \sim -1$. Hence, the critical point $(x_c,y_c)=(1,0.4572)$ corresponds to the de-Sitter phase that represents a future attractor.

\begin{table}[H]
\begin{center}\caption{Table shows the critical points and their behavior corresponding to Case I.}
\begin{tabular}{|c|c|c|c|c|}
\hline
Critical Points $(x_c,y_c)$ & Eigenvalues $\lambda_1$ and $\lambda_2$ & Nature of critical point  & $q$ & $\omega$ \\
\hline 
$(1,1)$ & $0.38 \:\: \text{and} \:\: 1.69$ & Unstable & $0.69$ & $0.12$ \\
\hline
$(1,0.4572)$ & $-3 \:\: \text{and} \:\: -1.69$ & Stable & $-1$ & $-1$ \\
\hline
\end{tabular}\label{Table-2a}
\end{center}
\end{table}

\begin{figure}[h]
\centering
\includegraphics[width=10cm,height=10cm]{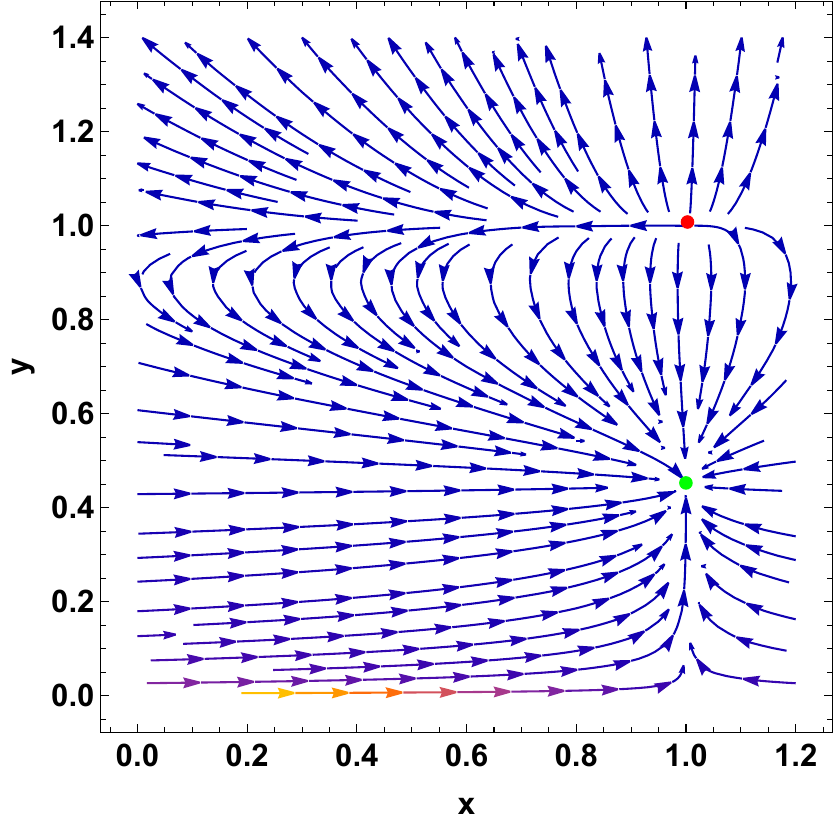}
\caption{Phase plot in the $x-y$ plane corresponding to Case I with a red dot and green dot denoting the past and future attractor, respectively, followed by the arrowhead representing the direction of the trajectories.}
\label{cha4f4}
\end{figure}

\justify From the phase plot presented in Figure \eqref{cha4f4} it is evident that the evolutionary trajectory of bulk viscous cosmological model emerges from the critical point $(1,1)$ as a past attractor and then converges to the critical point $(1,0.4572)$ which is nothing but a future attractor. Thus, model corresponding to Case I depicts the evolution of the Universe starting with an initial singularity and behaves like a de-Sitter phase in the far future.

\justify \textbf{Case II: When the viscosity coefficient depends on velocity but not on acceleration, i.e.  $\zeta =\zeta_{0}+\zeta_{1}\left( \frac{\dot{a}}{a}\right)  =\zeta_{0}+\zeta_{1}H$}.

\justify In this case, the autonomous system of equations \eqref{5bb} and \eqref{5cc} becomes
\begin{equation}\label{5m}
x'=\frac{(x-1)}{\alpha y} \left[ \bar{\zeta}_0 (1-y)+\bar{\zeta}_1 y \right]  \text{,}
\end{equation}
and
\begin{equation}\label{5n}
y'=\frac{(y-1)}{2\alpha} \left[ \bar{\zeta}_0 (1-y)+(\bar{\zeta}_1+3\alpha)y \right]  \text{.}
\end{equation}
The deceleration and the EoS parameter given in equations \eqref{5d} and \eqref{5e} reduce to
\begin{equation}\label{5o}
q=\frac{1}{2\alpha} \left[ \alpha+ \bar{\zeta}_1+ \frac{\bar{\zeta}_0 (1-y)}{y} \right]     \text{,}
\end{equation}
and
\begin{equation}\label{5p}
\omega= \frac{1}{6\alpha} \left[ 2\bar{\zeta}_1 + \frac{2\bar{\zeta}_0 (1-y)}{y} \right] \text{.}
\end{equation}
Now, in solving equations $x'=0$ and $y'=0$, we obtain the coordinates of the critical points $(x_c,y_c)$ corresponding to the autonomous equations \eqref{5m} and \eqref{5n} as,
\begin{equation}\label{5q}
(x_c,y_c)=(1,1) \:\: \text{and} \:\: (x_c,y_c)=(1,\frac{\bar{\zeta}_0}{\bar{\zeta}_0-\bar{\zeta}_1-3\alpha}) \text{.}
\end{equation}
We investigate the behavior of each of the critical points obtained in equation \eqref{5q}.
\justify \textbf{(i)} $(x_c,y_c)=(1,1)$ : In this case, the eigenvalues obtained for the reduced Jacobian matrix $J$ are, 
\begin{equation}\label{5r}
\lambda_1= \frac{\bar{\zeta}_1}{\alpha} \sim 0.38  \:\: \text{and} \:\: \lambda_2= \frac{(\bar{\zeta}_1+3\alpha)}{2\alpha} \sim 1.69 \text{.}
\end{equation}
Since, $\lambda_1>0$ and $\lambda_2>0$, therefore, the critical point $(1,1)$ is unstable. In addition, as in the previous case, $y_c=1$ implies either $H_0=0$ or $H \rightarrow \infty $ and hence the critical point $(x_c,y_c)=(1,1)$ represents the initial singularity characterized by $H \rightarrow \infty $, i.e. a past attractor since $H_0$ is non-zero. Furthermore, from equations \eqref{5o} and \eqref{5p}, we obtained $q \sim 0.69$ and $\omega \sim 0.12$.

\justify \textbf{(ii)} $(x_c,y_c)=(1,\frac{\bar{\zeta}_0}{\bar{\zeta}_0-\bar{\zeta}_1-3\alpha})=(1,0.4498)$ : In this case, the eigenvalues obtained for the reduced Jacobian matrix $J$ are, 
\begin{equation}\label{5s}
\lambda_1= -3  \:\: \text{and} \:\: \lambda_2= -\frac{(\bar{\zeta}_1+3\alpha)}{2\alpha} \sim -1.69
\end{equation}
Since, $\lambda_1<0$ and $\lambda_2<0$, therefore, the critical point $(1,0.4498)$ is stable. From equations \eqref{5o} and \eqref{5p}, we obtained $q \sim -1$ and $\omega \sim -1$. Hence, the critical point $(x_c,y_c)=(1,0.4498)$ corresponds to a de-Sitter type Universe that represents a future attractor.

\begin{table}[H]
\begin{center}\caption{Table shows the critical points and their behavior corresponding to Case II.}
\begin{tabular}{|c|c|c|c|c|}
\hline
Critical Points $(x_c,y_c)$ & Eigenvalues $\lambda_1$ and $\lambda_2$ & Nature of critical point  & $q$ & $\omega$ \\
\hline 
$(1,1)$ & $0.38 \:\: \text{and} \:\: 1.69$ & Unstable & $0.69$ & $0.12$ \\
\hline
$(1,0.4498)$ & $-3 \:\: \text{and} \:\: -1.69$ & Stable & $-1$ & $-1$ \\
\hline
\end{tabular}\label{Table-3a}
\end{center}
\end{table}

\begin{figure}[h]
\centering
\includegraphics[width=10cm,height=10cm]{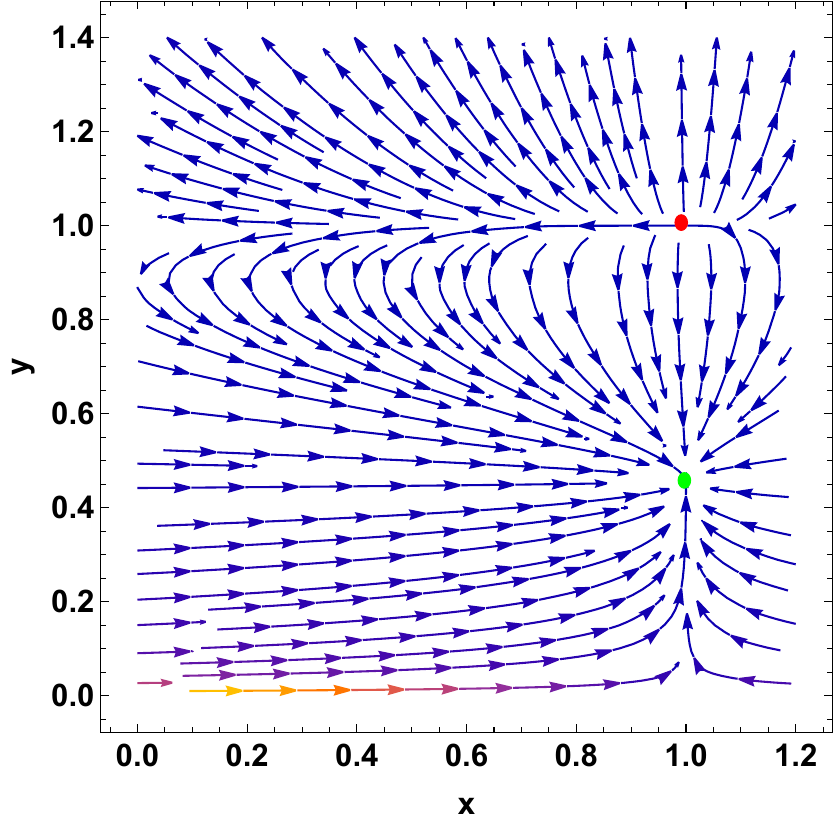}
\caption{Phase plot in the $x-y$ plane corresponding to Case II with a red dot and green dot denoting the past and future attractor, respectively, followed by the arrowhead representing the direction of the trajectories.}
\label{cha4f5}
\end{figure}

\justify From the phase plot presented in Figure \eqref{cha4f5}, it is clear that the evolutionary trajectory of our model emerges from the critical point $(1,1)$ and then converges to the critical point $(1,0.4498)$. Thus, our model with bulk viscosity corresponding to Case II represents the Universe evolving from an initial singularity to a de-Sitter type Universe in the far future, as we obtained in Case I.

\justify \textbf{Case III: when the viscosity coefficient does not depend on both velocity and acceleration, i.e., $\zeta =\zeta_{0}$}.

\justify In this case, the autonomous system of equations \eqref{5bb} and \eqref{5cc} becomes
\begin{equation}\label{5t}
x'=\frac{\bar{\zeta}_0(x-1)(1-y)}{\alpha y }   \text{,}
\end{equation}
and
\begin{equation}\label{5u}
y'=\frac{(y-1)}{2\alpha} \left[ \bar{\zeta}_0 (1-y)+3\alpha y \right] \text{.}
\end{equation}
The deceleration and the equation of the state parameter given in \eqref{5d} and \eqref{5e} reduce to
\begin{equation}\label{5v}
q=\frac{1}{2} + \frac{\bar{\zeta}_0 (1-y)}{2\alpha y}  \text{,}
\end{equation}
and
\begin{equation}\label{5w}
\omega=  \frac{\bar{\zeta}_0(1-y)}{3\alpha y}  \text{.}
\end{equation}
In solving equations $x'=0$ and $y'=0$, we obtained the coordinates of the critical points $(x_c,y_c)$ corresponding to the autonomous equations \eqref{5t} and \eqref{5u} as,
\begin{equation}\label{5x}
(x_c,y_c)=(x,1) \:\: \text{and} \:\: (x_c,y_c)=\left(1,\frac{\bar{\zeta}_0}{\bar{\zeta}_0-3\alpha}\right) \text{.}
\end{equation}
\justify \textbf{(i)} $(x_c,y_c)=(x,1)$ : \\
In this case, the eigenvalues obtained for the reduced Jacobian matrix $J$ are, 
\begin{equation}\label{5y}
\lambda_1= 0  \:\: \text{and} \:\: \lambda_2= \frac{3}{2} \text{.}
\end{equation}
It is evident that the first component $x$ of the critical point varies from 0 to 1 whereas $y_c=1$. Therefore, the critical point $(x_c,y_c)$ is not an isolated point; rather it is a line of critical points that indicates the initial state of the Universe since $H \rightarrow \infty $ As $\lambda_1=0$ and $\lambda_2>0$, therefore every critical point $(x,1)$ on the line is unstable, i.e. a past attractor. In addition, we obtained $q=\frac{1}{2} $ and $\omega =0$ using equations \eqref{5v} and \eqref{5w}. These values of EoS and deceleration parameter correspond to a decelerated matter dominated Universe.  

\justify \textbf{(ii)} $(x_c,y_c)=(1,\frac{\bar{\zeta}_0}{\bar{\zeta}_0-3\alpha})=(1,0.4015)$ : In this case, the eigenvalues obtained for the reduced Jacobian matrix $J$ are, 
\begin{equation}\label{5z}
\lambda_1= -3  \:\: \text{and} \:\: \lambda_2= -\frac{3}{2}  \text{.}
\end{equation}
Since, $\lambda_1<0$ and $\lambda_2<0$, therefore, the critical point $(1,0.4015)$ is stable. From equations \eqref{5v} and \eqref{5w}, we obtained $q \sim -1$ and $\omega \sim -1$. Hence, the critical point $(x_c,y_c)=(1,0.4015)$ corresponds to a de-Sitter type Universe that represents a future attractor.

\begin{table}[H]
\begin{center}\caption{Table shows the critical points and their behavior corresponding to Case III.}
\begin{tabular}{|c|c|c|c|c|}
\hline
Critical Points $(x_c,y_c)$ & Eigenvalues $\lambda_1$ and $\lambda_2$ & Nature of critical point  & $q$ & $\omega$ \\
\hline 
$(x,1)$ & $0 \:\: \text{and} \:\: \frac{3}{2}$ & Unstable & $\frac{1}{2}$ & $0$ \\
\hline
$(1,0.4015)$ & $-3 \:\: \text{and} \:\: -\frac{3}{2}$ & Stable & $-1$ & $-1$ \\
\hline
\end{tabular}\label{Table-4a}
\end{center}
\end{table}

\begin{figure}[H]
\centering
\includegraphics[width=10cm,height=10cm]{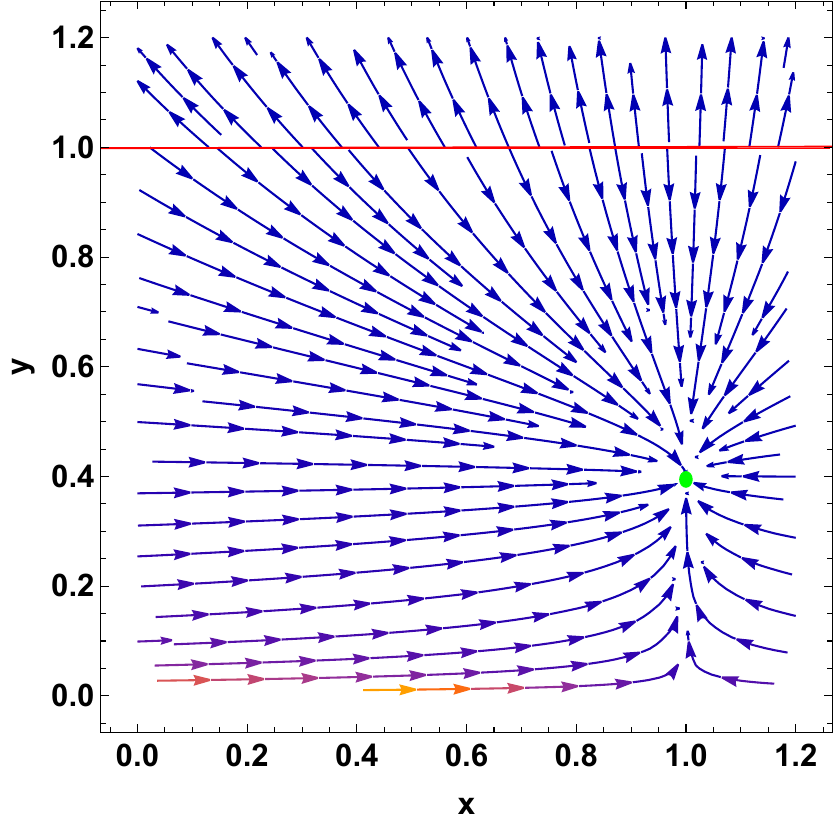}
\caption{Phase plot in the $x-y$ plane corresponding to Case III with the red line of critical points and green dot denoting the past and future attractor, respectively, followed by the arrowhead representing the direction of the trajectories.}
\label{cha4f6}
\end{figure}

\justify From the phase plot presented in Figure \eqref{cha4f6} it is clear that the evolution trajectory of our model emerges from the non isolated critical point $(x,1)$ that corresponds to a decelerated matter dominated Universe and then converges to the critical point $(1,0.4015)$ representing a future attractor. Thus, the cosmological model with bulk viscosity behaves like $\Lambda$CDM model. We found that the accelerated de-Sitter type phase comes purely from the $\bar{\zeta}_0$ case without any geometrical modification to GR. Furthermore, due to the constraint $x=1$, only the critical point $(1,1)$ that lies on the line of critical points $(x,1)$ is physical.

\section{Non-linear $f(Q)$ model }\label{cha4sec7}
\justify

We consider the following $f(Q)$ function for our analysis \cite{R49},
\begin{equation}\label{51}
f(Q)= -Q+\beta Q^2  \text{.}
\end{equation}
We consider only the case $\bar{p}=-3\zeta_{0}H= -\bar{\zeta}_0 H_0 H$. Then by using Friedmann equations we obtained the following first-order non-linear differential equation,
\begin{equation}\label{52}
2\dot{H} (36\beta H^2-1)+3H^2 (18\beta H^2-1) + \bar{\zeta}_0 H_0 H =0  \text{.}
\end{equation}
Now, we define the following dimensionless variables,
\begin{equation}\label{53}
x= \frac{-\rho}{3H^2(18 \beta H^2-1)}  \:\: \text{and} \:\:   y=\frac{1}{\frac{H_0}{H}+1} \text{.}
\end{equation}
The variable $y$ lies in the range $0 \leqslant y \leqslant 1$ and from the first Friedmann equation, we have the constraint $x=1$.
Then we obtained the following autonomous differential equations corresponding to our non-linear $f(Q)$ model, 

\begin{equation}\label{54}
x'=\frac{dx}{dN}=  \frac{\bar{\zeta}_0 (x-1)(1-y)^3}{y \left[ \bar{\beta}y^2-(1-y)^2 \right] }  \text{,}
\end{equation}
and
\begin{equation}\label{55}
y'=\frac{dy}{dN}=\frac{(y-1)}{2\left[2 \bar{\beta}y^2-(1-y)^2 \right]} \left[ 3y \lbrace \bar{\beta}y^2-(1-y)^2 \rbrace + \bar{\zeta}_0 (1-y)^3 \right]  \text{.}
\end{equation}

Here, $\bar{\beta}=18H_0^2 \beta $. Further, by using the definition of the EoS and deceleration parameter, we acquired

\begin{equation}\label{57}
q=-1+ \frac{1}{2y\left[2 \bar{\beta}y^2-(1-y)^2 \right]} \left[ 3y \lbrace \bar{\beta}y^2-(1-y)^2 \rbrace + \bar{\zeta}_0 (1-y)^3 \right]  \text{,}
\end{equation}
and
\begin{equation}\label{58}
\omega= -1+ \frac{1}{3y\left[2 \bar{\beta}y^2-(1-y)^2 \right]} \left[ 3y \lbrace \bar{\beta}y^2-(1-y)^2 \rbrace + \bar{\zeta}_0 (1-y)^3 \right] \text{.}
\end{equation}

Since, $\bar{\zeta}_0>0$, we fix  $\bar{\zeta}_0 =1$ and then we have investigated the stability of the autonomous system given by the equations \eqref{54}-\eqref{55} in the neighborhood of the critical points. In particular, for $\bar{\beta}=0$, this autonomous system is the same as that of Case III, and hence the dynamics. Therefore, we analyzed our non-linear $f(Q)$ model for parameter values $\bar{\beta}=-1$ and $\bar{\beta}=1$. The results obtained are presented in Table \ref{Table-5a}.

\begin{table}[H]
\begin{center}\caption{Table shows the critical points and their behavior corresponding to non-linear $f(Q)$ model.}
\begin{tabular}{|c|c|c|c|c|c|}
\hline
& Critical Points $(x_c,y_c)$ & Eigenvalues $\lambda_1$ and $\lambda_2$ & Nature of critical point  & $q$ & $\omega$ \\
\hline 
$\bar{\beta}=-1$&$\left( x,1 \right)$ & $\frac{3}{4}\:\: \text{and} \:\: 0$ & Unstable & $-\frac{1}{4}$ & $-\frac{1}{2}$ \\
&$\left( 1,0.234 \right)$ & $-3\:\:  \text{and} \:\: -1.62$ & Stable & $-1$ & $-1$ \\
\hline
& $\left(x,1\right)$ & $\frac{3}{4}\:\: \text{and} \:\: 0$ & Unstable & $-\frac{1}{4}$ & $-\frac{1}{2}$ \\
$\bar{\beta}=1$ & $\left( 1,0.283 \right)$ & $-3\:\:  \text{and} \:\: -1.15$ & Stable & $-1$ & $-1$ \\
& $\left( 1,0.426 \right)$ & $-3\:\:  \text{and} \:\: -9.6$ & Stable & $-1$ & $-1$ \\
\hline
\end{tabular}\label{Table-5a}
\end{center}
\end{table}

\begin{figure}[H]
\centering
\includegraphics[width=10cm,height=10cm]{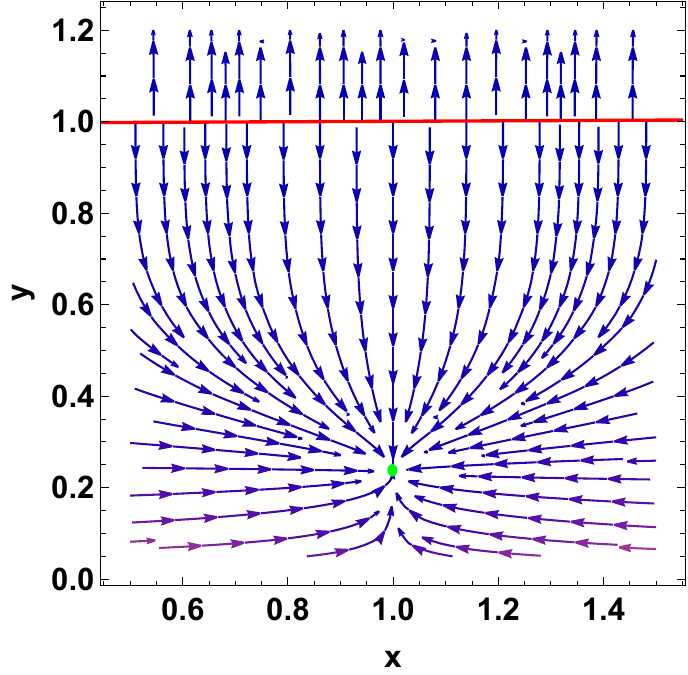}
\caption{Phase plot in the $x-y$ plane corresponding to the non-linear $f(Q)$ model ($\bar{\beta}=-1$) with the red line of critical points and green dot denoting the past and future attractor, respectively, followed by the arrowhead representing the direction of the trajectories.}
\label{cha4n1}
\end{figure}

\begin{figure}[H]
\centering
\includegraphics[width=10cm,height=10cm]{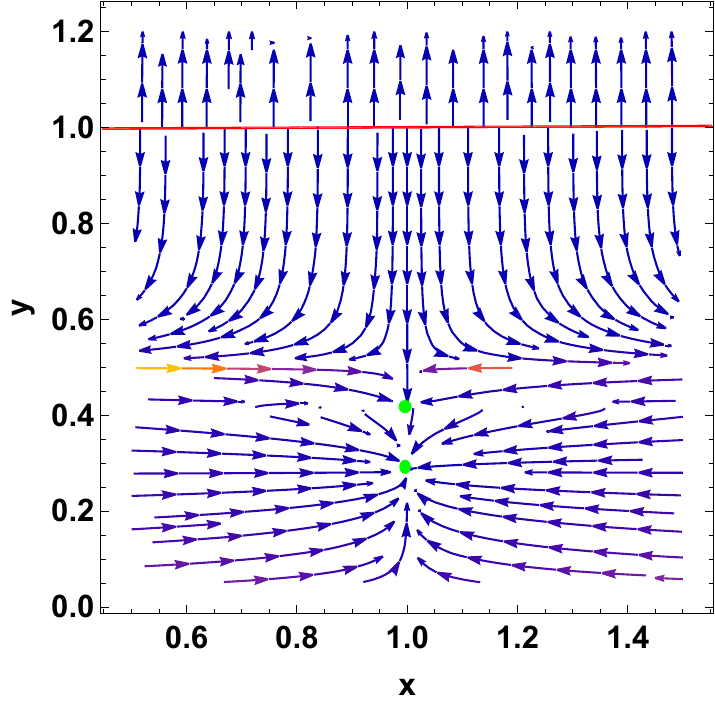}
\caption{Phase plot in the $x-y$ plane corresponding to the non-linear $f(Q)$ model ($\bar{\beta}=1$) with the red line of critical points and green dot denoting the past and future attractor, respectively, followed by the arrowhead representing the direction of the trajectories.}
\label{cha4n2}
\end{figure}

\begin{figure}[H]
\centering
\includegraphics[width=10cm,height=7cm]{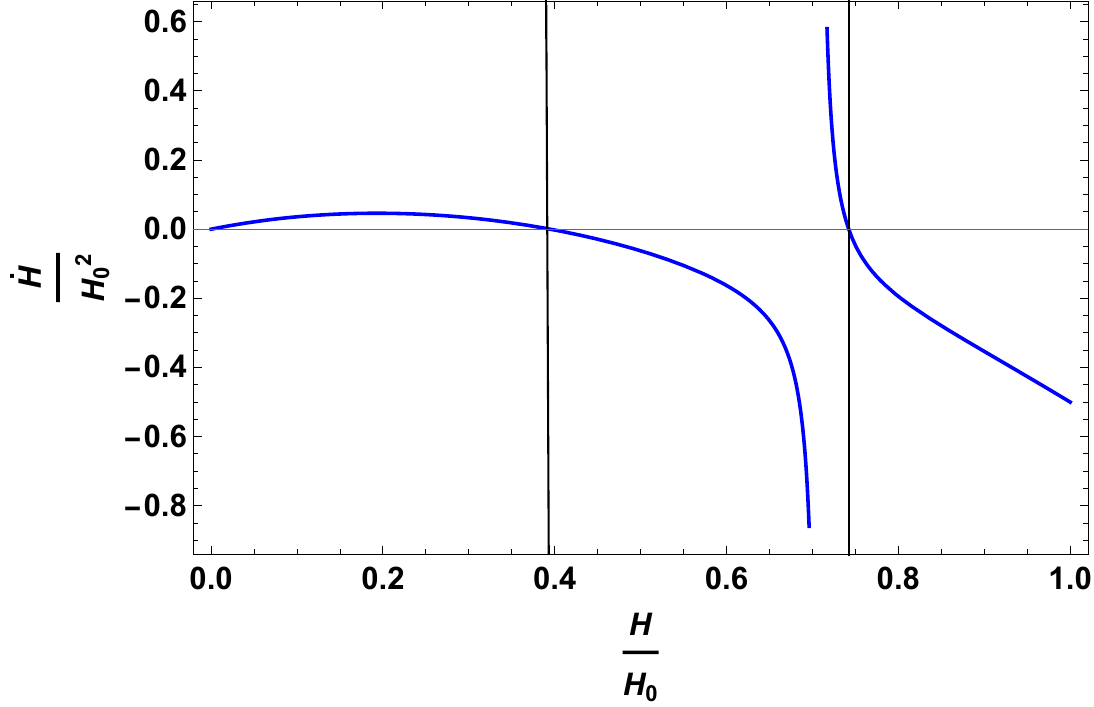}
\caption{$\dot{H}$ as a function of $\frac{H}{H_0}$ from equation \eqref{52} with $\bar{\beta}=\bar{\zeta}_0=1$, where vertical black lines indicating the fix points that corresponds to $\dot{H}=0$.}
\label{cha4add}
\end{figure}

From Table \ref{Table-5a} it is clear that there is only one de-Sitter type future attractor for the case $\bar{\beta}=-1$, while two future attractors correspond to the case $\bar{\beta}=1$. From the phase plot presented in the Figure \eqref{cha4n1} it is clear that the evolution trajectory of our non-linear model, for the case $\bar{\beta}=-1$, emerges from the non isolated critical points $(x,1)$ and then it converges to the critical point $(1,0.234)$ representing a de-Sitter type accelerated phase. In addition, note that, due to the constraint $x=1$, only the critical point $(1,1)$ that lies on the line of critical points $(x,1)$ is physical. Moreover, from the phase plot presented in the Figure \eqref{cha4n2} we found that the evolution trajectory of our non-linear model, for the case $\bar{\beta}=1$, emerges from the critical point $(1,1)$ and then settles down to two stable critical points $(1,0.283)$ and $(1,0.426)$ representing the de-Sitter type accelerated phase. The behavior of $\dot{H}$ as a function of $\frac{H}{H_0}$ from equation \eqref{52} with $\bar{\beta}=\bar{\zeta}_0=1$ is presented in Figure \eqref{cha4add}. We can observe that two regimes are separated by a line $H=\frac{H_0}{\sqrt{2\bar{\beta}}}$ of divergence $\dot{H}$ and that evolution only to the upper attractor $\left( 1,0.426 \right)$ from an initial big bang $y=1$ is possible. Thus, only the fixed point $\left( 1,0.426 \right)$ is a viable attractor from an initial big bang. Hence, evolution from the big bang $H \rightarrow \infty$ can occur only towards the right fix point at $H \approx 0.75 H_0$.

\section{Conclusions}\label{cha4sec8}

In this chapter, we have investigated the role of different bulk viscosity coefficients in cosmic evolution under the framework of symmetric teleparallel gravity. We have considered three different cases of bulk viscosity coefficients that are well known in the literature, namely $(i)\zeta =\zeta_{0}+\zeta_{1}\left( \frac{\dot{a}}{a}\right) +\zeta_{2}\left( \frac{{\ddot{a}}}{\dot{a}}\right) $, $(ii)\zeta =\zeta_{0}+\zeta_{1}\left( \frac{\dot{a}}{a}\right)$, and $(iii)\zeta =\zeta_{0}$. Then we calculate the value of the Hubble parameter in terms of redshift corresponding to all three cases assuming a linear $f(Q)$ model, specifically $f(Q)=\alpha Q$ where $\alpha \neq 0$.  We presented the observation constraints (see Table \ref{Table-1aa} and Figures \eqref{cha4f1}-\eqref{cha4f3}) on the model parameter and viscosity coefficients corresponding to all three cases using the combined $H(z)$+Pantheon+BAO dataset. Then we investigated the asymptotic behavior of the cosmological bulk viscous model by using phase space analysis. We derived the set of autonomous differential equations and the corresponding critical points for all three cases (see Table \ref{Table-2a},\ref{Table-3a}, and \ref{Table-4a}). In addition, we presented the phase space plot corresponding to all three cases (see Figures \eqref{cha4f4},\eqref{cha4f5}, and \eqref{cha4f6}). We found that the viscous fluid model represents a Universe evolving from a matter dominated decelerated epoch, which is a past attractor, to a stable de-Sitter accelerated epoch, which is a future attractor. In addition, we investigated the physical behavior of the pressure component in the presence of viscosity, the effective EoS, and the $r-s$ parameter. We found that the pressure component in the presence of bulk viscosity presented in figure \eqref{cha4f7} shows a negative behavior for all three cases. The plot for the effective EoS parameter presented in figure \eqref{cha4f8} shows that the current Universe is going through a period of accelerated expansion. Lastly, from figure \eqref{cha4f9}, we found that the trajectories of our cosmological viscous model lie in the quintessence region. In addition, these trajectories converge to the fixed point $\Lambda$CDM for all three cases, which coincides with the results obtained in the phase space analysis. We found that the accelerated de-Sitter type phase comes purely from the $\bar{\zeta}_0$ case without any geometric modification to GR. Moreover, we found that the late-time behavior of all three cases of viscosity coefficients is identical. Furthermore, we have considered a non-linear model $f(Q)$, specifically $f(Q)=-Q+\beta Q^2$, and then we analyzed the behavior of model using the dynamical approach presented in Table \ref{Table-5a}. We found that there is only one de-Sitter type future attractor for case $\bar{\beta}=-1$, whereas two future attractors correspond to case $\bar{\beta}=1$ (see Figures \eqref{cha4n1} and \eqref{cha4n2}). However, the upper attractor $\left( 1,0.426 \right)$ in Figure \eqref{cha4n2} is a viable attractor from an initial big bang, and therefore the evolution from the big bang $H \rightarrow \infty$ can occur only towards the right fix point at $H \approx 0.75 H_0$ (see Figure \eqref{cha4add}).  Moreover, the dynamics of the case $\bar{\beta}=0$ is the same as that of Case III. Hence, we conclude that the late-time behavior of the considered non-linear model $f(Q)=-Q+\beta Q^2$ with $\beta \leq 0$ is similar to the linear case, whereas for case $\beta > 0$ results are quite different.  


 \chapter{Asymptotic dynamics of viscous $f(Q)$ models: Comparative analysis of inverse and quadratic forms} 

 \label{Chapter5} 

 \lhead{Chapter 5. \emph{Asymptotic dynamics of viscous $f(Q)$ models: Comparative analysis of inverse and quadratic forms}} 
 \vspace{8 cm}
 * The work presented in this chapter is covered by the following two publications: \\
 
 \textit{Phase-space analysis of the viscous fluid cosmological models in the coincident $f(Q)$ gravity}, Physics of the dark Universe \textbf{43}, 101421 (2024).

 \clearpage

In this chapter, we consider a newly proposed parameterization of the viscosity coefficient $\zeta$, specifically $\zeta=\bar{\zeta}_0 {\Omega^s_m} H $, where $\bar{\zeta}_0 = \frac{\zeta_0}{{\Omega^s_{m_0}}} $ within the coincident $f(Q)$ gravity formalism. We consider a non-linear function $f(Q)= -Q +\alpha Q^n$, where $\alpha$ and $n$ are arbitrary model parameters, which is a power law correction to the STEGR scenario. We find an autonomous system by invoking the dimensionless density parameters as the governing phase-space variables. We discuss the physical significance of the model corresponding to the parameters choices $n=-1$ and $n=2$ along with the exponent choices $s=0, 0.5$ and $1.05$. We find that model I shows the stable de-Sitter type or stable phantom type (depending on the choice of exponent $s$) behavior with no transition epoch, whereas model II shows the evolutionary phase from the radiation epoch to the accelerated de-Sitter epoch via passing through the matter-dominated epoch. Hence, we conclude that model I provides a good description of the late-time cosmology but fails to describe the transition epoch, whereas model II modifies the description in the context of the early Universe and provides a good description of the matter and radiation era along with the transition phase. 

\section{Introduction}\label{sec1v}

Modified gravity plays an prominent role in enhancing our understanding of the evolution of the Universe, encompassing its early phase and later stages marked by accelerated expansion, often associated with DE \cite{CANT}. Moreover, it also serves as a potential resolution for observational conflicts \cite{COSI}. These theories involve modifying GR to introduce additional degrees of freedom, potentially offering corrections at both the background and perturbation levels. Various methods exist to construct such modifications of GR. One such gravitational modification is obtained by utilizing the non-metricity to produce an equivalent formulation of GR, widely recognized as symmetric teleparallel gravity \cite{NEST}. In this approach, a generic affine connection is utilized, characterized by both vanishing torsion and vanishing curvature in relation to the Levi-Civita connection. This involves relaxing the metricity condition corresponding to the generic affine connection. Recently, this formulation has been further modified to give $f(Q)$ gravity \cite{R51}. The extended symmetric teleparallel formalism has attracted interest within the cosmology community as a promising pathway to investigate novel physics beyond the established $\Lambda$CDM cosmology. Specific functional forms $f(Q)$ have been shown to resolve tension $\sigma8$ \cite{BARR}, while other particular forms facilitate a more precise description of cosmological observational data \cite{ANAG,NUNES}. Recently, notable cosmological implications of $f(Q)$ gravity in various contexts have appeared, for instance, neutrino physics \cite{NEOM}, black hole physics \cite{RODR,e7}, astrophysical objects \cite{SNEHA,ZINNAT}, quantum cosmology \cite{CAPE-1,PALIA-1}, cosmological perturbations \cite{ET}, BBN constraints \cite{ANAG-2}, phantom cosmology \cite{ANDER}, inflation \cite{CAPE-2}, and many others \cite{perturb,Hohmann2,DE-1,DE-2,PALIA-2}. 

It is well known that incorporating auxiliary variables can transform the cosmological set of field equations into a set of autonomous differential equations \cite{COPE}. This leads to a system $X' = f(X)$, where $X$ stands for the column vector encompassing auxiliary variables, and $f(X)$ represents the vector field. The examination of stability for the given autonomous system involves a multi-step process. First, the equilibrium points (or critical points) denoted by $X_c$ are estimated using the equation $X' = 0$. Following this, we assume linear perturbations near the critical point $X_c$, representing them as $ X = X_c + P $, where $P$ is the column vector representing perturbed auxiliary variables. Consequently, it is possible to establish the matrix equation $P = AP$ (up to first order), where A represents the matrix comprising the coefficients of the perturbed equations. The stability properties of each hyperbolic-type equilibrium point can be examined utilizing the eigenvalues of the coefficient matrix A. An equilibrium point $X_c$ of the given autonomous system is stable (unstable) or a saddle based on whether the real parts of the associated eigenvalues are negative (positive) or possess real parts with opposite signs. Various notable findings within the realm of modified gravity using the dynamical system approach have appeared in references \cite{DE-3,WOM,Mishra-2,HAMID,ADDP6}. 
The chapter is organized as follows. In sec \ref{cha5sec2}, we present the governing equations of motion corresponding to flat homogeneous and isotropic FLRW background. In sec \ref{cha5sec3}, we set up an autonomous system corresponding to a generic power law $f(Q)$ function along with a newly proposed parameterization of the viscosity coefficient. Further, we present our analysis for two toy models with corresponding phase diagrams and relevant cosmological parameters. Finally, in sec \ref{cha5sec4}, we discuss our findings.

\section{Equations of motion}\label{cha5sec2}

In order to investigate the cosmological consequences under the cosmological principle assumption, we start with the following flat FLRW metric \eqref{FLRW}.
To initiate this analysis, we begin with the teleparallel constraint associated with a flat geometry, which represents a purely inertial connection. Subsequently, a gauge transformation, parameterized by $\Lambda^\alpha_\mu$ \cite{R49}, can be performed,
\begin{equation}\label{3b}
 \Upsilon^\alpha_{\: \mu \nu}  = (\Lambda^{-1})^\alpha_{\:\: \beta} \partial_{[ \mu}\Lambda^\beta_{\: \: \nu ]} \text{.}
\end{equation}
As a result, the generic affine connection can be represented in the following way, 
\begin{equation}\label{3c}
\Upsilon^\alpha_{\: \mu \nu} = \frac{\partial x^\alpha}{\partial \zeta^\rho} \partial_\mu \partial_\nu \zeta^\rho \text{.}
\end{equation}
This representation takes advantage of an arbitrary element of the group $ GL(4,\mathbb{R}) $, which is defined by the transformation $ \Lambda^\alpha_{\: \: \mu}=\partial_\mu \zeta^\alpha$. It is essential to note that $ \zeta^\alpha $ represents an arbitrary vector field in this context. In addition, this suggests that it is feasible to eliminate the affine connection by employing a coordinate transformation. Such a coordinate transformation is well-known as the gauge coincident. Assuming the coincident gauge in the present formalism, the non-metricity scalar associated with the metric \eqref{FLRW} simplifies to $Q=6H^2$.

In the context of the dark matter fluid, the concept of bulk viscous pressure is proposed as a means to incorporate an effective pressure, enabling a broader range of phenomenological outcomes within a cosmological framework that extends beyond the conventional perfect fluid model. This effective pressure is defined as $\bar{p}=p-3\zeta H$, where $\zeta$ represents the bulk viscosity coefficient, ensuring compliance with the second law of thermodynamics under the condition that $\zeta > 0$. In this chapter, we consider the newly proposed parameterization of $\zeta$ as follows \cite{NVV},
\begin{equation}\label{3d}
\zeta=\zeta_0 H^{1-2s} H_0^{2s} \left(\frac{\rho_m}{\rho_{m_0}}\right)^s=\bar{\zeta}_0 {\Omega^s_m}H  \text{.}
\end{equation}
where $\bar{\zeta}_0 = \frac{\zeta_0}{{\Omega^s_{m_0}}} $. This parameterization offers some benefits; one notable advantage is that it encompasses widely recognized models, $ \zeta=\zeta(H)$ for the case $ s=0$ and $ \zeta \sim {{\rho}_m}^{\frac{1}{2}} $ for the case $s = \frac{1}{2}$.
The Friedmann like equations for the functional form $f(Q) = - Q + \Psi(Q)$ with respect to the metric \eqref{FLRW} can be given as,
\begin{equation}\label{3ff}
\Psi+Q-2Q\Psi_Q=2\rho  \text{,}
\end{equation}
and
\begin{equation}\label{3gg}
\dot{H}=\frac{p+\rho}{2(-1+\Psi_Q+2\Psi_{QQ})} \text{.}
\end{equation}
The Friedmann equations \eqref{3ff} and \eqref{3gg} can be recognized as STEGR cosmology with an additional DE fluid part. The evolution of the DE components emerging due to non-metricity is defined by
\begin{equation}\label{3h}
\rho_{de}=-\frac{\Psi}{2}+ Q\Psi_Q  \text{,}
\end{equation}
and
\begin{equation}\label{3i}
p_{de}=-\rho_{de}-2\dot{H} \left( \Psi_Q+2Q\Psi_{QQ} \right)  \text{.}
\end{equation}
where $\rho_{de}$ and $p_{de}$ represent the corresponding energy density and pressure term. The effective Friedmann equations possessing radiation and viscous type matter along with a contribution of geometrical DE component reads as,
\begin{equation}\label{3j}
3H^2= \rho_m + \rho_r + \rho_{de}  \text{,}
\end{equation}
and
\begin{equation}\label{3k}
\dot{H}=-\frac{1}{2} [\rho_m + \rho_r + \rho_{de} + \bar{p}_m + p_r+p_{de}]  \text{.}
\end{equation}
Moreover, the standard continuity equation can be obtained as,
\begin{equation}\label{3l}
\dot{\rho}_m + 3H\left(\rho_m- 3 \zeta H \right)=0  \text{,}
\end{equation}
\begin{equation}\label{3m}
\dot{\rho}_r + 4H \rho_r=0  \text{,}
\end{equation}
\begin{equation}\label{3n}
\dot{\rho}_{de} + 3H\left(\rho_{de}+p_{de}\right)=0  \text{.}
\end{equation}

\section{The cosmological model and phase-space analysis}\label{cha5sec3}

\justifying

\begin{figure}[H]
\centering
\includegraphics[width=5.6cm,height=5.6cm]{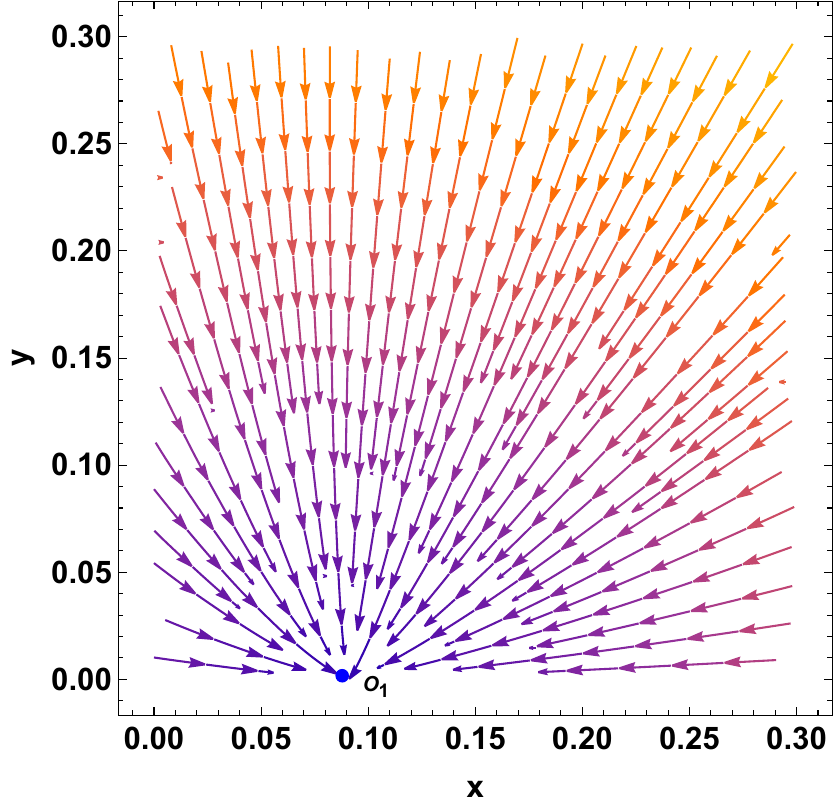}
\includegraphics[width=5.6cm,height=5.6cm]{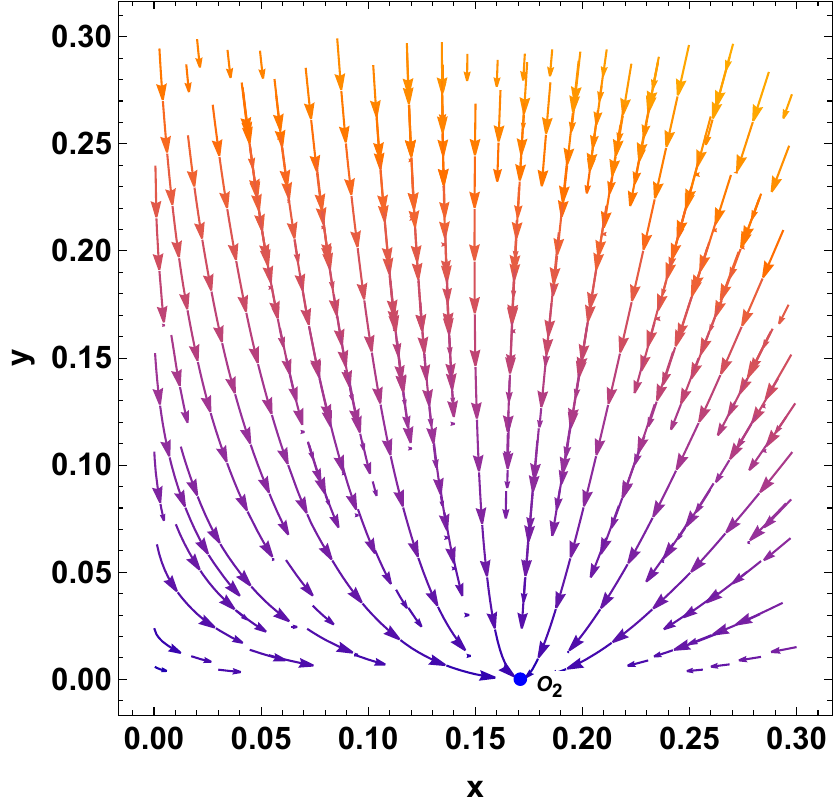}
\includegraphics[width=5.6cm,height=5.6cm]{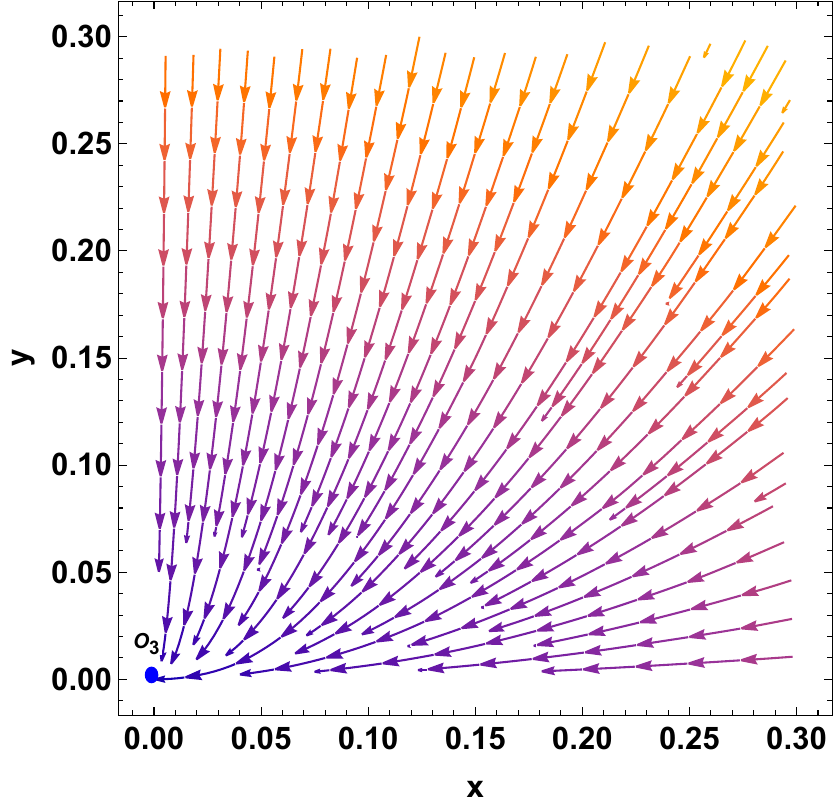}
\caption{Phase-space diagrams representing the behavior of trajectories for the model I corresponding to cases $(s,\bar{\zeta}_0)=(0,0.1)$, $(0.5,0.5)$ and $(1.05,0.01)$ respectively.}
\label{f1}
\end{figure}

\begin{figure}[h]
\centering
\includegraphics[width=6cm,height=4.5cm]{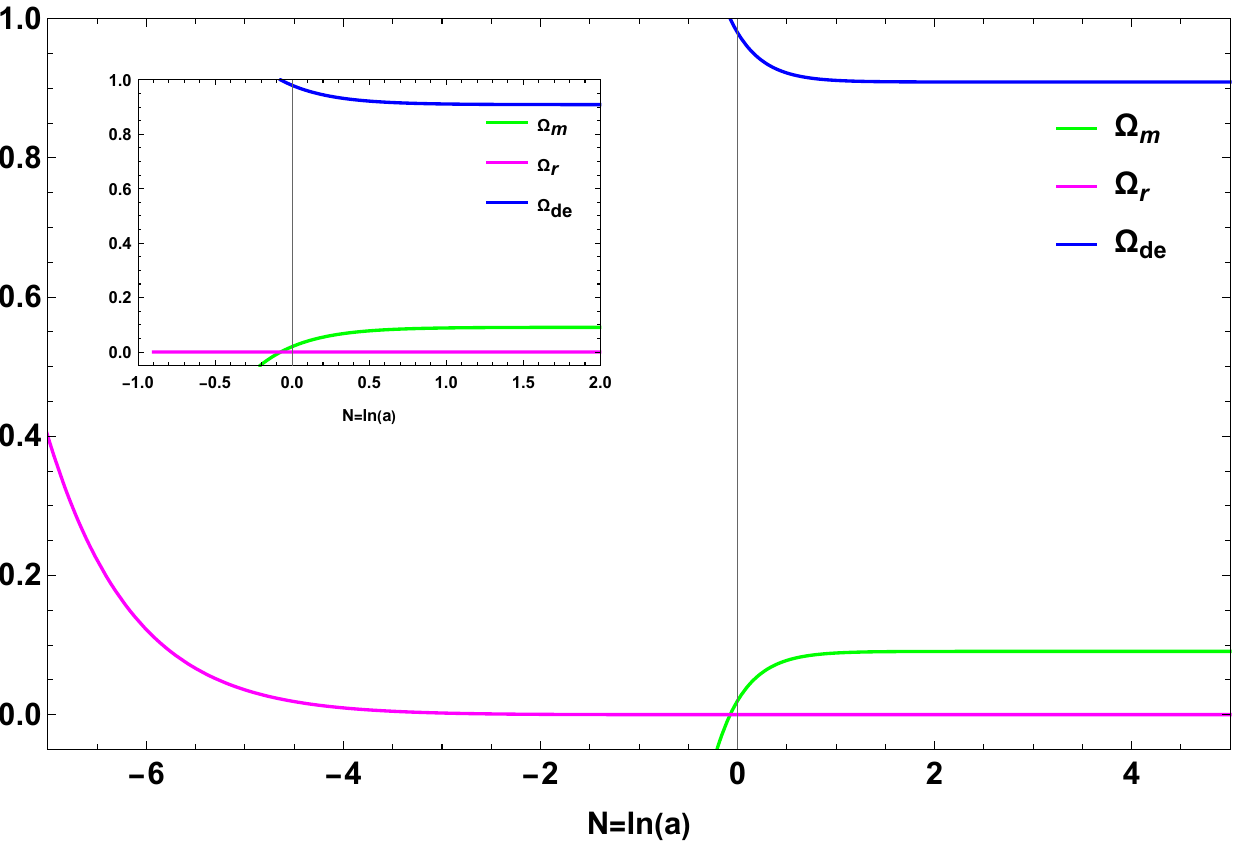}
\includegraphics[width=6cm,height=4.5cm]{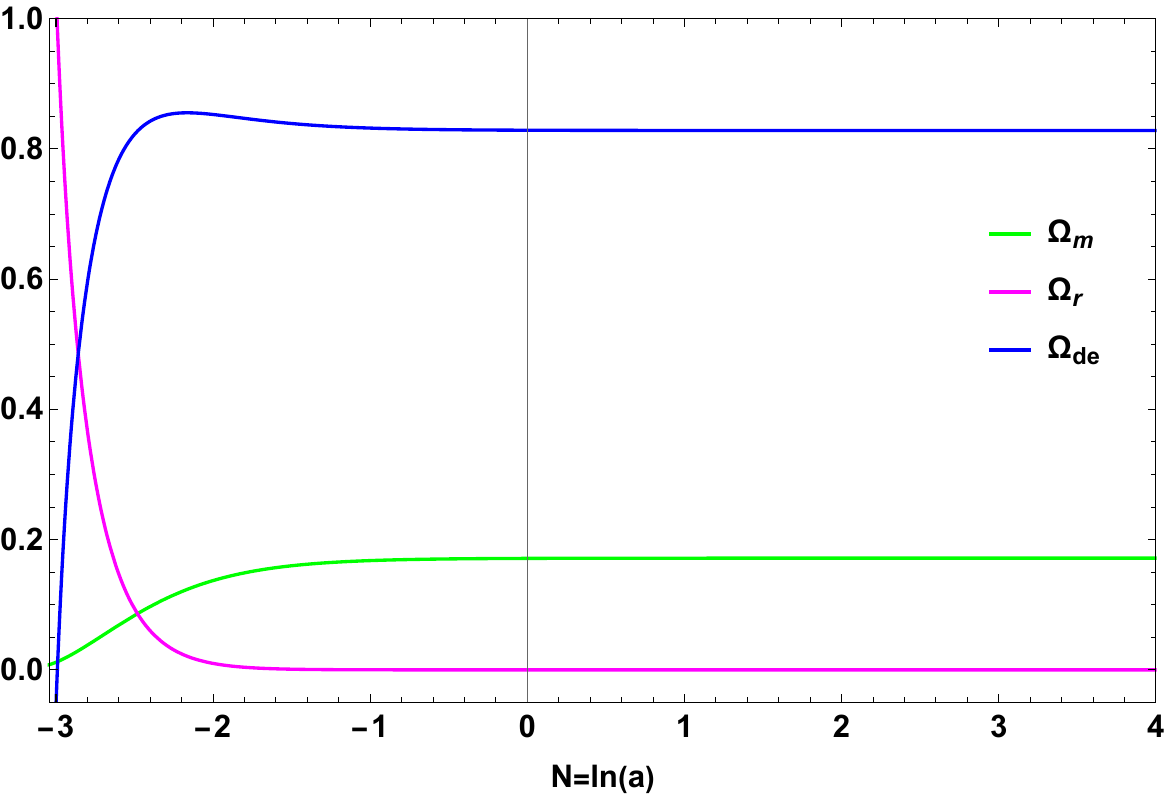}
\includegraphics[width=6cm,height=4.5cm]{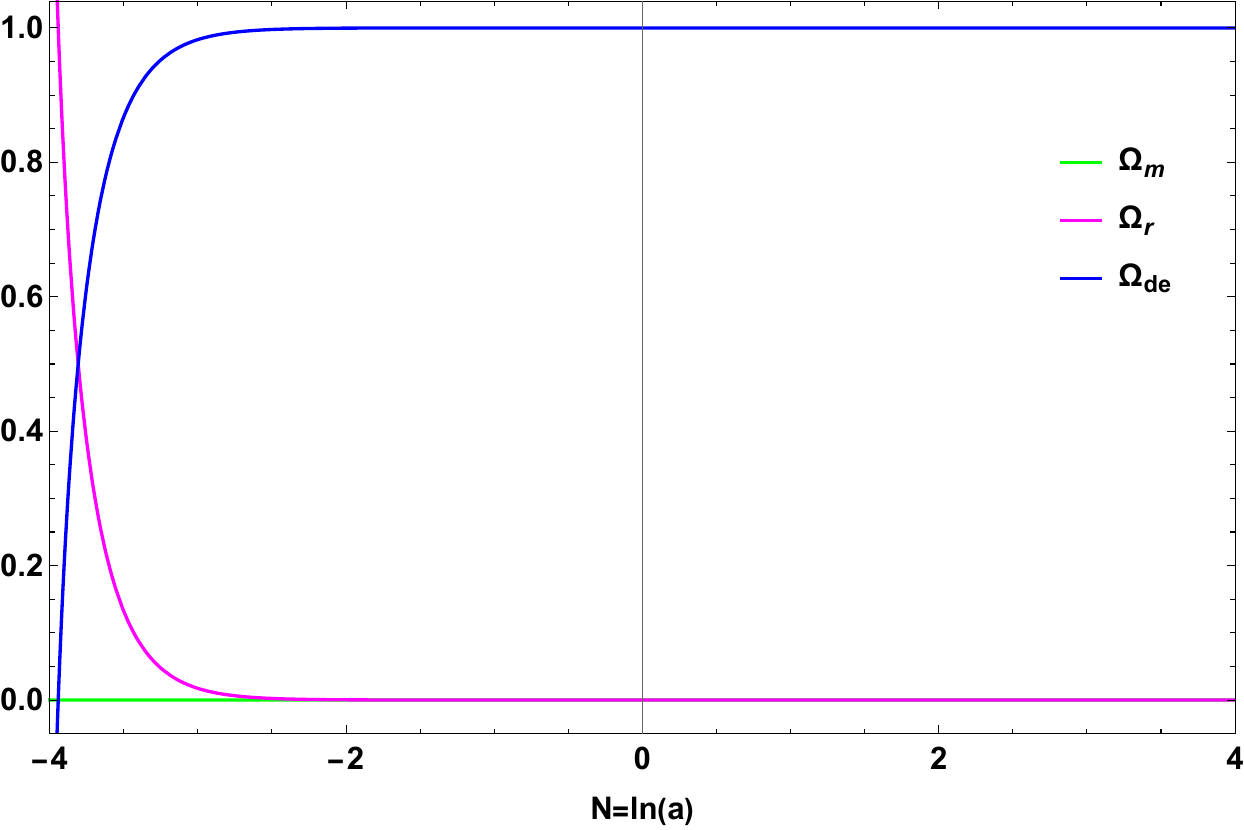}
\caption{Profile of the matter, radiation, and the DE density parameter for the model I corresponding to cases $(s,\bar{\zeta}_0)=(0,0.1)$, $(0.5,0.5)$ and $(1.05,0.01)$ respectively.}
\label{f2}
\end{figure}

\begin{figure}[h]
\centering
\includegraphics[width=6cm,height=4.5cm]{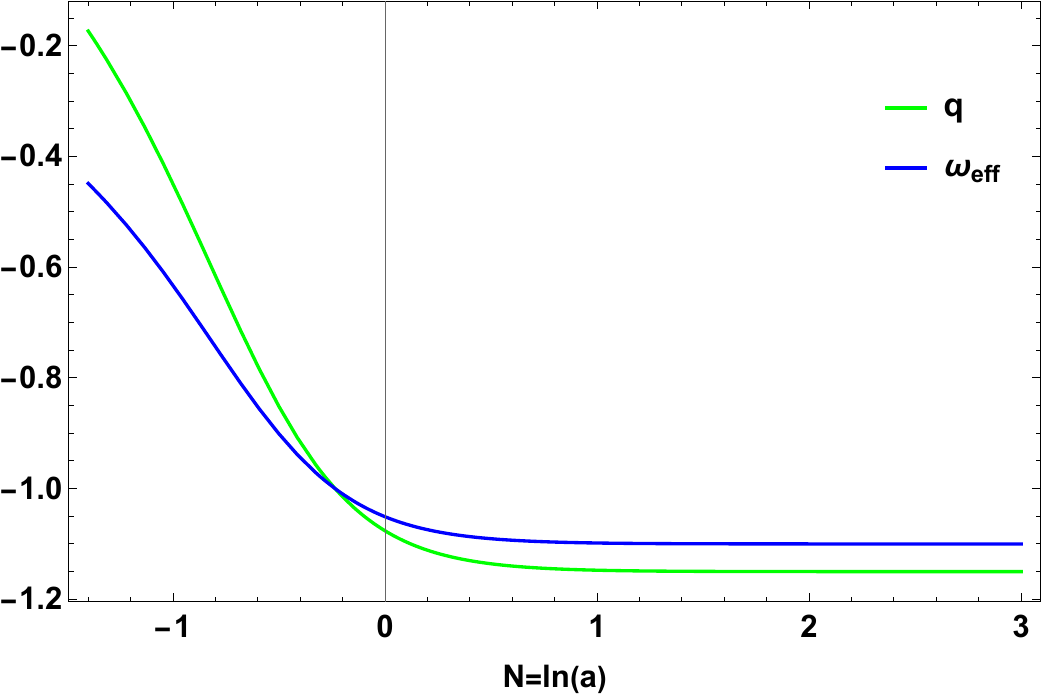}
\includegraphics[width=6cm,height=4.5cm]{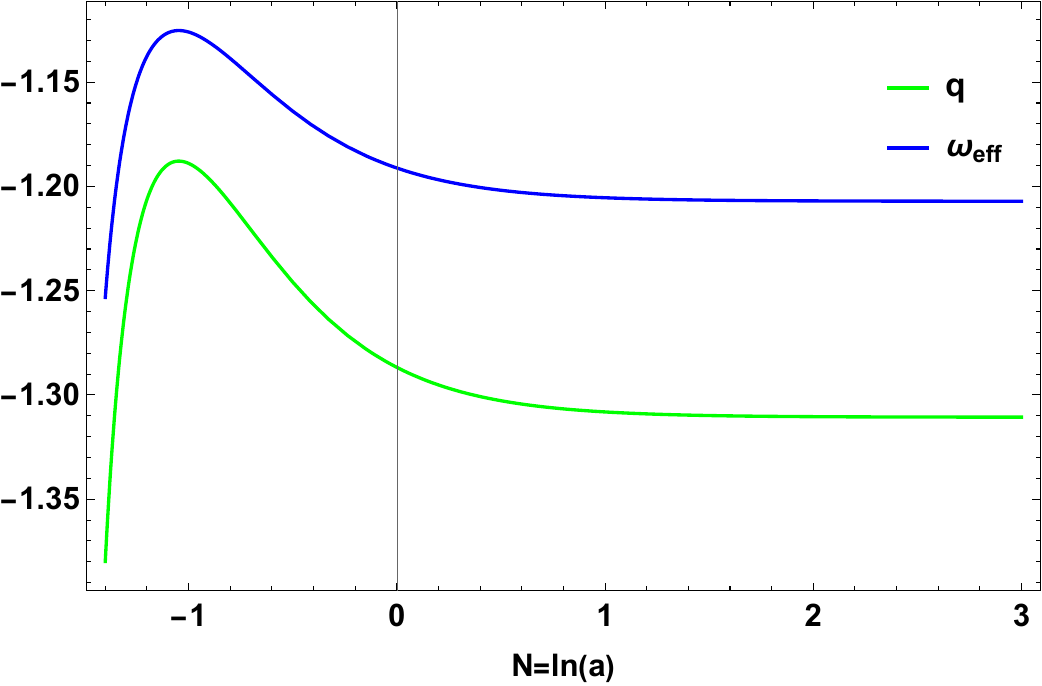}
\includegraphics[width=6cm,height=4.5cm]{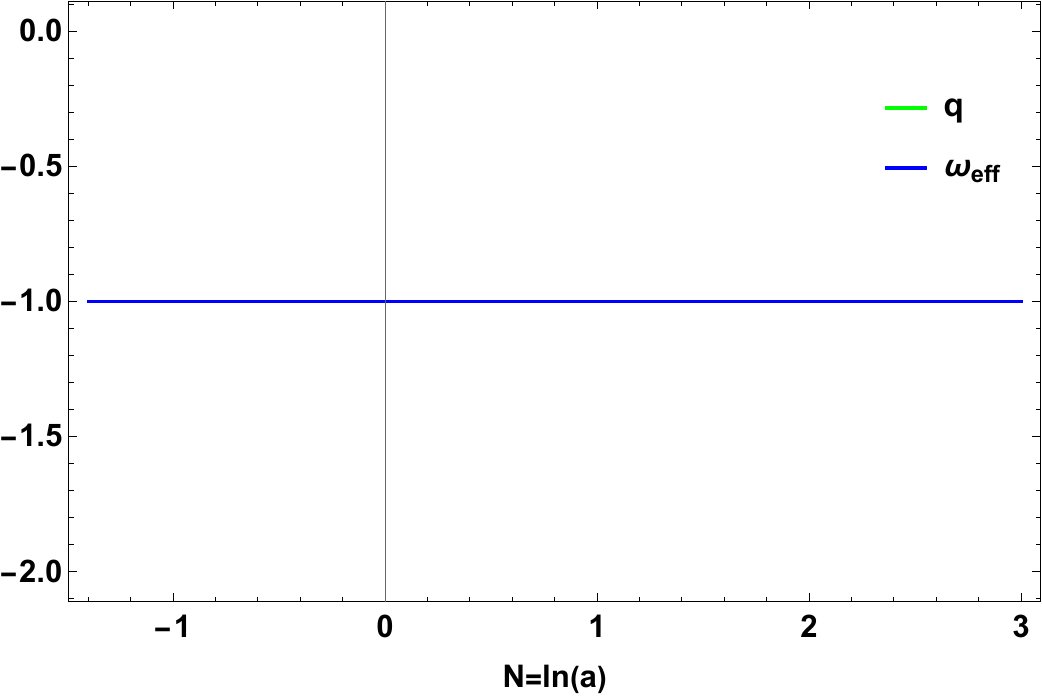}
\caption{Profile of the effective EoS and deceleration parameter for the model I corresponding to cases $(s,\bar{\zeta}_0)=(0,0.1)$, $(0.5,0.5)$ and $(1.05,0.01)$ respectively.}
\label{f3}
\end{figure}

In this section, we begin with the following dimensionless variables that encompass the entire evolution of the system's phase space. This variable allows us to transform the system's dynamics into the structure of an autonomous system, which makes it easier to understand the behavior of the system. The dimensionless variables considered are as follows
\begin{equation}\label{4a}
x=\Omega_m=\frac{\rho_m}{3H^2}, \:\: y=\Omega_r=\frac{\rho_r}{3H^2} \:\: \text{and} \:\: z=\Omega_{de}=\frac{\rho_{de}}{3H^2} \text{.}
\end{equation}
Using equation \eqref{4a}, the expression \eqref{3j} becomes
\begin{equation}\label{4b}
x+y+z=1  \text{.}
\end{equation}
Hence, it follows that $z=1-x-y$ and $0 \leq x,y,z \leq 1$. Again, using equations \eqref{3j} and \eqref{3k} with continuity equations, we obtain the following,
\begin{equation}\label{4c}
\frac{\dot{H}}{H^2}=\frac{-3x\left(1+\bar{\zeta}_0 x^{s-1}\right)-2y}{2(\Psi_Q+2Q \Psi_{QQ}-1)} \text{.}
\end{equation}
\justify We consider a power law function of the non-metricity, specifically, $\Psi(Q)=\alpha Q^n$ i.e. $f(Q)=-Q+\alpha Q^n$, where $\alpha$ and $n$ are arbitrary model parameters. Also, the assumed choice of the $f(Q)$ non-metricity function is suitable to close the governing dynamical system. Similarly to this choice, the expression \eqref{4c} becomes,
\begin{equation}\label{4d}
\frac{\dot{H}}{H^2}=\frac{-3x\left(1+\bar{\zeta}_0 x^{s-1}\right)-2y}{2\left[n(1-x-y)-1\right]} \text{.}
\end{equation}
The corresponding expressions for the deceleration and the effective EoS parameter reads as,
\begin{equation}\label{4e}
q=-1+ \frac{3x\left(1+\bar{\zeta}_0 x^{s-1}\right)+2y}{2\left[n(1-x-y)-1\right]} \text{,}
\end{equation}
and
\begin{equation}\label{4f}
\omega_{eff}=-1+\frac{3x\left(1+\bar{\zeta}_0 x^{s-1}\right)+2y}{3\left[n(1-x-y)-1\right]} \text{.}
\end{equation}
We acquired the following autonomous system of equations, with respect to the e-folding time $N=ln(a)$, for the assumed cosmological settings consisting of the power law $f(Q)$ function with the cosmic fluid incorporates radiation and viscous matter, 
\begin{equation}\label{4g}
x'=x\left[3\bar{\zeta}_0 x^{s-1}-3+\frac{\left[3x\left(1+\bar{\zeta}_0 x^{s-1}\right)+2y\right]}{\left[n(1-x-y)-1\right]}\right],
\end{equation}
\begin{equation}\label{4h}
y'=\frac{y}{\left[n(1-x-y)-1\right]}\left[3x\left(1+\bar{\zeta}_0 x^{s-1}\right)-4n(1-x-y)+2y+4\right].
\end{equation}
\justify From the presented set of equations it is evident that computing the equilibrium points of the system explicitly for the arbitrary choice of parameters is difficult. Hence, for further investigation, we will explore two specific $f(Q)$ models for some well known choice of the exponent, as follows: 
\justify \textbf{Model I: $f(Q)=-Q+\alpha Q^{-1}$ (i.e. $n=-1$) :} The considered model is characterized by the value $n < 1$ (i.e., a correction to STEGR) that can provide modifications to the late-time cosmological phenomenon, potentially influencing the emergence of DE \cite{R49}. Similarly to this model, we investigate the dynamical system presented in \eqref{4g}-\eqref{4h} for the aforementioned well-known cases of the exponent $s$, specifically $(s,\bar{\zeta}_0) = (0,0.1), (0.5,0.5)$ and $(1.05,0.01)$. For each of the cases, we linearize the system in the neighborhood of the critical points obtained and examine the corresponding stability. The outcome of the investigation corresponding to each case is presented in the following Table \eqref{Table-1} and the corresponding phase-space diagrams in Figure \eqref{f1}. Moreover, the behavior of the corresponding dimensionless density parameters, along with the effective equation of the state and deceleration parameters, is presented in Figures \eqref{f2} and \eqref{f3}.

\begin{table}[H]
\centering
\caption{Table shows the critical points and their behavior corresponding to model I.}
\label{Table-1}

\begin{tabular}{|c|c|c|c|c|c|}
\hline
Cases $(s,\bar{\zeta}_0)$ & Critical Points $(x_c,y_c)$ 
& Eigenvalues & Nature of critical point & $q$ & $\omega_{\text{eff}}$ \\
\hline

$(0,0.1)$ & $O_1(0.0909,0)$ 
& $-4.3,\,-3.457$ 
& Stable & $-1.15$ & $-1.1$ \\
\hline

$(0.5,0.5)$ & $O_2(0.1715,0)$ 
& $-4.621,\,-2.320$ 
& Stable & $-1.00374$ & $-1.00249$ \\
\hline

$(1.05,0.01)$ & $O_3(0,0)$ 
& $-4,\,-3$ 
& Stable & $-1$ & $-1$ \\
\hline

\end{tabular}\label{Table-1}
\end{table}

\textbf{Case I $(s=0,\bar{\zeta}_0=0.1)$}: In relation to case I, we obtained a single critical point $O_1(0.0909,0)$ with the associated eigenvalues $(\lambda_1,\lambda_2)=(-4.3,-3.457)$ and $(q,\omega_{eff})=(-1.15,-1.1)$. Hence, $O_1$ is a stable equilibrium point representing the phantom type accelerated expansion of the Universe with no transition epoch (see left panel of the Figure \eqref{f1}). The same is reflected in the evolutionary description of the cosmological parameters shown in Figures \eqref{f2} and \eqref{f3} (left panel).

\textbf{Case II $(s=0.5,\bar{\zeta}_0=0.5)$}: In relation to case II, we obtained a single critical point $O_2(0.1715,0)$ with the associated eigenvalues $(\lambda_1,\lambda_2)=(-4.621,-2.320)$ and the value of deceleration and the EoS parameters are $(q,\omega_{eff})=(-1.00374,-1.00249)$. Hence, $O_2$ is a stable equilibrium point representing the phantom type accelerated expansion of the Universe with no transition epoch (see middle panel of Figure \eqref{f1}). The same is reflected in the evolutionary description of cosmological parameters shown in Figures \eqref{f2} (right panel above) and \eqref{f3} (middle panel).

\textbf{Case III $(s=1.05,\bar{\zeta}_0=0.01)$}: In relation to case III, we obtained a single critical point $O_3(0,0)$ with the associated eigenvalues $(\lambda_1,\lambda_2)=(-4,-3)$ and $(q,\omega_{eff})=(-1,-1)$. Hence, $O_3$ is a stable equilibrium point representing the de-Sitter type accelerated expansion of the Universe with no transition epoch (see right panel of Figure \eqref{f1}). The same is reflected in the evolutionary description of cosmological parameters shown in Figures \eqref{f2} (below) and \eqref{f3} (right panel).

\begin{figure}[H]
\centering
\includegraphics[width=5.5cm,height=5.5cm]{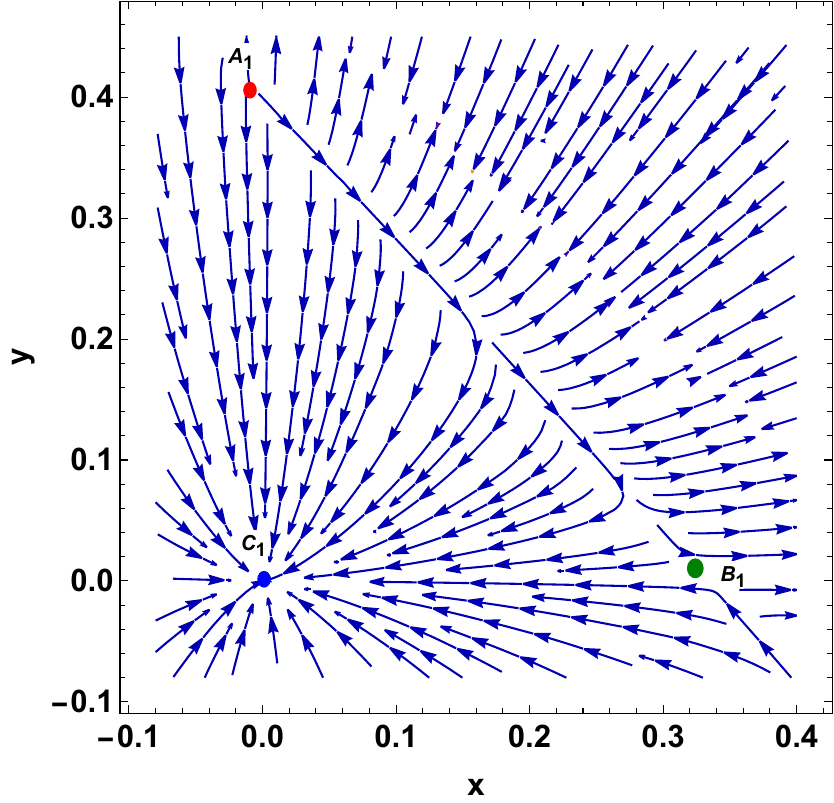}
\includegraphics[width=5.5cm,height=5.5cm]{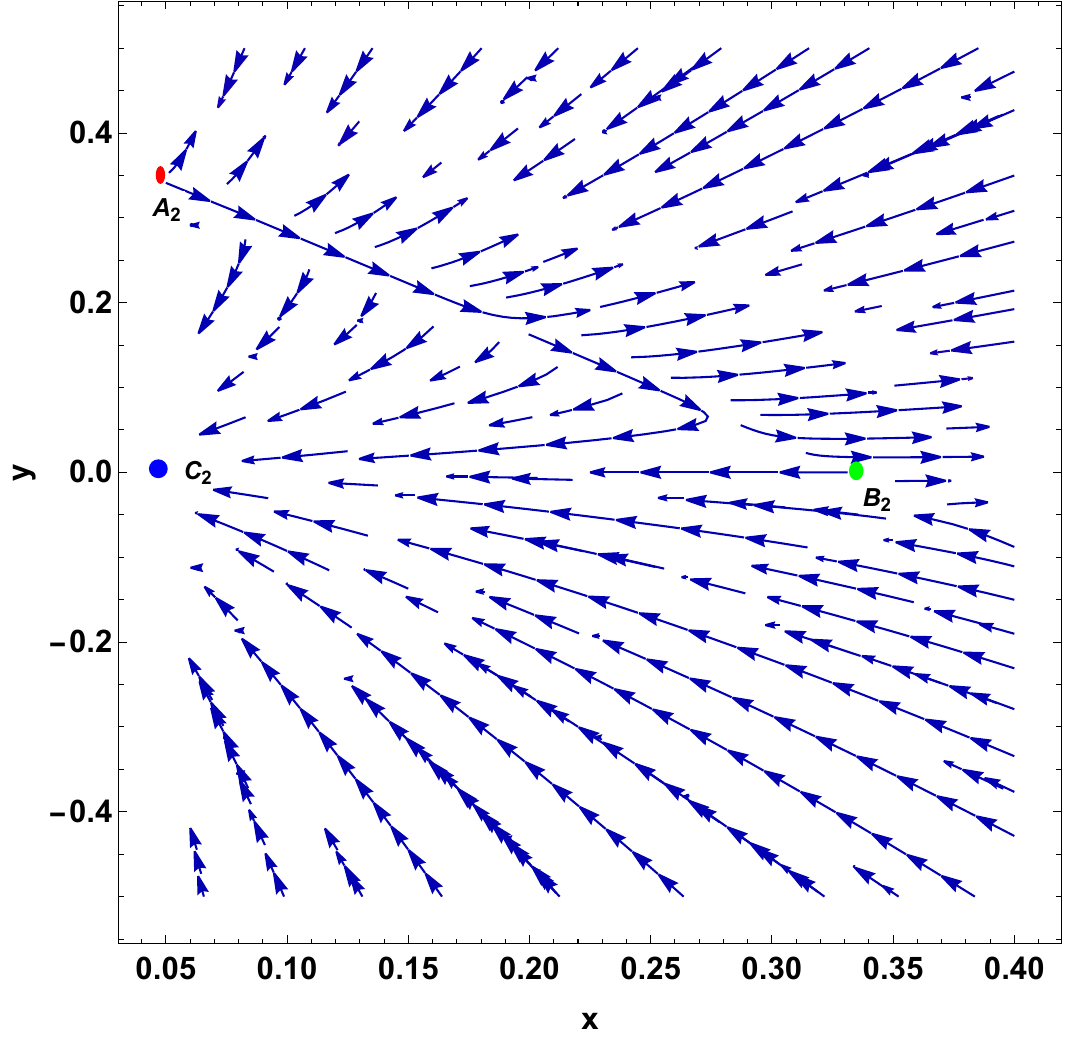}
\includegraphics[width=5.5cm,height=5.5cm]{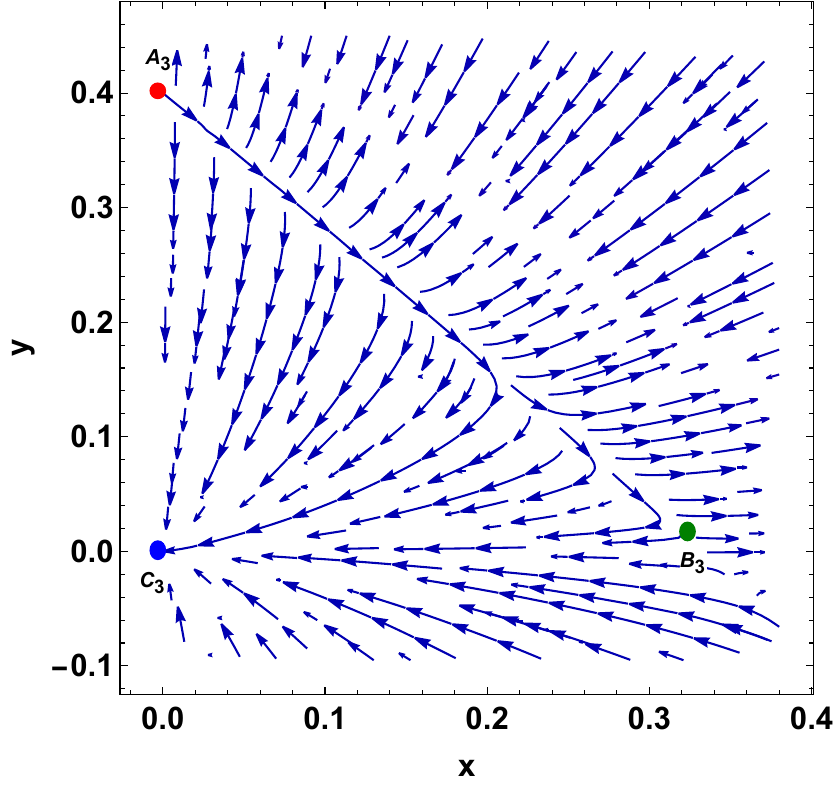}
\caption{Phase-space diagrams representing the behavior of trajectories for the model II corresponding to cases $(s,\bar{\zeta}_0)=(0,0.001)$, $(0.5,0.005)$, and $(1.05,0.01)$ respectively.}
\label{f4}
\end{figure}

\begin{figure}[H]
\centering
\includegraphics[width=6cm,height=5cm]{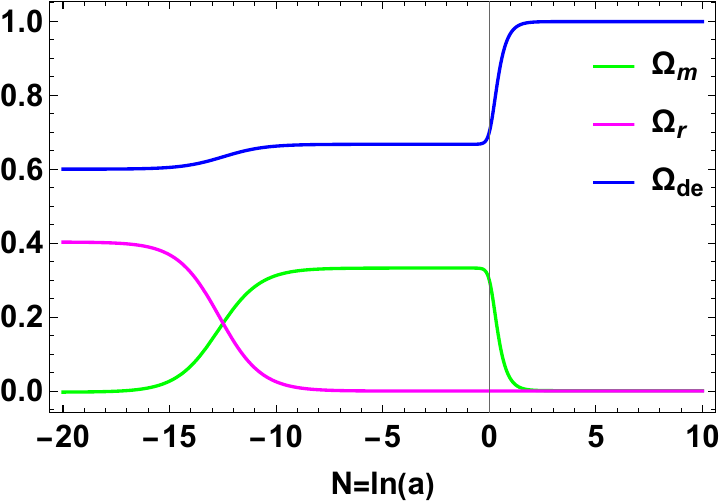}
\includegraphics[width=6cm,height=5cm]{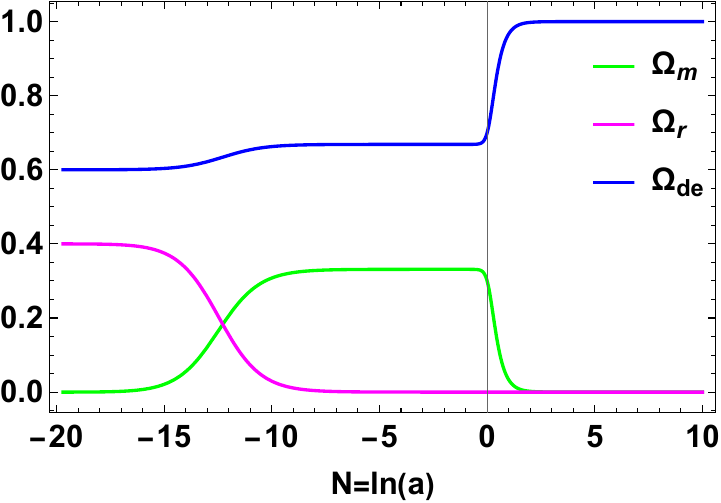}
\includegraphics[width=6cm,height=5cm]{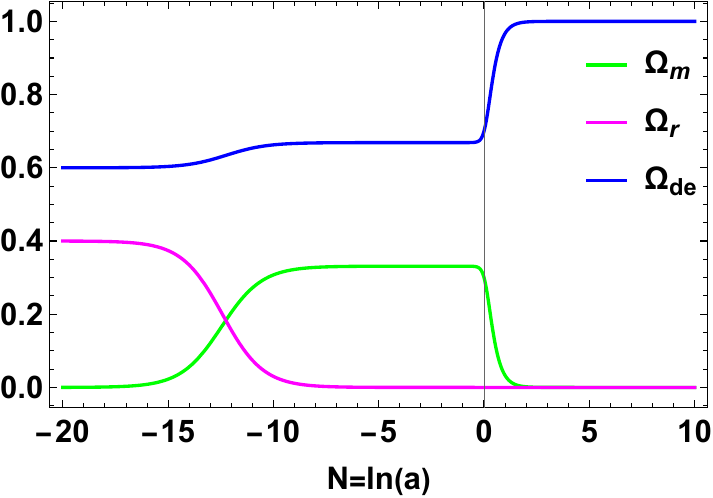}
\caption{Profile of the matter, radiation, and the DE density parameter for the model II corresponding to cases $(s,\bar{\zeta}_0)=(0,0.001)$, $(0.5,0.005)$, and $(1.05,0.01)$ respectively.}
\label{f5}
\end{figure}

\begin{figure}[H]
\centering
\includegraphics[width=6cm,height=5cm]{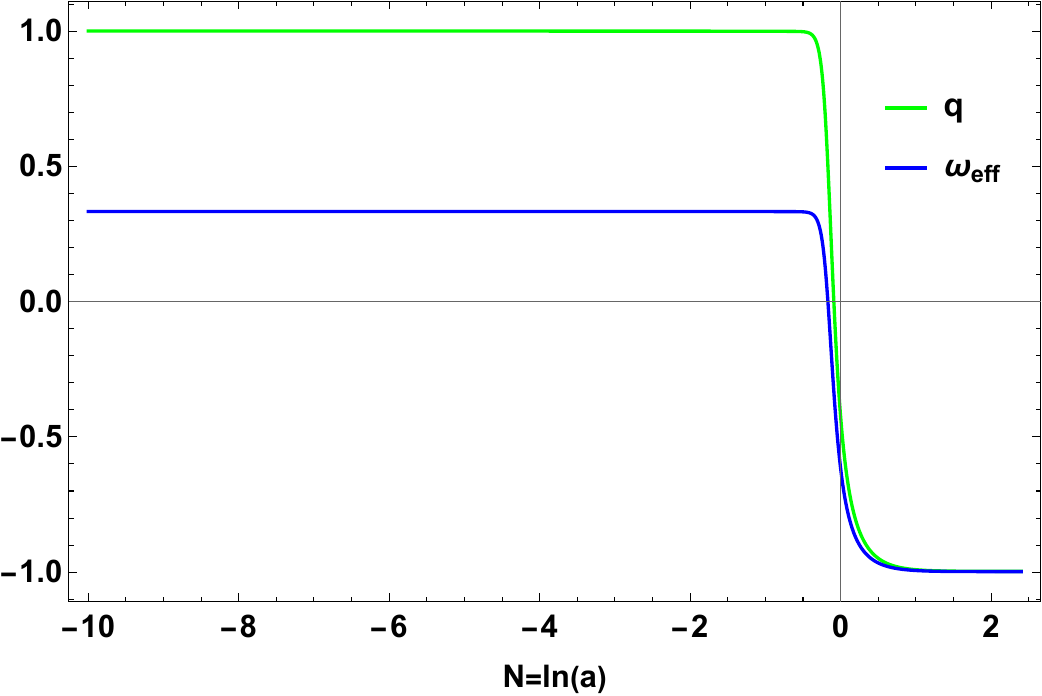}
\includegraphics[width=6cm,height=5cm]{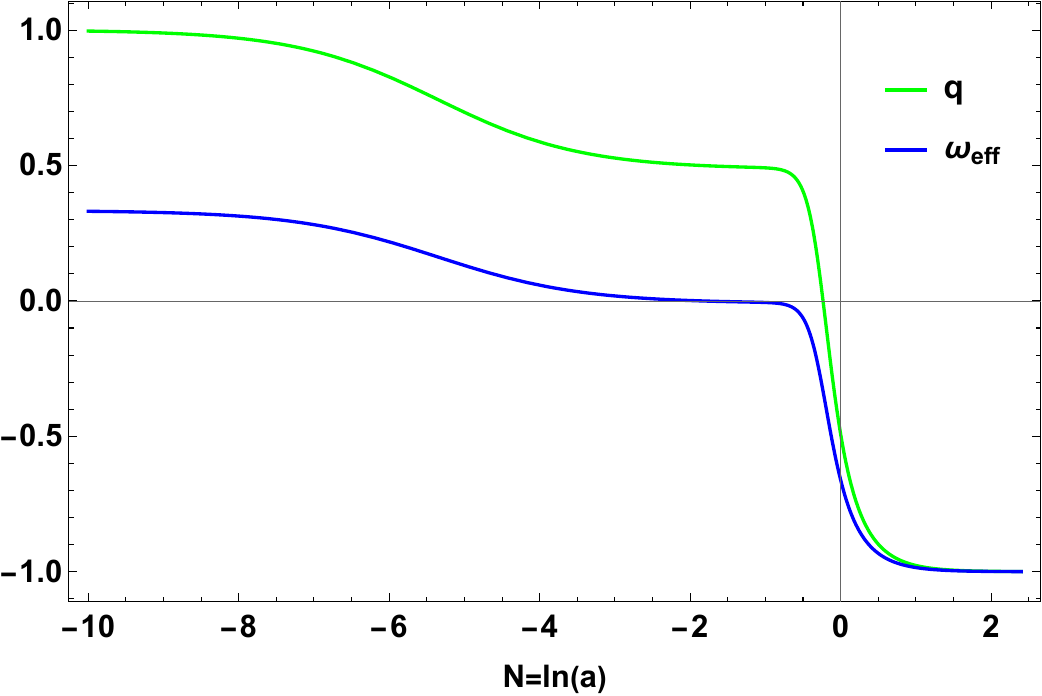}
\includegraphics[width=6cm,height=5cm]{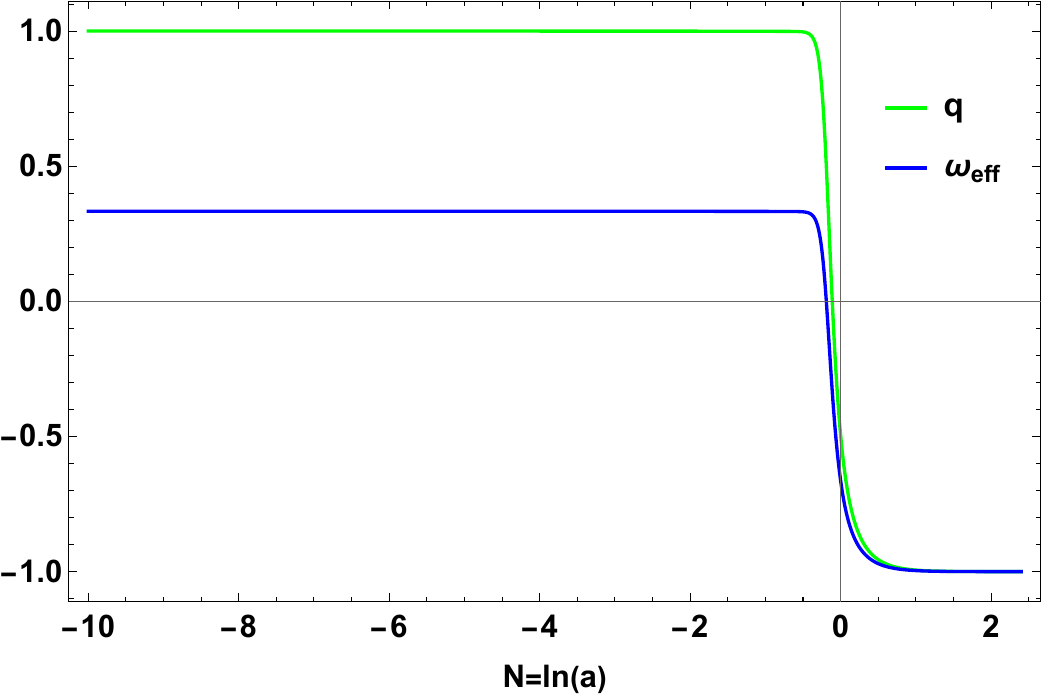}
\caption{Profile of the effective EoS and deceleration parameter for the model II corresponding to cases $(s,\bar{\zeta}_0)=(0,0.001)$, $(0.5,0.005)$, and $(1.05,0.01)$ respectively.}
\label{f6}
\end{figure}

\justify \textbf{Model II: $f(Q)=-Q+\alpha Q^{2}$ (i.e. $n=2$) :} The considered model is characterized by a value $n > 1$ (i.e., a correction to STEGR) that can provide modifications to the early Universe phenomenon, offering potential inflationary solutions \cite{R49}. Similarly to this model, we investigate the dynamical system presented in \eqref{4g}-\eqref{4h} for the aforementioned well-known cases of the exponent $s$, specifically $(s,\bar{\zeta}_0) = (0,0.001), (0.5,0.005)$ and $(1.05,0.01)$. The outcome of the investigation corresponding to each case is presented in the following Table \eqref{Table-2} and the corresponding phase-space diagrams in Figure \eqref{f4}. Moreover, the behavior of the corresponding dimensionless density parameters, along with the effective equation of the state and deceleration parameters, is presented in Figures \eqref{f5} and \eqref{f6}.

\begin{table}[H]
\centering
\caption{Table shows the critical points and their behavior corresponding to model II.}
\label{Table-2}

\begin{tabular}{|c|c|c|c|c|c|}
\hline
Cases $(s,\bar{\zeta}_0)$ & Critical Points $(x_c,y_c)$ 
& Eigenvalues & Nature of critical point 
& $q$ & $\omega_{\text{eff}}$ \\
\hline

{$(0,0.001)$}
& $A_1(0,0.403)$ 
& $19.976,\;1.008$ 
& Unstable & $1$ & $\frac{1}{3}$ \\
& $B_1(0.332,0)$ 
& $8.919,\;-1.009$ 
& Saddle & $0.495$ & $-0.003$ \\
& $C_1(0.001,0)$ 
& $-3.993,\;-2.990$ 
& Stable & $-0.996$ & $-0.997$ \\
\hline

{$(0.5,0.005)$}
& $A_2(0.02,0.4)$ 
& $19.85,\;1$ 
& Unstable & $1$ & $0.3337$ \\
& $B_2(0.331,0)$ 
& $8.79,\;-1.02$ 
& Saddle & $0.487$ & $-0.0086$ \\
& $C_2(0,0)$ 
& $-4,\;-1.5$ 
& Stable & $-0.999$ & $-0.999$ \\
\hline

{$(1.05,0.01)$}
& $A_3(0,0.4)$ 
& $20,\;1$ 
& Unstable & $1$ & $\frac{1}{3}$ \\
& $B_3(0.333,0)$ 
& $9.144,\;-0.9716$ 
& Saddle & $0.514$ & $0.00946$ \\
& $C_3(0,0)$ 
& $-4,\;-3$ 
& Stable & $-1$ & $-1$ \\
\hline

\end{tabular}
\end{table}

\textbf{Case I $(s=0,\bar{\zeta}_0=0.001)$}: In relation to case I, the critical points obtained are $A_1(0,0.403)$, $B_1(0.332,0)$ and $C_1(0.001,0)$ with the associated eigenvalues $(\lambda_1,\lambda_2)=( 19.976,1.008)$,\\  $( 8.919, -1.009)$ and $(-3.993,-2.990)$ respectively. Moreover, the associated $(q,\omega_{eff})$ are $(1,\frac{1}{3})$, $(0.495,-0.003)$ and $(-0.996,-0.997)$ respectively. Hence, the equilibrium point $A_1$ is unstable, $B_1$ is saddle, and $C_1$ is stable, representing the radiation, matter, and de-Sitter phase, respectively. It is evident from the direction of phase-space trajectories (see left panel of the Figure \eqref{f4}), that the evolutionary trajectory of the model emerges from the radiation epoch and then converges to the de-Sitter accelerated epoch via passing through the matter-dominated epoch i.e. having the evolution phase $A_1 \rightarrow B_1 \rightarrow C_1$. The evolutionary description of the cosmological parameters shown in Figures \eqref{f5} and \eqref{f6} (left panel) reflects the same.

\textbf{Case II $(s=0.5,\bar{\zeta}_0=0.005)$}: In relation to case II, the critical points obtained are $A_2(0.02,0.4)$, $B_2(0.331,0)$ and $C_2(0,0)$ with the associated eigenvalues $(\lambda_1,\lambda_2)=(19.85,1)$,\\
$(8.79,-1.02)$ and $(-4,-1.5)$ respectively. Moreover, the associated $(q,\omega_{eff})$ are $(1,0.337)$, $(0.487,-0.0086)$ and $(-0.999,-0.999)$ respectively. Hence, the equilibrium point  $A_2$ is unstable, $B_2$ is saddle, and $C_2$ is stable, representing the radiation, matter, and de-Sitter phase, respectively. It is evident from the direction of phase-space trajectories (see middle panel of the Figure \eqref{f4}), that the evolutionary trajectory of the model emerges from the radiation epoch and then converges to the de-Sitter accelerated epoch via passing through the matter-dominated epoch i.e, having the evolution phase $A_2 \rightarrow B_2 \rightarrow C_2$. The evolutionary description of the cosmological parameters shown in Figures \eqref{f5} and \eqref{f6} (middle panel) reflects the same.

\textbf{Case III $(s=1.05,\bar{\zeta}_0=0.01)$}: corresponding to case III, the critical points obtained are $A_3(0,0.4)$, $B_3(0.333,0)$ and $C_3(0,0)$ with the associated eigenvalues $(\lambda_1,\lambda_2)=(20,1)$,$(9.144,-0.971)$ and $(-4,-3)$, respectively. Moreover, the associated $(q,\omega_{eff})$ are $(1,\frac{1}{3})$, $(0.514,0.00946)$ and $(-1,-1)$ respectively. Hence, the equilibrium point $A_3$ is unstable , $B_3$ is saddle and $C_3$ is stable, representing the radiation, matter, and de-Sitter phase, respectively. It is evident from the direction of phase-space trajectories (see the right panel of the Figure \eqref{f4}), the evolutionary trajectory of the model emerges from the radiation epoch and then converges to the de-Sitter accelerated epoch via passing through the matter-dominated epoch i.e. having the evolution phase $A_3 \rightarrow B_3 \rightarrow C_3$.  The evolutionary description of the cosmological parameters shown in Figures \eqref{f5} and \eqref{f6} (right panel) reflects the same.

\section{Conclusions}\label{cha5sec4}

The study of modified gravity incorporating non-metricity has garnered substantial attention in recent times, spanning various contexts such as black holes, wormholes, late-time observational constraints, and more. Meanwhile, viscous fluid cosmology has gained prominence for its explanation of the early stages of the Universe and for its approach to late-time acceleration. In this work, we have explored the function of the viscosity coefficient in the progression of cosmic evolution within the coincident $f(Q)$ gravity formalism. We begin with a non-linear function, specifically $f(Q)=-Q+ \Psi(Q) = -Q +\alpha Q^n$, where $\alpha$ and $n$ are arbitrary model parameters, and a newly proposed parameterization of the viscosity coefficient $\zeta$, specifically $ \zeta=\Bar{\zeta}_0 {\Omega^s_m} H $, where $\bar{\zeta}_0 = \frac{\zeta_0}{{\Omega^s_{m_0}}} $. The power law model can describe the desired thermal history of the Universe, both at the background and perturbation levels \cite{WOM}. We obtained the set of autonomous differential equations by invoking the dimensionless density parameters as the governing phase-space variables corresponding to the assumed generic $f(Q)$ function. The proposed functional form involves a power law correction to the STEGR scenario and is of considerable importance in both early- and late-time cosmological contexts \cite{R49}. Specifically, the $f(Q)$ function under consideration with the value $n > 1$ can provide modifications to the early Universe phenomenon, whereas the value $ n < 1 $ can provide modifications to the late-time cosmological phenomenon, potentially influencing the emergence of DE. Moreover, the assumed parameterization of the viscosity coefficient encompasses widely recognized models, $ \zeta=\zeta(H)$ for the case $ s=0$ and $ \zeta \sim {{\rho}_m}^{\frac{1}{2}} $ for the case $s = \frac{1}{2}$. Thus, for our analysis, we consider two toys that incorporate both corrections to the STEGR case with the aforementioned choices of the exponent $s$.

We consider the parameter choices for model I (i.e. $f(Q)=-Q+\alpha Q^{-1}$ ) as $(s,\bar{\zeta}_0) = (0,0.1), (0.5,0.5)$ and $(1.05,0.01)$, while for model II (i.e. $f(Q)=-Q+\alpha Q^2$ ) as $(s,\bar{\zeta}_0) = (0,0.001), (0.5,0.005)$ and $(1.05,0.01)$. The corresponding critical points and their behavior are presented in Tables \eqref{Table-1} and \eqref{Table-2}, and the corresponding phase-space diagrams are presented in Figures \eqref{f1} and \eqref{f4}. Moreover, the evolutionary description of cosmological parameters such as density, deceleration, and the effective EoS is presented in Figures \eqref{f2} and \eqref{f3} for model I and in Figures \eqref{f5} and \eqref{f6} for model II. We found that corresponding to model I, we obtained a stable de-Sitter type or stable phantom type (depending on the choice of exponent $s$) accelerated expansion of the Universe with no transition epoch. Moreover, models with $n < 0$ have a smaller value of $f_{\sigma8}$ in the case of a perfect fluid distribution, and hence models with $n < 0$ are not favored by the data \cite{WOM}. The same is reflected in our analysis, as this case fails to describe cosmological epochs other than accelerated de-Sitter expansion. Further, for model II, we obtained the evolutionary phase from the radiation epoch to the accelerated de-Sitter epoch via passing through the matter-dominated epoch. Hence, we conclude that model I provides a good description of the late-time cosmology but fails to describe the transition epoch, whereas model II modifies the description in the context of the early Universe and provides a good description of the matter as well as radiation era along with the transition phase. In addition, the analysis and results of the present investigation favor the aforementioned results of the study \cite{R49} in the context of STEGR corrections. Moreover, one can reproduce the dynamics of $\Lambda$CDM by considering different $f(Q)$ functions, such as the exponential case \cite{e12}. Furthermore, one can always find a stable de-Sitter accelerated expansion for any choice of parameter $n$ $(n \neq 1)$ utilizing the non-trivial connections \cite{ADDP6}.


\chapter{Concluding remarks and future perspectives} 

\label{Chapter6} 

\lhead{Chapter 6. \emph{Concluding remarks and future perspectives}} 


 \clearpage
 
In this thesis, we have examined various bulk viscous cosmological models by taking different bulk viscous coefficients to effectively describe
the late-time evolution of the Universe within a non-Riemannian geometric framework, specifically modified symmetric teleparallel gravity formulated in terms of non-metricity. A chapter-by-chapter summary of the analyzes performed and the key results obtained is presented below.

\section{Concluding remarks}

In chapter \ref{Chapter1}, we begin with an introductory overview of the foundations of GR, where gravitation is interpreted as the geometry of curved spacetime. We then review several important cosmological solutions within GR and discuss the strengths and limitations of the standard cosmological model. Furthermore, we introduce the basic concepts of non-Riemannian geometric frameworks, including teleparallel and symmetric teleparallel gravity, along with their extended formulations. Finally, we discuss the introduction of bulk viscosity in cosmology.
 
In chapter \ref{Chapter2}, we analyze the power law $f(Q)$ model along with the viscosity coefficient $\zeta= \zeta_0 \rho H^{-1} + \zeta_1 H $ in the framework of $f(Q)$ gravity. In addition, we derived the analytical solutions of the corresponding field equations for a flat FLRW metric. The free parameters of the obtained solutions have been constrained using the CC+Pantheon+SH0ES sample. We performed Bayesian statistical analysis to estimate the posterior probability utilizing the likelihood function and the MCMC random sampling technique. Moreover, we performed a statistical analysis using the AIC and BIC to assess the robustness of our MCMC analysis. We obtain $\Delta AIC=0.515 $ and $\Delta BIC=15.856 $. We also examined the evolutionary behavior of some prominent cosmological parameters. The effective EoS parameter predicts the accelerating behavior of the expansion phase of the Universe. The value of the EoS parameter in the present redshift $(z=0)$ is $\omega_0 \approx -0.7378 $ corresponding to the combined CC+Pantheon+SH0ES sample. Further, we investigate the behavior of the $f(Q)$ model using parameters $r-s$ and $Om(z)$ and find that our assumed viscous fluid $f(Q)$ model embodies quintessence-like behavior and can successfully describe the late-time scenario.

In chapter-\ref{Chapter3}, we consider the power law model $f(Q)$, specifically $f(Q)=\alpha Q^n$, along with a cosmic matter fluid that has viscous effects with the transport coefficient $\zeta=\zeta_0 \sqrt{\Omega}+\zeta_1 \Omega H$. In addition, we obtained the analytic solution for the field equations. The estimated mean values of the free parameters in the analytic solution are obtained by using the combined CC+Pantheon+ dataset. In addition, we investigate the asymptotic nature of the assumed viscous cosmological $f(Q)$ model by invoking phase-space analysis. We conclude that the assumed viscous $f(Q)$ model with the
obtained constraints on parameters successfully predicts an evolution of the Universe from
a matter-dominated decelerated epoch to stable accelerated de-Sitter epoch. Thus, the considered
cosmological scenario can be a good alternative to describe the late-time phenomenon of the
Universe.

In chapter-\ref{Chapter4}, we examined the evolution of a flat FLRW Universe filled with non-relativistic bulk viscous matter within the framework of linear $f(Q)$ gravity, specifically $f(Q)=\alpha Q$. The bulk viscosity was assumed to be time dependent and parameterized as $\zeta =\zeta_{0}+\zeta_{1}\left( \frac{\dot{a}}{a}\right) +\zeta_{2}\left( \frac{{\ddot{a}}}{\dot{a}}\right) =\zeta_{0}+\zeta_{1}H+\zeta_{2}\left( \frac{\dot{H}}{H}+H\right) $, with its components proportional to the expansion rate and acceleration of the Universe. We derived analytical solutions for the model and subsequently constrained the free parameters using observational data, including $57$ $H(z)$ measurements, $1048$ Pantheon supernova samples, and six BAO data points. The resulting parameter constraints indicate that the proposed model is in good agreement with the observational data.
Furthermore, by employing phase-space analysis, we showed that the viscous fluid model describes a Universe that evolves from a matter-dominated decelerated phase, acting as a past attractor to a stable de-Sitter accelerated phase, which serves as a future attractor. In addition, we examined the statefinder $r-s$ diagnostics for all three cases and found that the viscous cosmological model resides in the quintessence regime and asymptotically approaches the $\Lambda$CDM scenario at late times. Lastly, we have considered a non-linear model $f(Q)=-Q+\beta Q^2$ and then we analyzed the behavior of the model using a dynamical approach and found that there is only one de-Sitter type future attractor for the case $\bar{\beta}=-1$ whereas two future attractors correspond to $\bar{\beta}=1$. Moreover, the dynamics of the case $\bar{\beta}=0$  is the same as that of case III. We conclude that the
late-time behavior of the considered non-linear model $f(Q)=-Q+\beta Q^2$ with $\beta \le 0$ is similar to the linear case, whereas for the case $\beta \ge 0$ the results are quite different.

In chapter-\ref{Chapter5}, we have explored the function of the viscosity coefficient in the progression of cosmic evolution within the coincident $f(Q)$ gravity formalism. We begin with a non-linear function, specifically $f(Q)=-Q+\Psi(Q)=-Q+\alpha Q^n$, along with the bulk viscous coefficient $\zeta=\bar{\zeta}_0 {\Omega^s_m}H $, where $\bar{\zeta}_0 = \frac{\zeta_0}{{\Omega^s_{m_0}}} $. 
We consider the parameter choices for model I (i.e. $f(Q)=-Q+\alpha Q^{-1}$) as $(s,\bar{\zeta}_0) = (0,0.1), (0.5,0.5)$ and $(1.05,0.01)$, while for model II (i.e. $f(Q)=-Q+\alpha Q^2$ ) as $(s,\bar{\zeta}_0) = (0,0.001), (0.5,0.005)$ and $(1.05,0.01)$. We found that corresponding to model I, we obtained the stable de-Sitter type or stable phantom type (depending on the choice of exponent $s$) accelerated expansion of the Universe with no transition epoch. Further, for model II, we obtained the evolutionary phase from the radiation epoch to the accelerated de-sitter epoch via passing through the matter-dominated epoch. Hence, we conclude that model I provides a good description of the late-time cosmology but fails to describe the transition epoch, whereas model II modifies the description in the context of the early Universe and provides a good description of the matter as well as the radiation era along with the transition phase.

\section{Future perspectives}

This study investigates the role of bulk viscosity in cosmic evolution within modified gravity theories. Current research, including the work presented here, demonstrates that incorporating bulk viscosity into these modified frameworks can effectively account for the observed accelerated expansion without requiring a cosmological constant $\Lambda$. Future investigations could productively explore inflationary models, Big Bang nucleosynthesis bounds, and the formation of large-scale structures within this combined framework of modified gravity and bulk viscous effects.






\addtocontents{toc}{\vspace{2em}} 

\backmatter


\label{References}
\lhead{\emph{References}}


\begin{thebibliography}{100}




































\bibitem{B2} S. Carroll, ``Spacetime and Geometry: An Introduction to GR" (\textit{Addison Wesley, Boston}) (2004).

\bibitem{R26} A. Einstein, \textit{Ann. Phys.} \textbf{354}, 769(1916).

\bibitem{R27}H. P. Robertson, \textit{Astrophys. J.} \textbf{82}, 284 (1935).

\bibitem{R28} A. G. Walker, \textit{Proc. Lon. Math. Soc.} \textbf{42}, 90127 (1937).

\bibitem{R29} A. Einstein, \textit{Sitzber. Preuss. Akad. Wiss.} \textbf{142}, 87 (1917).

\bibitem{R30} A. Einstein, \textit{Sitzber. Preuss. Akad. Wiss.} \textbf{235}, 37 (1931).

\bibitem{R31} Y. B. Zel\'dovich, \textit{JETP Lett.} \textbf{6}, 316 (1967).

\bibitem{R32} S. Weinberg, \textit{Phys. Rev. Lett.} \textbf{59}, 2607 (1987).

\bibitem{R2} Planck Collaboration, N. Aghanim, et al., \textit{Astron. Astrophys.} \textbf{641}, 67 (2020).

\bibitem{R33} E. Abdalla et al., \textit{J. High En. Astrophys.} \textbf{2204}, 002 (2022).

\bibitem{R34} CANTATA collaboration, ``Modified Gravity and Cosmology: An Update by the CANTATA Network" (\textit{Springer})  (2021).

\bibitem{R35} D. M. Scolnic et al., \textit{Astrophys. J.} \textbf{938}, 113 (2022). 

\bibitem{R36} J. B. Jiménez, L. Heisenberg, and T. S. Koivisto, \textit{Universe} \textbf{5}, 173 (2019).

\bibitem{R37} T. Ortin, ``Gravity and Strings" (\textit{Cambridge University Press}) (2015).

\bibitem{R38} D. Iosifidis, \textit{Class. Quant. Grav.} \textbf{36}, 085001 (2019).

\bibitem{R39} A. Einstein, \textit{Sitzber. Preuss. Akad. Wiss.} \textbf{17}, 217-221 (1928).

\bibitem{R40} R. Weitzenb$\ddot{o}$ck, ``Invarianten Theorie" (\textit{Nordhoff, Groningen}) (1923).

\bibitem{R41} J. W. Maluf, \textit{Ann. Phys.} \textbf{525}, 339 (2013).

\bibitem{R42} H. A. Buchdahl, \textit{Mon. Not. Roy. Astro. Soc.} \textbf{150}, 1 (1970).

\bibitem{R43} A. A. Starobinsky, \textit{Phys. Lett. B} \textbf{91}, 99 ( 1980).

\bibitem{R44} R. Ferraro, F. Fiorini, \textit{Phys. Rev. D} \textbf{75}, 084031 (2007).

\bibitem{R45} N. Tamanini, C. G. Boehmer, \textit{Phys. Rev. D} \textbf{86}, 044009 (2012).

\bibitem{R46} M. Krssak, E. N. Saridakis, \textit{Class. Quant. Grav.} \textbf{33}, 115009 (2016). 

\bibitem{R47} Y. F. Cai et al., \textit{Rept. Prog. Phys.} \textbf{79}, 106901 (2016).

\bibitem{R47aa} T. P. Sotiriou, B. Li, J. D. Barrow, \textit{Phys. Rev. D} \textbf{83}, 104030 (2011).

\bibitem{R47ab} B. Li, T. P. Sotiriou, J. D. Barrow, \textit{Phys. Rev. D} \textbf{83}, 064035 (2011).

\bibitem{R47ac} A. Paliathanasis, J. D. Barrow, P.G.L. Leach, \textit{Phys. Rev. D} \textbf{94}, 023525 (2016).

\bibitem{R48} F. W. Hehl et al., \textit{Phys. Rep.} \textbf{258}, 1 (1995).

\bibitem{R49} J. B. Jim\'enez et al., \textit{Phys. Rev. D} \textbf{101}, 103507 (2020).

\bibitem{R50} J. M. Nester, H. J. Yo, \textit{Chin. J. Phys.} \textbf{37}, 113 (1999).

\bibitem{R51} J. B. Jim\'enez, L. Heisenberg,  T. Koivisto, \textit{Phys. Rev. D} \textbf{98}, 044048 (2018).

\bibitem{IB-1} I. Brevik, \textit{Entropy} \textbf{14}, 2302 (2012).

\bibitem{IB-2} I. Brevik,  O. Gron, \textit{Astrophys. Space Sci.} \textbf{347}, 399 (2013).

\bibitem{IB-3} I. Brevik et al., \textit{Int. J. Mod. Phys. D} \textbf{26}, 1730024 (2017).

\bibitem{IB-4} I. Brevik, A. N. Makarenko,  A. V. Timoshkin, \textit{Int. J. Geom. Methods Mod.} \textbf{16}, 1950150 (2019)

\bibitem{IB-5} I. Brevik,  B. D. Normann, \textit{Symmetry} \textbf{12}, 1085 (2020).

\bibitem{JM} N. D. J. Mohan, A. Sasidharan,  T. K. Mathew, \textit{Eur. Phys. J. C} \textbf{77}, 849 (2017).

\bibitem{AVS} A. V. Astashenok, S. D. Odintsov,  A. S. Tepliakov, \textit{Nucl. Phys. B} \textbf{974}, 115646 (2022).

\bibitem{MAT} A. Sasidharan,  T. K. Mathew, \textit{Eur. Phys. J. C} \textbf{75}, 348 (2015).

\bibitem{C.E.} C. Eckart, \textit{Phys. Rev.} \textbf{58}, 919 (1940).

\bibitem{W.I.} W. Israel, J. M. Stewart, \textit{Phys. Lett. B} \textbf{58},
213 (1976).

\bibitem{W.I.-2} W. Israel, \textit{Ann. Phys.} (N.Y.) \textbf{100}, 310
(1976).

\bibitem{W.I.-3} W. Israel, J. M. Stewart, \textit{Proc. R. Soc. Lond. B} 
\textbf{365}, 43 (1979).


\bibitem{R52} M. Li, X.-D. Li, S. Wang, Y. Wang, ``Dark Energy" (\textit{Peking University Press-World Scientific Advanced Physics Series} (2014).

\bibitem{R53} J. A. Snyman, ``Practical Mathematical Optimization: An Introduction to Basic optimization Theory and Classic and New Gradient Based Algorithms" (\textit{Springer}) (2005).

\bibitem{R54} K. Pearson, \textit{Nature} \textbf{72}, 294 (1905).

\bibitem{R55} D. MacKay, ``Information Theory, Inference, and Learning Algorithms" (\textit{Cambridge University Press}) (2003).

\bibitem{R56} D. F. Mackey et al., 	\textit{Publ. Astron. Soc. Pac.} \textbf{125}, 306 (2013).

\bibitem{da49} S. Bahamonde et al., 	\textit{Phys. Reports} \textbf{775}, 1 (2018).

\bibitem{da50} S. Strogatz., “Nonlinear dynamics and chaos: with applications to physics, biology,
chemistry, and engineering,” \textit{CRC Press} (2018).

\bibitem{da51} S. Wiggins., “Introduction to applied nonlinear dynamical systems and chaos,”  vol. 2, \textit{Springer Science $\And$ Business Media,} (2003).


















\bibitem{m1} Y. Xu et al., \textit{Eur. Phys. J. C} \textbf{79}, 708 (2019).

\bibitem{m2} Y. Xu et al., \textit{Eur. Phys. J. C} \textbf{80}, 449 (2020).

\bibitem{FOPK} A. Hazarika et al., \textit{Phys. Dark Univ.} \textbf{50}, 102092 (2025).

 \bibitem{ADD1} O. Sokoliuk,  A. Baransky, \textit{Astron. Nachr.} \textbf{343}, 220003 (2022). 

\bibitem{ADD2} A. Dixit, D. C. Maurya, A. Pradhan, \textit{Int. J. Geom. Methods Mod. Phys.} \textbf{19}, 12 (2022).

\bibitem{ADD3} M. M. Gohain, K. Bhuyan, \textit{Phys. Dark Univ.} \textbf{43}, 101424 (2024).

\bibitem{ADD4} M. Koussour et al., \textit{Chin. J. Phys.} \textbf{90}, 97 (2024).

\bibitem{ADD5} M. Koussour et al., \textit{Phys. Dark Univ.} \textbf{45}, 101527 (2024).

\bibitem{ADD6} M. Koussour et al., \textit{Mod. Phys. Lett. A} \textbf{39}, 2450023 (2024).

\bibitem{Hohmann2} M. Hohmann,  \textit{Phys. Rev. D} \textbf{104}, 124077 (2021). 

\bibitem{fQfT1} F. D'Ambrosio, L. Heisenberg, S. Kuhn, \textit{Class. Quantum Grav.} \textbf{39}, 025013 (2022).

\bibitem{LGG} L. Gomez, G. Palma, A. Rincon, N. Cruz, E. Gonzalez, \textit {Eur. Phys. J. Plus} \textbf{138}, 738 (2023).

\bibitem{ADD10} V. A Pai,  T. K. Mathew, \textit{Class. Quantum Grav.} \textbf{41}, 31 (2024).

\bibitem{LZ} R. Lazkoz et al., \textit{Phys. Rev. D} \textbf{100}, 104027 (2019)

\bibitem{RSU} R. Solanki et al., \textit{Universe} \textbf{9(1)}, 12 (2023).

\bibitem{Mackey/2013} D. F. Mackey et al., \textit{Publ. Astron. Soc. Pac.} \textbf{125}, 306(2013).

\bibitem{RS} R. Solanki et al., {\it Phys. Dark Univ.} {\bf 32}, 100820 (2021).


\bibitem{cantata} CANTATA collaboration, Modified Gravity and Cosmology: An
Update by the CANTATA Network, 2105.12582.

\bibitem{modifiedgrav} Timothy Clifton et al., \textit{Physics Reports} 
\textbf{513}, 1 (2012).




\bibitem{ADDP1} D. A. Gomes et al., \textit{Phys. Rev. Lett.} \textbf{132}, 141401 (2024).

\bibitem{ADDP2} L. Heisenberg, M. Hohmann, and S. Kuhn, arXiv:2311.05495[gr-qc] (2023).

\bibitem{Hohmann1} M. Hohmann et al, \textit{Phys. Rev. D} \textbf{99}, 024009 (2019).




\bibitem{jimenez2} J.B. Jim\'enez, L. Heisenberg, T. S. Koivisto, \textit{JCAP} {\bf 1808}, 039 (2018).

\bibitem{lcdm} F.K. Anagnostopoulos, S. Basilakos, E. N. Saridakis, \textit{Phys. Lett. B}, \textbf{822} (2021).

\bibitem{accfQ3} L. Atayde, N. Frusciante, \textit{Phys. Rev. D} \textbf{104}, 6 (2021).



\bibitem{fQfT3} S. Capozziello, V. De Falco,  C. Ferrara, \textit{Eur. Phys. J. C} \textbf{82}, 865 (2022).

\bibitem{gde} B. J. Theng, T. H. Loo, A. De, \textit{Chin. J. Phys.} \textbf{77}, 1551 (2022).

 \bibitem{ad/ec} A. De,  L.T. How, \textit{Phys. Rev. D}, \textbf{106},
 048501 (2022).

\bibitem{ad/bianchi} A. De, T. H. Loo, \textit{Class. Quantum Grav.} \textbf{40}, 115007 (2023).

\bibitem{signa} N. Frusciante, \textit{Phys. Rev. D} \textbf{103}, 0444021
(2021).

 \bibitem{perturb} W. Khyllep, A. Paliathanasis,  J. Dutta, \textit{Phys.
 Rev. D} \textbf{103}, 103521 (2021).






\bibitem{vs2} A. Sasidharan et al., \textit{Eur. Phys. J. C} \textbf{78}, 628 (2018)

\bibitem{vsform} J. Yang, R. H. Lin, X. H. Zhai,  \textit{Eur. Phys. J. C} \textbf{82}, 1039 (2022).

\bibitem{ADDP5} A. Paliathanasis, \textit{Gen. Relativ. Gravit.} \textbf{55}, 130 (2023).

\bibitem{ADDP6} A. Paliathanasis, \textit{Phys. Dark Univ.} \textbf{41}, 101255 (2023).

\bibitem{ADDP7} A. Paliathanasis, \textit{Phys. Dark Univ.} \textbf{42}, 101355 (2023).

\bibitem{CYU} H. Yu, B. Ratra,  F.Y. Wang, \textit{Astrophys. J.} \textbf{856}, 3 (2018)

\bibitem{Brout} D. Brout et al., \textit{Astrophys. J.} \textbf{938}, 110 (2022).

\bibitem{CPCP1} C. P. Singh,  P. Kumar, \textit{Eur. Phys. J. C} \textbf{74}, 3070 (2014).

\bibitem{CPCP2} G.C. Samanta,  R. Myrzakulov, \textit{Chin. J. Phys.} \textbf{55}, 1044(2017).

\bibitem{CPCP3} L. Heisenberg, \textit{Phys. Rep.} \textbf{1066}, 1(2024).

\bibitem{O1} N. Aghanim et al., \textit{Astron. Astrophys.} \textbf{641}, A6 (2020).

\bibitem{O2} G. Hinshaw et al., \textit{Astrophys. J. Suppl.} \textbf{208}, 19 (2013).

\bibitem{O3} M. M. Ivanov, M. Simonovic, M. Zaldarriaga, \textit{J. Cosmol. Astropart. Phys.} \textbf{05}, 042 (2020).

\bibitem{O4} K. Wang, Q. G. Huang, \textit{J. Cosmol. Astropart. Phys.} \textbf{06}, 045 (2020).

\bibitem{O5} T. M. C. Abbott et al. \textit{Phys. Rev. D} \textbf{98}, 043526 (2018).

\bibitem{O6} M. A. Troxel et al. \textit{Phys. Rev. D} \textbf{98}, 043528 (2018).

\bibitem{O7} D. M. Scolnic, et al., \textit{Astrophys. J.} \textbf{859}, 101 (2018).


\bibitem{SCAPE} S. Capozziello,  R. D'Agostino, \textit{Phys. Lett. B} \textbf{832}, 137229 (2022).

\bibitem{RCN} R. D'Agostino,  R. C. Nunes, \textit{Phys. Rev. D} \textbf{106}, 124053 (2022).

\bibitem{Riess} A.G. Riess et al., \textit{Astron. J.} \textbf{116}, 1009 (1998). 

\bibitem{Perlmutter} S. Perlmutter et al., \textit{Astrophys. J.} \textbf{517}, 565 (1999).


\bibitem{D.J.} D. J. Eisenstein et al., \textit{Astrophys. J.} \textbf{633},
560 (2005).

\bibitem{W.J.} W. J. Percival at el., \textit{Mon. Not. R. Astron. Soc.} 
\textbf{401}, 2148 (2010).

\bibitem{C.L.} C.L. Bennett et al., \textit{Astrophys. J. Suppl.} \textbf{148}, 119 (2003).

\bibitem{D.N.} D.N. Spergel et al., [WMAP Collaboration], {\it Astrophys. J. Suppl.} {\bf 148}, 175 (2003).

\bibitem{R.R.} R. R. Caldwell, M. Doran, \textit{Phys. Rev. D} \textbf{69},
103517 (2004).

\bibitem{Z.Y.} Z. Y. Huang et al., \textit{J. Cosmol. Astropart. Phys.} \textbf{0605}, 013 (2006).

\bibitem{COP} E. J. Copeland, M. Sami, S. Tsujikawa, \textit{Int. J. Mod. Phys. D} \textbf{15}, 1753 (2006).

\bibitem{L.A.} L. Amendola et al., {\it Phys. Rev. D} {\bf 75}, 083504 (2007).

\bibitem{SA} S. Appleby,  R. Battye, {\it Phys. Lett. B} {\bf 654}, 7 (2007).

\bibitem{SM1} S. Mandal et al., \textit{ Phys. Rev. D} {\bf 102}, 124029(2020).

\bibitem{SM2} S. Mandal et al., \textit{ Phys. Rev. D} {\bf 102}, 024057(2020).


\bibitem{e4} B. J. Barros et al., \textit{Phys.Dark Univ.} \textbf{30}, 100616 (2020). 

\bibitem{n1} I. Soudi et al.,  \textit{Phys. Rev. D} \textbf{100}, 044008 (2019). 

\bibitem{e7} F. D Ambrosio et al., \textit{Phys. Rev. D} \textbf{105}, 024042 (2022).

\bibitem{ND} N. Dimakis, A. Paliathanasis, T. Christodoulakis, \textit{Class. Quantum Grav.} \textbf{38}, 225003 (2021).


\bibitem{e8} G. Mustafa et al., \textit{Phys. Lett. B} \textbf{821}, 136612 (2021).

\bibitem{e1} F. K. Anagnostopoulos, S. Basilakos, E. N.Saridakis, \textit{Phys. Lett. B} \textbf{822}, 136634 (2021).

\bibitem{bb1} F. D. Paolis, M. Jamil, A. Qadir, \textit{Int. J. Theor. Phys.} \textbf{49}, 621 (2010).

\bibitem{hh1} K. Hu, T. Katsuragawa, T. Qiu  \textit{Phys. Rev. D} \textbf{106}, 044025 (2022). 

\bibitem{rs1} R. Solanki, P. K. Sahoo, \textit{Annalen Phys.} \textbf{534}, 2200076 (2022). 



\bibitem{ZHH} Z. Hassan, S. Mandal, P. K. Sahoo, \textit{Forts. Phys.} \textbf{69}, 2100023 (2021). 

\

 



\bibitem{DZ1} A. Sasidharan, T. K. Mathew, \textit{J. High Energy Phys.} \textbf{06}, 138 (2016).

\bibitem{DZ2} J. Ren, X. H. Meng, \textit{Phys. Lett. B} \textbf{633}, 1 (2006).

\bibitem{Planck} Planck Collaboration, \textit{Astron. Astrophys.} \textbf{641}, A6 (2020).  
\bibitem{Scolnic/2018} D.M. Scolnic et al., \textit{Astrophys. J.} \textbf{859}, 101(2018).

\bibitem{BMS} R. Kessler,  D. Scolnic, \textit{Astrophys. J.} \textbf{836},56 (2017).


\bibitem{BAO1} C. Blake et al., \textit{Mon. Not. Roy. Astron. Soc.} \textbf{418},
1707 (2011).

\bibitem{BAO6} R. Giostri et al., \textit{J. Cosmol. Astropart. Phys.} \textbf{1203},
027 (2012).

\bibitem{V.S.} V. Sahni et al., \textit {JETP Lett.} \textbf {77}, 201 (2003).













\bibitem{CANT} E. N. Saridakis et al., \textit{Modified Gravity and Cosmology: An Update by the CANTATA Network}, Springer (2021).

\bibitem{COSI} E. Abdalla et al., \textit{JHEAp} \textbf{34}, 49 (2022).

\bibitem{NEST} J. M. Nester, H. J. Yo, \textit{Chin. J. Phys.} \textbf{37}, 113 (1999).



\bibitem{BARR} B. J. Barros, T. Barreiro, T. Koivisto, N. J. Nunes, \textit{Phys. Dark Univ.} \textbf{30}, 100616 (2020).

\bibitem{ANAG} F. K. Anagnostopoulos, S. Basilakos, E. N. Saridakis, \textit{Phys. Lett. B} \textbf{822}, 136634 (2021).

\bibitem{NUNES} J. Ferreira, T. Barreiro, J. Mimoso, N. J. Nunes, \textit{Phys. Rev. D} \textbf{105}, 123531 (2022).

\bibitem{NEOM} L. Atayde, N. Frusciante, \textit{Phys. Rev. D} \textbf{107}, 124048 (2023).

\bibitem{RODR}  J. T. S. S. Junior,  M. E. Rodrigues, \textit{Eur. Phys. J. C} \textbf{83}, 475 (2023).


\bibitem{SNEHA} S. Pradhan, S. Mandal, P. K. Sahoo, \textit{Chin. Phys. C} \textbf{47}, 055103 (2023).

\bibitem{ZINNAT} Z. Hassan, S. Ghosh, P. K. Sahoo, K. Bamba, \textit{Eur. Phys. J. C} \textbf{82}, 1116 (2022).

\bibitem{CAPE-1} F. Bajardi, S. Capozziello, \textit{Eur. Phys. J. C} \textbf{83}, 531 (2023).

\bibitem{PALIA-1}  N. Dimakis, A. Paliathanasis, T. Christodoulakis, \textit{Class. Quantum Grav.} \textbf{38}, 225003 (2021).

\bibitem{ET} S. Sahlu, E. Tsegaye, \textit{arXiv}, arXiv:2206.02517 (2022).

\bibitem{ANAG-2}  F. K. Anagnostopoulos, V. Gakis, E. N. Saridakis, S. Basilakos, \textit{Eur. Phys. J. C} \textbf{83}, 58 (2023).

\bibitem{ANDER} A. Lymperis, \textit{JCAP} \textbf{11}, 018 (2022).

\bibitem{CAPE-2} S. Capozziello, M. Shokri, \textit{Phys. Dark Univ.} \textbf{37}, 101113 (2022).



\bibitem{DE-1} G. Subramaniam, A. De, T. H. Loo, Y. K. Goh, \textit{Fortschr. Phys.} \textbf{2023}, 2300038 (2023).

\bibitem{DE-2} G. Subramaniam, A. De, T. H. Loo, Y. K. Goh, \textit{Phys. Dark Univ.} \textbf{41}, 101243 (2023).

\bibitem{PALIA-2} N. Dimakis, M. Roumeliotis, A. Paliathanasis, P. S. Apostolopoulos, T. Christodoulakis,  \textit{Phys. Rev. D} \textbf{106}, 123516 (2022).


\bibitem{COPE} E. J. Copeland, A. R. Liddle, D. Wands, \textit{Phys. Rev. D} \textbf{57}, 4686 (1998).

\bibitem{DE-3} H. Shabani, Avik De, T. H. Loo, \textit{Eur. Phys. J. C}, \textbf{83}, 535 (2023).

\bibitem{WOM} W. Khyllep, J. Dutta, E. N. Saridakis, K. Yesmakhanova, \textit{Phys. Rev. D} \textbf{107}, 044022 (2023).

\bibitem{Mishra-2} S. A. Kadam, B. Mishra, J. L. Said, \textit{Eur. Phys. J. C} \textbf{82}, 680 (2022).

\bibitem{HAMID} H. Shabani,  M. Farhoudi, \textit{Phys. Rev. D} \textbf{88}, 044048 (2013).




\bibitem{NVV} N. Cruz et al., \textit{Phys. Dark Univ.}, \textbf{42},101351 (2023).

 
\bibitem{e12} C. G. Boehmer, E. Jensko, R. Lazkoz, \textit{Universe}, \textbf{9},  166 (2023).






























































































































































   
























 



















\end{thebibliography}
\cleardoublepage
\pagestyle{fancy}

\label{Publications}

\lhead{\emph{List of Publications}}
\chapter{List of Publications}
\section*{Thesis publications}
\begin{enumerate}

\item \textbf{Dheeraj Singh Rana} and P.K. Sahoo, ``Cosmological constraints in symmetric teleparallel gravity with bulk viscosity”, \textcolor{blue}{General Relativity and Gravitation} \textbf{56}, 82 (2024).

\item \textbf{Dheeraj Singh Rana}, Raja Solanki and P.K. Sahoo, ``Power Law $f(Q)$
 Cosmology with Bulk Viscous Fluid”, \textcolor{blue}{Annalen der Physik} \textbf{536}, 2400072 (2024).

\item  Raja Solanki,\textbf{Dheeraj Singh Rana}, Sanjay Mandal and P.K. Sahoo, ``Viscous Fluid Cosmology in Symmetric Teleparallel Gravity”, \textcolor{blue}{Fortschritte der Physik} \textbf{71},  2200202 (2023).

\item \textbf{Dheeraj Singh Rana}, Raja Solanki and P.K. Sahoo, ``Phase-space analysis of the viscous fluid cosmological models in the coincident $f(Q)$ gravity”, \textcolor{blue}{Physics of the dark Universe} \textbf{43}, 101421 (2024).

\end{enumerate}

\section*{Other publications}
\begin{enumerate}

\item \textbf{Dheeraj Singh Rana}, S.S Mishra, Aaqid Bhat, and P.K. Sahoo, ``Viscous Cosmology in $f(Q,L_m)$ Gravity: Insights from CC, BAO, and GRB Data”, \textcolor{blue}{Universe} \textbf{11}, 242 (2025).

\item \textbf{Dheeraj Singh Rana},  Raja Solanki and P.K. Sahoo, ``Phase-space analysis of the viscous fluid cosmological models in the coincident $f(T)$ gravity”, \textcolor{blue}{Annals of Physics} \textbf{492}, 170567 (2026).

\end{enumerate}

\cleardoublepage
\pagestyle{fancy}
\lhead{\emph{Conferences/Workshops/Schools}}

\chapter{Conferences/Workshops/Schools}
\section*{Papers presented}
 \begin{enumerate}
\item Presented research paper entitled “\textit{Phase-space analysis of the viscous fluid cosmological models in the coincident $f(Q)$ gravity}” at the “\textbf{International Conference on Gravitation, Astrophysics and Cosmology}”
jointly organized by \textbf{the Tensor Society of India and GLA University, Mathura}, during the period \textcolor{blue}{14th-16th June, 2024}.

\item Presented research paper entitled “\textit{Phase-space analysis of the viscous fluid cosmological models in the coincident $f(Q)$ gravity}” at the “\textbf{31st International Conference of International Academy of Physical Sciences}” organized by the \textbf{Pt. Ravishankar Shukla University, Raipur} during \textcolor{blue}{20th -21th December, 2024}.
 \end{enumerate}
 \section*{Attended}
 \begin{enumerate}
 
\item Attended the International Workshop on “\textit{Astronomical Data Analysis with Python}” organized by \textbf{Maulana Azad National Urdu University, Hyderabad, India} during \textcolor{blue}{5th - 8th  September, 2023}.

\item Participated in the International Conference on “\textit{89th Annual Conference of the Indian Mathematical Society}” organized by \textbf{Department of Mathematics, Birla Institute of Technology and Science, Pilani, Hyderabad Campus, Telangana, India} during \textcolor{blue}{22nd - 25th December, 2023}.

 \end{enumerate}

\cleardoublepage
\pagestyle{fancy}
\lhead{\emph{Biography}}

\chapter{Biography}

\section*{Brief biography of the candidate:}
\textbf{Mr. Dheeraj Singh Rana} earned his Bachelor's degree from the Pt. L.M.S Govt. Post Graduate College, Rishikesh in 2016, followed by a Master's degree in Mathematics from the same College in 2018.  He qualified for the Council of Scientific and Industrial Research (CSIR), the National Eligibility Test (NET) for Lectureship (LS) in 2021, the Junior Research Fellowship (JRF) in 2022, and the Graduate Aptitude Test in Engineering (GATE) in 2022. During his four-year research career, he has published 5 research articles in highly reputed international journals Physics of the Dark Universe, etc. In addition, he presented his work at several national conferences showcasing his dedication and contributions to the field of mathematics.

\section*{Brief biography of the supervisor:}
\textbf{ Prof. Pradyumn Kumar Sahoo} is a distinguished academic with over 24 years of experience in Applied Mathematics, Cosmology, Astrophysical Objects, the General Theory of Relativity and Modified Theories of Gravity. He earned his Ph.D. from Sambalpur University, Odisha, in 2004 and joined the Department of Mathematics at BITS, Pilani, Hyderabad Campus, in 2009 as an Assistant Professor. He currently serves as a Professor in the Department and is an Associate Member of the Inter-University Center for Astronomy and Astrophysics (IUCAA), Pune. In recognition of his academic contributions, he received the “Prof. S. Venkateswaran Faculty Excellence Award” from BITS, Pilani in 2022. In 2023, he was recognized as a Distinguished Referee for Europhysics Letters (EPL) by the European Physical Society (EPS) and named an Outstanding Reviewer for the Canadian Journal of Physics, highlighting his critical role in maintaining the accuracy and quality of scientific research. He was also awarded a visiting professor fellowship at Transilvania University of Brașov, Romania. A survey by researchers from Stanford University has ranked him among the top 2\% of scientists globally in the field of Nuclear and Particle Physics over the past five years. During his prolific career, Prof. Sahoo has published more than 270 research articles in reputed national and international journals and has participated in numerous conferences worldwide as an invited speaker. He has also visited the European Organization for Nuclear Research (CERN) in Geneva, Switzerland, as a visiting scientist. Beyond his research, he has led and contributed to multiple sponsored projects, including initiatives funded by the University Grants Commission (UGC), DAAD Research Internships in Science and Engineering (RISE) Worldwide, the Council of Scientific and Industrial Research (CSIR), the National Board for Higher Mathematics (NBHM) and the Anusandhan National Research Foundation (ANRF), under India’s Department of Science and Technology (DST). His active involvement extends to serving as an expert reviewer for government research projects and as an editorial board member for several prominent journals. Through his research collaborations at national and international levels, Prof. Sahoo has made significant contributions to advancing knowledge in his field.

\end{document}